\documentclass[aps,prx,twocolumn,superscriptaddress,nofootinbib,floatfix,letterpaper]{revtex4-1}
\usepackage{enumitem}
\usepackage{makecell}
\usepackage{array}

\newcolumntype{?}{!{\vrule width 1pt}}

\newcommand{\calU}{\mathcal{U}}
\newcommand{\ZZ}{\mathbb{Z}}
\newcommand{\Inn}{\operatorname{Inn}}

\renewcommand{\Vec}{{\cV\text{ec}}}
\newcommand{\Rep}{{\cR\text{ep}}}
\newcommand{\bulk}{\mathsf{bulk}}
\renewcommand{\tilde}[1]{\widetilde{#1}}

\usepackage{euscript}
\newcommand\eG{\EuScript{G}}
\newcommand\eM{\EuScript{M}}
\newcommand\sM{\mathsf{M}}

\begin{document}

\begin{titlepage}
\title{
Categorical symmetry and non-invertible anomaly\\
in symmetry-breaking and topological phase transitions
}

\author{Wenjie Ji}
\affiliation{Department of Physics, Massachusetts Institute of
Technology, Cambridge, Massachusetts 02139, USA}

\author{Xiao-Gang Wen}
\affiliation{Department of Physics, Massachusetts Institute of
Technology, Cambridge, Massachusetts 02139, USA}

\begin{abstract} 

For a zero-temperature Landau symmetry breaking transition in $n$-dimensional
space that completely breaks a finite symmetry $G$, the critical point at the
transition has the symmetry $G$.  In this paper, we show that the critical
point also has a dual symmetry -- a \emph{$(n-1)$-symmetry} described by a
higher group when $G$ is Abelian or an \emph{algebraic $(n-1)$-symmetry} beyond
higher group when $G$ is non-Abelian. In fact, any $G$-symmetric system, when
restricted to symmetric sub-Hilbert space, can be viewed as a boundary of
$G$-gauge theory in one higher dimension. The conservation of gauge charge and
gauge flux in the bulk $G$-gauge theory gives rise to the symmetry and the dual
symmetry respectively.  So any $G$-symmetric \emph{system} actually has a
larger symmetry called \emph{categorical symmetry}, which is a combination of
the symmetry and the dual symmetry.  However, part (and only part) of the
categorical symmetry must be spontaneously broken in any gapped phase of the
system, but there exists a gapless state where the categorical symmetry is not
spontaneously broken. Such a gapless state corresponds to the usual critical
point of Landau symmetry breaking transition.  The above results remain valid
even if we expand the notion of symmetry to include \emph{higher symmetries}
and \emph{algebraic higher symmetries}.  Thus our result also applies to
critical points for transitions between topological phases of matter.  In
particular, we show that there can be several critical points for the
transition from the 3+1D $Z_2$ gauge theory to a trivial phase.  The critical
point from Higgs condensation has a categorical symmetry formed by a $Z_2$
0-symmetry and its dual -- a $Z_2$ 2-symmetry, while the critical point of the
confinement transition has a categorical symmetry formed by a $Z_2$ 1-symmetry
and its dual -- another $Z_2$ 1-symmetry.

\end{abstract}


\maketitle

\end{titlepage}

{\small \setcounter{tocdepth}{1} \tableofcontents }

\section{Introduction}

Consider a Landau symmetry breaking transition \cite{L3726,L3745} in a quantum
system in $n$-dimensional space at zero temperature that completely breaks a
finite on-site symmetry $G$.  The critical point at the transition is a gapless
state with $G$-symmetry.  When $n=1$, it is well known that the 1+1D gapless
critical point has two decoupled sectors at low energies: right-movers and
left-movers \cite{Ght9108028,CFT12}.  Thus the critical point has a low energy
emergent symmetry $G\times G$.  In this paper, we would like to show that a
similar symmetry ``doubling'' phenomenon also appears for critical points of
Landau symmetry breaking phase transitions in all other dimensions.   In this
paper we will use $n$d to refer to space dimensions and $(n+1)$D to refer to
spacetime dimensions.

In general, the quantum critical point always connects two $n$d phases: a
symmetric phase with no ground state degeneracy and a symmetry breaking phase
with $|G|$ degenerate ground states.  The symmetric phase is characterized by
its point-like excitations that carry irreducible representations of $G$.  The
collection of those point-like excitations, plus their trivial braiding and
non-trivial fusion properties, give us a \emph{local fusion $n$-category}
denoted by $n\Rep(G)$ (see, for example,
\Ref{LW170404221,LW180108530,KZ200308898,KZ200514178} and Appendix \ref{cat}).
The conservation of those point-like excitations is encoded by the non-trivial
fusion of the irreducible representations $R_q$'s \begin{align} R_{q_1} \otimes
R_{q_2} = \bigoplus_{q_3} N_{q_1q_2}^{q_3} R_{q_3}, \end{align} which reflects
the $G$-symmetry.  In fact, due to Tannaka duality, the local fusion
$n$-category $n\Rep(G)$ completely characterizes the symmetry group $G$.  Since
the critical point touches the symmetric phase, the ground state at the
critical point is also symmetric under $G$.

The symmetry breaking phase has ground states labeled by the group elements
$g\in G$, $|\Psi_g\>$, that is not invariant under the symmetry transformation:
$U_h |\Psi_g\>=|\Psi_{hg}\>,\ g,h \in G$.  We can always consider a symmetrized
ground state $|\Psi_0\> =\sum_g |\Psi_g\>$ that is invariant under the symmetry
transformation: $U_h |\Psi_0\>=|\Psi_0\>$.  The symmetry breaking phase has
domain wall excitations and we only consider the symmetrized states with domain
walls. The domain walls can also fuse in a non-trivial way, which form a local
fusion $n$-category, denoted by $n\Vec_G$ (see \Ref{KZ200308898,KZ200514178} and
Appendix \ref{cat}).  The non-trivial fusion also leads to a ``conservation''
of domain-wall excitations.  

It turns out that the ``conservation'' of $(n-1)$-dimensional domain-wall
excitations can be viewed as a result of \emph{algebraic higher
symmetry}\cite{KZ200308898,KZ200514178} generated by closed string operators
$W_q(S^1)$ that commute with the lattice Hamiltonian
\begin{align}
 W_q(S^1) H = H W_q(S^1),
\end{align}
for all the closed loops $S^1$ (see Section \ref{ahsymmG}).  \Ref{KZ200514178}
conjectured that there is a generalization of Tannaka duality: a local fusion
$n$-category $\cR$ (such as $n\Vec_G$) completely characterizes an
\emph{algebraic higher symmetry} (in the present case, an \emph{algebraic
$(n-1)$-symmetry}).  When the symmetry group is Abelian, such an algebraic
$(n-1)$-symmetry is a \emph{$(n-1)$-symmetry} described by a higher group.
This special case was discussed in \Ref{GW14125148}.  When the symmetry group
is non-Abelian, such an algebraic $(n-1)$-symmetry is beyond  higher group
description and is not a $(n-1)$-symmetry.

To contrast higher symmetry (described by higher group) and algebraic higher
symmetry (beyond higher group), we note that for group-like higher symmetry,
the string operators satisfy a group-like algebra
\begin{align}
 W_{q_1}(S^1) W_{q_2}(S^1) = W_{q_1\cdot q_2}(S^1),
\end{align}
while for algebraic higher symmetry, the string operators satisfy a more
general algebra
\begin{align}
 W_{q_1}(S^1) W_{q_2}(S^1) = \sum_{q_3} N_{q_1q_2}^{q_3} W_{q_3}(S^1).
\end{align}

The notion of \emph{algebraic higher symmetry} has appeared in
\Ref{FSh0607247,DR10044725,CY180204445,TW191202817,KZ191213168} for 1+1D
conformal field theory (CFT) via non-invertible defect lines and for lattice
models in general dimensions in \Ref{KZ200308898,KZ200514178}. This generalizes the
notion of higher form symmetry
\cite{KT13094721,GW14125148,TK151102929,CT171104186,KR180505367,L180207747,W181202517,WW181211955,WW181211967}
(or higher symmetry
\cite{K032,W0303,LW0316,NOc0605316,NOc0702377,Y10074601,B11072707}).\footnote{Higher
form symmetry and higher symmetry are similar. They have only a small
difference: The action of higher form symmetry becomes an identity when acts on
contractable closed subspaces, while the action of higher symmetry may not be
an identity when acts on contractable closed subspaces. Higher symmetry is a
symmetry in a lattice model. Higher symmetry reduces to higher form symmetry in
gapped ground state subspaces (\ie in the low energy effective topological
quantum field theory).} A higher symmetry is described by a higher group, while
an algebraic higher symmetry is beyond higher groups.  The charged excitations
(the charge objects) of an algebraic higher symmetry is in general
characterized by a \emph{local fusion higher category} (see \Ref{KZ200514178} and
Appendix \ref{cat}).

We see that the $G$-symmetry breaking phase, with domain wall excitations forming $n\Vec_G$, actually has an algebraic
$(n-1)$-symmetry, denoted as $G^{(n-1)}$.  Since the critical point touches the
symmetry breaking phase, the critical point also has the algebraic
$(n-1)$-symmetry $G^{(n-1)}$.  Therefore, the gapless state, at the critical
point of $G$-symmetry breaking transition, has both the $G$ 0-symmetry and the
algebraic $(n-1)$-symmetry $G^{(n-1)}$.  We call this combined symmetry a
\emph{categorical symmetry}.  In fact, the categorical symmetry even appears
off the critical point, but in a spontaneously broken form.

There is a \emph{holographic way} to see the appearance of categorical symmetry
in any $G$-symmetric system. We note that when restricted to the symmetric
sub-Hilbert space, a $n$d $G$-symmetric system can be viewed as a system that
has a non-invertible gravitational anomaly, \ie can be viewed as the boundary
of the $G$-gauge theory in one higher dimension\cite{KW1458,JW190513279}. The bulk
gauge charges, when brought to the boundary, are the excitations of the $G$
$0$-symmetry, and the gauge fluxes, brought to the boundary, are excitations of
the algebraic $(n-1)$-symmetry.  The conservation of the gauge charges and
gauge fluxes in $G$-gauge theory in the bulk leads to the $G$ $0$-symmetry and
algebraic $(n-1)$-symmetry $G^{(n-1)}$ respectively, in our $G$-symmetric
system that corresponds to the boundary (for details, see Section \ref{IsTO}).
Therefore, the $G$-symmetric \emph{system} actually has a larger symmetry --
the \emph{categorical symmetry}, which is a combination of the symmetry (from
the conservation of gauge charges) and the algebraic $(n-1)$-symmetry (from the
conservation of gauge flux), with non-trivial mutual statistics between  gauge
charges and gauge flux.  Such a categorical symmetry is fully characterized by
the $G$-gauge theory in one higher dimension.

Since the gapped boundaries of $G$-gauge theory always come from the
condensation of the gauge charges and/or the gauge flux, the gapped ground
state of $G$-symmetric system always breaks the categorical symmetry partially,
either the $G$-symmetry, or the algebraic $(n-1)$-symmetry, or some other
mixtures of the two symmetries.  A state with the full categorical symmetry
(\ie both the $G$-symmetry and the algebraic $(n-1)$-symmetry) must be gapless.
We show that such a gapless state describes the critical point of the Landau
symmetry breaking transition.  

More generally, all possible gapless states in a $G$-symmetric system are
classified by gapless boundaries of $G$-gauge theory in one higher dimension.
This universal emergence of categorical symmetry at critical point, and its
origin from non-invertible gravitational anomaly (\ie topological order in one
higher dimension), may help us to systematically understand gapless states of
matter.

It is worthwhile to point out that a structure similar to categorical symmetry
was found previously in Anti-de Sitter/Conformal field theory (AdS/CFT)
correspondence\cite{M9831,Wh9802150,KWh9905104,HO181005338}, where a global
symmetry $G$ in a conformal field theory (CFT) at the boundary is related to an
appearance of a gauge theory of group $G$ in the bulk. In this paper, we stress
that the categorical symmetry encoded by the bulk is actually a bigger symmetry
than the usual symmetry at the boundary.  We point out that the $G$-symmetric
gapless critical theory at the $(n+1)$D  boundary actually have both the
0-symmetry $G$ and the dual algebraic $(n-1)$-symmetry, which together form the
categorical symmetry.  For a more detailed discussion, see Section \ref{sum}.

In the following, we begin with studying a few concrete examples, to show the
appearance of categorical symmetries in Landau symmetry breaking transitions
and in topological phase transitions, as well as neighboring gapped states that
partially break the categorical symmetries spontaneously.  In Section \ref{Z2models},  \ref{Z2modelsaly} and \ref{BB}, we discuss the
example of models associated with the $Z_2$ group, in $1+1$D, $2+1$D, without
and with 't Hooft anomaly. We introduce the patch operators, as a main tool to
detect local charges in sub-Hilbert space that is symmetric under global
symmetries.  We show how to use the patch operators to describe categorical
symmetry.  In Section \ref{ahsymmG}, we discuss the categorical symmetries in
the lattice models with general finite group $G$ in any $n$ spacial dimension.
In Section \ref{eAHS}, we discuss the emergence of algebraic higher symmetry.
In Section \ref{eCS}, we discuss the emergence of categorical symmetry for a
set of low energy excitations, and how the categorical symmetry can constrain
on the possible phases and phase transitions induced by those low energy
excitations.  In particular, we discuss how the emergent categorical symmetry
determines the duality relations between low energy effective field theories. In Section \ref{sum}, we summarize our results and point out its implication for a particular AdS/CFT dual.

\section{$Z_2$ and dual $\t Z_2$ symmetries in 1+1D Ising
model}\label{Z2models}

\subsection{Duality transformation in 1+1D Ising model}

A common scenario happening at the critical point between the symmetric phase and
the symmetry breaking phase is the emergent symmetry. The example we know the best
is the Ising transition in one dimension. The paramagnetic phase is $Z_2$
symmetric, and the ferromagnetic phase spontaneously breaks the $Z_2$ symmetry.
The critical point is $Z_2$ symmetric. More than that, it also has an
additional $Z_2$ symmetry
\cite{BPZ8433,ZF8515,CY180204445,VP190506969,JW190901425}. 

To see both $Z_2$ symmetries, we consider Ising model on a ring of $L$ sites,
where on each site $i$ there are spin-up and spin-down state in the Pauli-$Z_i$
basis. So the Hilbert space is $\cV=\otimes_{i=0}^{L-1}
\{|\uparrow_i\rangle,|\downarrow_i\rangle\}$. Each state in the Hilbert space
can be also labeled in an alternative way, that is via the absence or presence
of a domain wall (DW) in the dual lattice of $L$ links. On each link
$i+\frac{1}{2}$, a domain wall means the spins on $i$ and $i+1$ are
anti-parallel. It follows that each basis state $|\psi\rangle$ of $\cV$ and its
$Z_2$ partner $|\psi'\rangle$ are labeled by the same kink variable on the dual
lattice. So the $Z_2$ symmetric state in $\cV$ is labeled uniquely by the DW
variable.  Moreover, a configuration of odd number of domain walls on links
cannot be mapped to any configuration of spins on sites.  Thus each DW variable
with an even number of DW's labels a unique $Z_2$ symmetric state.  Therefore,
we say that the $Z_2$ symmetric Hilbert space of spins on the sites is in
one-to-one correspondence with the Hilbert space of even number of DWs on the
links. Each of the Hilbert space is of dimension $2^{L-1}$. 

Next, we demonstrate an isomorphism between a set of operators on sites and
that on links. 
\begin{align}\label{Z2gauging}
X_iI_{i+1}\rightarrow \tilde{X}_{i-\frac{1}{2}}\tilde{X}_{i+\frac{1}{2}},~~~Z_{i}Z_{i+1}\rightarrow \tilde{I}_{i-\frac{1}{2}} \tilde{Z}_{i+\frac{1}{2}}.
\end{align}
Here, we use the following notation,
\begin{align}
X=\begin{pmatrix}
0 & 1 \\ 1 & 0
\end{pmatrix},~~Z=\begin{pmatrix}
1 & 0 \\ 0 & -1
\end{pmatrix},~~I=\begin{pmatrix}
1 & 0 \\ 0 & 1
\end{pmatrix}.
\end{align}
Physically,
the spin-flipping $X_i$ is the same as creating two DW's on $i-\frac{1}{2}$ and
$i+\frac{1}{2}$ links, represented by  $
\tilde{X}_{i-\frac{1}{2}}\tilde{X}_{i+\frac{1}{2}}$. The Ising coupling term
$Z_iZ_{i+1}$ also measures the energy cost of a domain wall, which is
represented by $\tilde{Z}_{i+\frac{1}{2}}$. Formally, the two sets of operators
$\{X_i, Z_{i}Z_{i+1}\}$ and
$\{\tilde{X}_{i-\frac{1}{2}}\tilde{X}_{i+\frac{1}{2}},
\tilde{Z}_{i+\frac{1}{2}}\}$ are two representations of the same
set of operators $\{W_i, W_{i+\frac{1}{2}}\}$ defined by the operator algebra,
for $i=\frac{1}{2},1,\cdots, L$,
 \begin{align}\label{localopalg}
 \text{(i)}~~&W_{i}^2=1,~\nonumber \\
\text{(ii)}~~&W_iW_{i+\frac{1}{2}}=-W_{i+\frac{1}{2}}W_i,\\
\text{(iii)}~~&[W_i,W_j]=0,~ |i-j|\geq 1. \nonumber 
 \end{align}
We also have a further global condition, 
\begin{align}\label{globalopalg}
\begin{split}
U_{Z_2}=\prod_{i=1,2,\cdots,L}W_i=1\,,\\
U_{\tilde{Z}_2}=\prod_{i=\frac{1}{2},\frac{3}{2},\cdots L-\frac{1}{2}}W_i=1\,.
\end{split}
\end{align} 
 
We have two $2^{L-1}$ dimensional representations of $W_i$'s, satisfying these relations (\ref{localopalg}) and (\ref{globalopalg}). 
In particular, the following Hamiltonian,
\begin{align}
\label{HW}
H=-\sum_{i=1,2\cdots,L} \left(BW_i +JW_{i+\frac{1}{2}}\right),
\end{align}
has the same spectrum independent of the representation. In the ``spin representation'', the Hamiltonian reduces to 
\begin{align}
\label{HIs}
\begin{split}
&H^\text{Is}=-\sum_{i=1,2,\cdots,L} \left(BX_i+JZ_{i}Z_{i+1}\right),\\
&U_{Z_2}= \prod_{i=1,2,\cdots L}X_i=1,~~U_{\tilde{Z}_2}=Z_{L+1}Z_1=1.
\end{split}
\end{align}
In the ``DW representation'', the Hamiltonian reduces to 
\begin{align}
\label{HDW}
\begin{split}
&H^\text{DW}=-\sum_{i=1,2,\cdots,L} \left(B\tilde X_{i-\frac{1}{2}}\tilde{X}_{i+\frac{1}{2}}+J\tilde{Z}_{i+\frac{1}{2}}\right),\\
&U_{Z_2}= \tilde{X}_{L+\frac{1}{2}}\tilde{X}_\frac{1}{2}=1,~~U_{\tilde{Z}_2}=\prod_{i=1,2,\cdots L}\tilde{Z}_{i-\frac{1}{2}}=1.
\end{split}
\end{align}

The unitary transformation (\ref{Z2gauging}) between the ``spin
representation'' and  ``DW representation'' is also known as $Z_2$ gauging. In
the current case, it is also the same as Kramers-Wannier duality. 

\begin{figure}[t]
\includegraphics[scale=0.5]{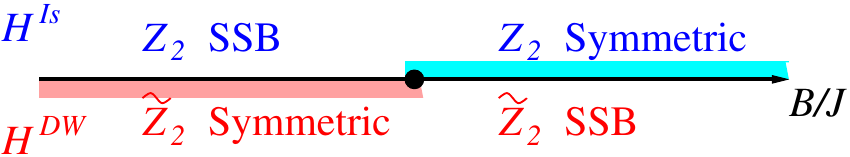}
\caption{
The same Ising model can be describe by $H^\text{Is}$ or by $H^\text{DW}$.  The
$Z_2$ symmetry is explicit in the $H^\text{Is}$ description, while the $\t Z_2$
dual symmetry is explicit in the $H^\text{DW}$ description.  The Ising model
has both the $Z_2$ symmetry and  $\t Z_2$ dual symmetry. The ground state
usually spontaneously breaks one of the symmetries, except at the $B=J$ critical
point, where both the symmetry and the dual symmetry (\ie the full $Z_2\vee \t Z_2$ categorical symmetry) are not  spontaneously
broken.
}
\label{IsDW}
\end{figure}

The Ising model has two exact $Z_2$ symmetries.  However, in our
spin and DW representations, we do not see them simultaneously.  In the spin
representation, we see one $Z_2$ symmetry generated by
\begin{align}
\label{UZ2}
U_{Z_2} = \prod_{i=1,\cdots,L} X_i,  
\end{align}
which is denoted as $Z_2$.  In the DW representation, we see the other $Z_2$
symmetry generated by
\begin{align}
\label{tUZ2}
U_{\t Z_2} = \prod_{i=0,\cdots,L} \t Z_{i+\frac12},  
\end{align}
which is denoted as $\t Z_2$.  $Z_2$ and $\t Z_2$ are two different $Z_2$
symmetries, as one can see from their different charge excitations.  Despite we
only see one symmetry in one formulation, the Ising model actually have both
symmetries.  The combination of the two symmetries is the so-called categorical
symmetry, which is denoted as $Z_2\vee \t Z_2$.  Certainly, the critical
model at  $|J|=|B|$ also have the $Z_2\vee \t Z_2$ categorical symmetry
(see Fig. \ref{IsDW}).  It is interesting to note that, in the ground state of
the Ising model, either $Z_2\vee \t Z_2$ categorical symmetry is
spontaneously broken partially (for example, one of the $Z_2$ is spontaneously
broken) or the ground state is gapless\cite{L190309028}.  This indicates that
$Z_2$ symmetry and the dual $\t Z_2$ symmetry are not independent.  There must
be a special relation between them. To reveal it, we need to discuss the charge excitations of the symmetries.

\subsection{Patch symmetry transformation}
\label{ps1}

In the above, we argue that the lattice model \eqn{HW} (or \eqn{HIs} or
\eqn{HDW}) in the symmetric sub-Hilbert space has two $Z_2$ symmetries generated by
$U_{Z_2}$ and $U_{\t Z_2}$.  But in the symmetric sub-Hilbert space, the two
operators are identity operator $U_{Z_2}=U_{\t Z_2}=1$.  The two $Z_2$
transformations are do-nothing transformations.  So what does it mean that
the lattice model in symmetric sub-Hilbert space has two $Z_2$ symmetries?  In
this section, we are going to solve this problem by introducing patch symmetry
transformations.  This is a new and better way to view symmetries in local
quantum systems. Such kind of operators have been studied mostly for continuous global symmetries, known in the literature \cite{Harlow:2018jwu} as \emph{splittable symmetry operators}, dating back to 1980s \cite{Doplicher:1982cv,Doplicher:1983if,Buchholz:1985ii}. Its one dimensional version also appears in \cite{L190309028}.

We notice that even though $U_{Z_2}$ and $U_{\tilde{Z}_2}$ act as identity in
the symmetric sub-Hilbert space, the sub-Hilbert space is not consisted of only a vacuum state. Rather, spin flip as well as domain wall excitations are still present, and they are subject to mod-2 conservations. So the effect of two $Z_2$ symmetries is still
there within the symmetric sub-Hilbert space.  For example, there exists a
state on the ring containing two well-separated spin excitations, each carrying
the $U_{Z_2}$-charge $1$ while the total $U_{Z_2}$-charge is $0$ mod 2.  How to
measure the local $U_{Z_2}$-charge via a symmetry transformation operator?

To
diagnose the conserved local charges of $U_{Z_2}$ and
$U_{\tilde{Z}_2}$ symmetries, subjective to conservation laws, we now introduce two sets of \emph{patch symmetry operators}\footnote{We hope the name is intuitive even when generalized to higher dimensions.} for
the simple model \eqn{HIs} or \eqn{HDW}.

The first patch symmetry is generated by the following transformations,
\begin{align}
\label{patchsym1}
U_{Z_2}(i,j)= \prod_{l=i}^j X_l,~~~ \text{for~~}j-i\geq 0,
\end{align}
which are required to act within the symmetric sub-Hilbert space.  They also
have the properties that for any $ j-i\geq 1, j'-i'\geq 0$,
\begin{align}
\begin{split}
&U_{Z_2}(i,k) = U_{Z_2}(i,j) U_{Z_2}(j+1,k) ,\\
&U_{Z_2}(i,j)^2 =1,\ \ \ [U_{Z_2}(i,j), U_{Z_2}(i',j')]=0.\\
\end{split}
\label{z2patchalg}
\end{align}

In condensed matter physics, a symmetry is simply a constraint on the lattice
Hamiltonian $H$.  We usually describe such a constraint as a constraint on the Hamiltonian $H$ as a whole. Under such constraint, the Hamiltonian $H$ is allowed to be local
or non-local.  This actually is a drawback of the standard formulation of the
symmetry, since its does not care about the locality of the Hamiltonian.  In
our new description of symmetry, using patch operators, we assume the
Hamiltonian $H$ to be sum of local operators $H=\sum_x O_x$.  Then the
constraint on the  Hamiltonian $H$ is expressed in terms of the constraint on
the local terms $O_x$.  In other words, a system is said to have the patch
symmetry, if it has the following properties: each local term in the
Hamiltonian
commutes with all the  patch symmetry operators, as long as the local term is
far away from the boundary of the patch operator.  For example, if $H_{l}$ is a
term on ${l,l+1}$, and if it commutes with all patch symmetry operators
$U_{Z_2}(i,j)$ with $i,j\neq l,l+1$, this term is symmetric under the patch
symmetry. If every term in the Hamiltonian has this property, we say the
Hamiltonian has the patch symmetry.  

Each patch symmetry operator $U_{Z_2}(i,j)$ measures the $Z_2$ spin excitations
in its ``bulk'', the sites covered by the patch, from site $i$ to site $j$. 
More precisely, a local operator $\psi_k$ carries $Z_2$-charge $1$
if it satisfies
\begin{align}
 U_{Z_2}(i,j) \psi_k = -\psi_k U_{Z_2}(i,j) , \ \ \ \ i \ll k \ll j.
\end{align}
$\psi_k \psi_l$ creates two  $Z_2$-charges at $k$ and $l$ within the symmetric
sub-Hilbert space. 

In \eqn{patchsym1}, the patch symmetry operator is given in the
spin-representation. In the DW representation, it will has a form
\begin{align}
\label{patchsym1DW}
 U_{Z_2}(i,j)=\tilde{X}_{i-\frac{1}{2}}\tilde{X}_{j+\frac{1}{2}},
\end{align}
which has a trivial ``body of the string'' but only two end points.  By the
definition given above about a system symmetric under the patch symmetry, and
due to the trivial bulk, any local Hamiltonian in the DW representation is
symmetric under $U_{Z_2}(i,j)$.
We may also take the patch operators
$U_{Z_2}(i,j)$ as creating a pair of $\tilde{Z}_2$-charged excitations (\ie a
pair of $Z_2$ domain walls) at the ending links $i-\frac{1}{2}$ and
$j+\frac{1}{2}$. 

The second patch symmetry is generated by the following operators,
in the spin- and DW-representations
\begin{align}
U_{\tilde{Z}_2} (i,j)=Z_{i}Z_j, \ \ \ \
U_{\tilde{Z}_2} (i,j)=\prod_{l=i}^{j-1}\tilde{Z}_{l+\frac{1}{2}},
\end{align}
which have the properties that they act within the symmetric sub-Hilbert space, for any $j-i\geq 1, j'-i'\geq 1$, 
\begin{align}
\begin{split}
U_{\t Z_2}(i,k) &= U_{\t Z_2}(i,j) U_{\t Z_2}(j,k) ,\\
U_{\tilde{Z}_2}(i,j)^2&=1, \ \ \
 [U_{\tilde{Z}_2}(i,j), U_{\tilde{Z}_2}(i',j')]=0.
\end{split}
\label{dualz2patchalg}
\end{align}

We see that any
local Hamiltonian in spin-representation has the $U_{\t Z_2}$ symmetry.
However, for the Hamiltonian in the ``DW representation'',  the second patch
symmetry gives rise to a non-trivial constraint on the Hamiltonian.
These symmetry operators serve to measure local domain wall excitations. We can
see this in the spin representation, \ie when $Z_i$ and $Z_j$ have opposite
sign, \ie there is domain between $i$ and $j$, then $U_{\tilde{Z}_2}
(i,j)=Z_{i}Z_j=-1$.

In summary, we identify two patch symmetries, each is generated by a set of
commuting operators.  The two kinds of patch symmetries act non-trivially even
in the symmetric sub-Hilbert space, and can impose constraints on Hamiltonians.
The symmetric Hamiltonians ensure the mod-2 conservation of the $Z_2$-charges
and domain walls.

The patch symmetry transformations also allow us to identify a special new
property -- the ``mutual statistics'' between charges of the two global symmetries (or patch
symmetries), given by the following relation, for $ i\ll i'\ll j\ll j'$,
\begin{align}
\label{mutualsta}
 U_{Z_2}(i,j) U_{\tilde{Z}_2}(i', j') &=- U_{\tilde{Z}_2}(i', j')U_{Z_2}(i,j). 
\end{align}
In other ways, a local charge under the $Z_2$ global symmetry is created at
each end point of a $\tilde{Z}_2$ patch operator.  If a $Z_2$ symmetry patch
and a $\tilde{Z}_2$ symmetry patch partially overlap, the single charge at one
end point can be measured by the $Z_2$ patch symmetry operator.  All such
statements remain true if exchanging $Z_2$ and $\tilde{Z}_2$. We call such
properties the ``mutual statistics'' between charges of the two patch
symmetries.  The collection of all  patch symmetries is nothing but the
\emph{categorical symmetry}. The property (\ref{mutualsta}) justifies the
symmetry to be $Z_2\vee\t Z_2$, rather than $Z_2\times \t Z_2$.

We see that the categorical symmetry in a system can be fully described by a
set of patch operators, without the need to go to one higher dimension.
Alternatively, later in section \ref{IsTO}, we describe the categorical
symmetry in terms of the topological order and the associated long-range
entanglement, by viewing the system  as a boundary of a topological order in
one higher dimension.  The above result confirms that the categorical symmetry
is indeed a property of the system itself.

Note that $\<U_{\tilde{Z}_2}(i, j)\>$ also turns out to be the correlation of order
parameters of the $Z_2$-symmetry, while $\<U_{Z_2}(i, j)\>$  happens to be the
correlation of order parameters of the $\t Z_2$-symmetry.  If there is a state,
where both the $Z_2$-symmetry and the  $\t Z_2$-symmetry are spontaneously
broken, then $U_{\tilde{Z}_2}(i, j)$ and $U_{Z_2}(i, j)$ behave like
$c$-numbers for the state. This will contradict with \eqn{mutualsta}.
Therefore, the $Z_2$-symmetry and the  $\t Z_2$-symmetry cannot be both
spontaneously broken. A more rigorous proof was given in \Ref{L190309028}.

\subsection{A model where both $Z_2$ symmetry and dual $\t Z_2$ symmetry are
explicit}

The Ising model in its spin representation \eqn{HIs} or in its DW
representation \eqn{HDW} only shows one of the $Z_2$ and dual $\t Z_2$
symmetries explicitly.  In this section, we will discuss the third representation
of the Ising model, where both the $Z_2$ and dual $\t Z_2$ symmetries appear
explicitly.

Consider a model with spin-up and down states defined on $N$ sites as well as on $N$ links. So we begin with $2^{2N}$ states. The model has the following Hamiltonian, 
\begin{align}
\label{HZ2Z2}
H=&
-\sum_{i} 
\left(B\tilde{X}_{i-\frac{1}{2}}X_i\tilde{X}_{i+\frac{1}{2}} +J \t Z_{i+\frac12}\right) \nonumber\\
&+
U (1-Z_{i}\t Z_{i+\frac12}Z_{i+1}) .
\end{align}
We only consider the low energy limit $U\rightarrow \infty$ limit, as well as under the
projection $U_{Z_2}=\prod_i X_i=1$. (Note that the $B$-term and the $J$-term commute with the constraint $U$-term and $U_{Z_2}$.) In the restricted low energy sector, we are left with $2^{N-1}$ states.
The above Hamiltonian is conventionally known as describing $Z_2$ matter field
coupled to $Z_2$ gauge field in 1+1D dual lattice.\footnote{ Here,
$\tilde{Z}_{i+\frac{1}{2}}$ and $\tilde{X}_{i+\frac{1}{2}}$ are matter field
and momenta on the sites of dual lattice.  $Z_{i}$ and $X_i$ are gauge field
and momenta on the links of dual lattice.  The Gauss law $Z_{i}\t
Z_{i+\frac12}Z_{i+1}=1$ is imposed as a dynamical constraint. } The Ising model \eq{HIs} with the $Z_2$ symmetric sub-Hilbert space turns out to be equivalent as the low energy effective theory
of the above model with $B,\ J>0$ and $U\rightarrow +\infty$. There are more than one way to proof this. In Appendix \ref{Z2equivmodels}, we give a proof using the stabilizer formalism in quantum information.  In the
next subsection, we will show the same Hamiltonian together with the same
projected sub-Hilbert space arises as the boundary theory of the $Z_2$
topological order. 

The $Z_2\times
\t Z_2$ symmetry (or more precisely, the $Z_2\vee \t Z_2$ categorical
symmetry) is explicit in the above model, which is generated by $U_{Z_2}$ in
\eqn{UZ2} and $U_{\t Z_2}$ in \eqn{tUZ2} as an on-site symmetry of the model.
Even though the $Z_2\times \t Z_2$ symmetry is on site in the model \eq{HZ2Z2},
it is anomalous when restricted in the low energy sector in the sense that in
$U=+\infty$ limit, the model cannot have gapped ground state that breaks the
full $Z_2\times \t Z_2$ symmetry.  (Certainly, when $U<J,B$,  the model can
have a gapped ground state that breaks the full $Z_2\times \t Z_2$ symmetry,
such as when $J=B=1, U=0$.) In $U=+\infty$ limit, only the gapless state at
$J=B$ has the full $Z_2\times \t Z_2$ symmetry that is not spontaneously
broken. 

What are the patch symmetry transformations for the $Z_2$ and $\t Z_2$
symmetry?  The first guess are
\begin{align}
 U_{Z_2}(i,j) = \prod_{k=i}^j X_k, \ \ \ \ \
 U_{\t Z_2}(i,j) = \prod_{k=i}^j \t Z_{k+\frac12}.
\end{align}
But $U_{Z_2}(i,j) = \prod_{k=i}^j X_k$ does not act within the low energy
sub-Hilbert space (in $U=+\infty$ limit).  To get around, we modify it at
the boundary, which leads to
\begin{align}
\label{TT}
\begin{split}
 U_{Z_2}(i,j) &= 
\t X_{i-\frac12} \Big(\prod_{k=i}^j X_k\Big) \t X_{j+\frac12} 
, \\
 U_{\t Z_2}(i,j) &= 
 \Big(\prod_{k=i}^j \t Z_{k+\frac12} \Big) 
.
\end{split}
\end{align}

The two sets of patch symmetry transformations still satisfy the algebras \eqn{z2patchalg} and \eqn{dualz2patchalg} as previously. Most importantly, the two sets of patch symmetry transformations have a $\pi$
``mutual statistics'' described by \eqn{mutualsta}. It is this property of the patch operators that we mean the symmetry we consider is $Z_2\vee \t Z_2$, but not $Z_2\times\t Z_2$. One can also easily check
that the local terms in the Hamiltonian \eq{HZ2Z2} commute with the patch
symmetry transformations when far away from the boundary.  So the Hamiltonian
\eq{HZ2Z2} has the $Z_2$ and $\t Z_2$ patch symmetries. In short, although the global symmetry transformation of $Z_2$ and $\tilde{Z}_2$ commute, yet the patch operators, which create charges at their end points, do not commute in the low energy sub-Hilbert space limit when $U\rightarrow\infty$.

We would like to remark that the same symmetry can be described by different yet equivalent choices of patch
symmetry transformations. Two  sets of patch symmetry transformations
are equivalent if the patch symmetry transformations only differ by ``local
neutral operators'' at the boundary of the patches.  Here a ``local neutral
operator'' is a local operator that commutes with all patch symmetry
transformations whose boundary is far away from the  operator.  The ``mutual
statistics'' of the patch symmetry transformations is not affected by those
local neutral operators. For example, we may instead choose $U_{Z_2}(i,j)=\tilde{Y}_{i-\frac{1}{2}}\left(\prod_{k=i}^j X_k\right)\tilde{Y}_{j+\frac{1}{2}}$ in (\ref{TT}).

\subsection{Symmetric sector of 1+1D Ising model as boundary of 2+1D $Z_2$
topological order}

\label{IsTO}

In the previous section, we discuss the anomaly property of $Z_2\vee \t
Z_2$ categorical symmetry, the mutual statistics of $Z_2$ and $\t Z_2$ charges in the low energy limit. This forbids a symmetric gapped ground state within the low energy sector.
In this section, we will show that the anomaly property of the $Z_2\vee \t
Z_2$ categorical symmetry is actually an effect of non-invertible gravitational
anomaly \cite{JW190513279}. More precisely, the theory with the categorical symmetry can be a boundary theory of a $Z_2$ topological order in one higher dimension. The charges and their mutual statistics \eqn{mutualsta} of the symmetry is 
determined by the bulk topological order.
To see the non-invertible gravitational anomaly in the 1+1D
Ising model, we concentrate on the so-called \emph{symmetric sub-Hilbert space}
$\cV_\text{symm}$ that is invariant under the $U_{Z_2}$ transformation.  The
space $\cV_\text{symm}$ of a $L$-site system does not have tensor product
expansion of the form $\otimes_{i=1}^L \cV_i $
\begin{align}
 \cV_\text{symm} \neq \otimes_{i=1}^L \cV_i .
\end{align}

Thus the symmetric sector of the Ising model can be viewed as having  a
non-invertible gravitational anomaly \cite{JW190513279}.  Indeed, the symmetric
sector of the Ising model can be viewed as a boundary of 2+1D $Z_2$ topological
order (the topological order characterized by $Z_2$ gauge theory), and thus has
a 1+1D non-invertible gravitational anomaly characterized by  2+1D $Z_2$
topological order \cite{JW190513279}.

A 2+1D $Z_2$ topological order has four types of excitations $\one,e,m,f$.
Here $\one$ is the trivial excitation, and $e,m,f$ are topological excitation
with mutual $\pi$-statistics between any two different ones.  $\one,e,m$ are bosons and $f$ is a
fermion.  They satisfy the following fusion relations,
\begin{align}
 e\otimes e =\one, \ \
 m\otimes m =\one, \ \
 f\otimes f =\one, \ \
 e\otimes m =f .
\end{align}

Let us construct the boundary effective
theory for the $m$ condensed boundary of $Z_2$ topological order.  Such a
boundary contains a gapped excitation that corresponds to the $e$-type
particle.  One might expect a second boundary excitation corresponding to the
$f$-type particle.  However, since $m$ is condensed on the boundary, the
$e$-type particle and the $f$-type particle are actual equivalent on the
boundary.  The simplest boundary effective lattice Hamiltonian that describes
the gapped $e$-type particles has a form (on a ring)
\begin{align}
H = - B \sum_i  X_i , \ \ \ \ B > 0.
\end{align}
Here a spin $X_i=1$
corresponds to an empty site and a spin $X_i=-1$ corresponds to a
site occupied with an $e$-type particle (with $2B$ as its energy gap).
However, the boundary Hilbert space does not have a direct product
decomposition $\otimes_i \cV_i$, due to the constraint
\begin{align}
 \prod_i X_i =1,
\end{align}
since the number of the  $e$-type particles on the boundary must be even
(assume the bulk has no topological excitations).  This is a reflection of
non-invertible gravitational anomaly. 
A more general boundary effective theory may have a form
\begin{align}
\label{tIsing}
H^\text{Is}_P = - B \sum_{i=1}^L  X_i - J \sum_{i=1}^L Z_i Z_{i+1}
,
\end{align}
where $Z_{L+1}\equiv Z_1$ and $Z_i Z_{i+1}$ creates a pair of the $e$-type
particles, or move an $e$-type particle from one site to another.  

In the above, we have shown that a boundary of 2+1D $Z_2$ topological order can
be described by \eqn{tIsing}.  The low energy sector of the model \eqn{HZ2Z2}
also describes the boundary of the 2+1D $Z_2$ gauge theory with $Z_2$-charge
$e$ and $Z_2$-vortex $m$,  where $e$ and $m$ particle has low energies on the
boundary.  An $e$ particle on the boundary corresponds to $X_i=-1$ and a $m$
particle corresponds to $\t Z_{i+\frac12}=-1$ in \eqn{HZ2Z2}.  The $Z_2\times
\t Z_2$ symmetry is the mod-2 conservation of $e$ and $m$ particles. This
explains the $Z_2\times \t Z_2$ symmetry in the symmetric sector of the Ising
model.  Note that, on the boundary, we may have a $e$ or $m$ condensation.  The
condensations may spontaneously break the $Z_2\times \t Z_2$ symmetry in the
ground state.  However, the model itself (given by the Hamiltonian and the sub-Hilbert space) always has $Z_2\times \t Z_2$
symmetry.

It is more precise to refer the  $Z_2\times \t Z_2$ symmetry as $Z_2\vee
\t Z_2$ \emph{categorical symmetry}.  This is because the $Z_2$ and $\t Z_2$
symmetries are not independent.  The $Z_2$-charge (the $e$ particle) and the
$\t Z_2$-charge (the $m$ particle) have a $\pi$ mutual statistics, when viewed
as particles in one higher dimension.  This gives rise to \eqn{mutualsta}.  The
term $Z_2\vee \t Z_2$ categorical symmetry includes such non-trivial
``mutual statistics'' between the $Z_2$ and $\t Z_2$ symmetry.  

The mutual $\pi$-statistics between $e$ and $m$ in the 2+1D bulk is encoded at
boundary by requiring the $Z_2$ domain wall to carry $\t Z_2$ charge and the
$\t Z_2$ domain wall to carry $Z_2$ charge.  This non-trivial mutual statistics
has a highly non-trivial effect:  in a gapped ground state, one and only one of
$Z_2$ and $\t Z_2$ symmetries must be spontaneously broken \cite{L190309028}.
Thus, a symmetric state that does not break the $Z_2\vee \t Z_2$
categorical symmetry must be gapless.  This is a consequence of 1+1D
non-invertible gravitational anomaly \cite{JW190513279} characterized by 2+1D
$Z_2$ topological order (\ie $Z_2$ gauge theory).  

To summarize, in the above, we considered the boundary of 2+1D $Z_2$ topological
order.  We argued that a boundary (gapped or gapless), \emph{as a system} (with
the symmetric sub-Hilbert space), always has a $Z_2\vee \t Z_2$
categorical symmetry. In contrast, a gapped boundary, \emph{as a state}, has a
partially spontaneous broken categorical symmetry, while one of the gapless
boundaries, \emph{as a state}, has the full categorical symmetry.  Next, let us
discuss patch symmetry transformations for the boundary of 2+1D $Z_2$
topological order, that describe the $Z_2\vee \t Z_2$ categorical
symmetry.  

\begin{figure}[t]
\includegraphics[scale=0.6]{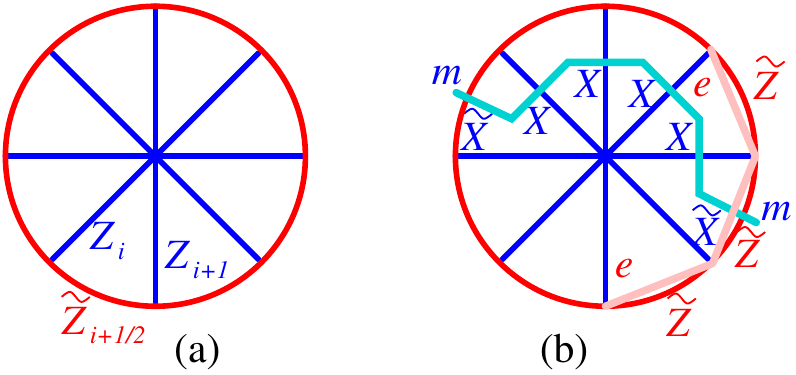}
\caption{
The reduced lattice, where  spin-$\frac12$ degrees of freedom live on the links. 
}
\label{wheel}
\end{figure}

We start with the bulk Hamiltonian for the $Z_2$ topological order on a square
lattice, where spin-$\frac12$ degrees of freedom live on the links: 
\begin{align}
\label{Hbulk}
H^{\text{bulk}}=
- U\sum_s \prod_{s\subset \partial l}X_l
- U\sum_p \prod_{l\subset \partial p} Z_l
,
\end{align}
where $s$ labels the sites, $l$  the links, and $p$  the plaquettes.  $\sum_s$
sums over all the sites in the bulk (\ie off the boundary), and $\sum_p$ sums
over all the plaquettes.  On the boundary, we can have any local Hamiltonian
$H^\text{bdy}$.  The boundary Hamiltonian remains finite as we take the $U\to
\infty$ limit.  $H^{\text{bulk}}$ in $U\rightarrow \infty$ limit is a
fixed-point Hamiltonian, and we can simplify it via tensor network
renormalization \cite{VC0466,V0705,LN0701,GW0931}. In the end, the model
\eqn{Hbulk} can be reduced to the one on the lattice in Fig. \ref{wheel}a
\cite{CGL1204}.  And the Hamiltonian is still
given by \eqn{Hbulk}, which describes the dynamics of boundary degrees of
freedom.

Now we will show that the wheel model with the Hamiltonian in \eqn{Hbulk} on lattice Fig.  \ref{wheel}a (plus extra
boundary terms) and the sub-Hilbert space with no bulk excitations is the same as the minimally coupled model with the Hamiltonian  in the symmetric
sub-Hilbert space satisfying $\prod_i X_i=1$.

We consider the $Z_2$ topological order on the wheel, the fixed-point lattice of a disk. There are in total $2N$ links, $N$ on the boundary, and $N$ inside. That is in total $2^{2N}$ states to start with. We consider the subspace that has no bulk excitations. That means the star and plaquette term satisfy 
\begin{align}
\prod_{j=1}^N X_j=1,~~Z_{i}\tilde{Z}_{i+\frac{1}{2}}Z_{i+1}=1,
\label{wheelconstraint}
\end{align}
for $i=1,\cdots N$. These reduce the Hilbert space we consider to be of dimension $2^{N-1}$. Now we consider the boundary Hamiltonian. Any term in it should first commute with \eqn{wheelconstraint}. Second, it describes the dynamics of $e$ and $m$ excitations on the boundary. In particular, we have
\begin{align}
H^{\text{bdy}}=-B\sum_i \tilde{X}_{i-\frac{1}{2}}X_i\tilde{X}_{i+\frac{1}{2}}-J\sum_i \tilde{Z}_i,
\end{align}
where the first term create pairs of $m$-particles, and the second term create pairs of $e$-particles. We may as well write the no bulk flux excitation as a dynamical constraint, and the boundary Hamiltonian is
\begin{align}
\label{Hbdy}
H^{\text{bdy}}=&-B\sum_i \tilde{X}_{i-\frac{1}{2}}X_i\tilde{X}_{i+\frac{1}{2}}-J\sum_i \tilde{Z}_i\nonumber\\
&+U\sum_i(1- Z_{i}\tilde{Z}_{i+\frac{1}{2}}Z_{i+1}).
\end{align}
And this is the same as \eqn{HZ2Z2}.

The upshot is the $Z_2$ minimally coupled model with a global $Z_2$ constraint is equivalent to a boundary Hamiltonian of $\ZZ_2$ topological order when the bulk has no topological excitations. The ground state subspace projection from the bulk to the boundary Hamiltonian is the same as the gauge and global constraint to the 1d minimally coupled model.\footnote{In fact, take the $J=0$ limit of either the model \eqn{HZ2Z2} or the wheel model restricted to the boundary \eqn{Hbdy}. The ground state is a symmetry protected topological (SPT) state protected by $Z_2\times \tilde{Z}_2$. However, under the $Z_2\vee Z_2$ symmetry specified by the patch operators \eqn{TT}, the ground state is in the same time a $m$ condensed state. This is because the patch operator $U_{Z_2}$ takes a constant value in the ground state. This is the string operator creating a pair of $m$, as illustrated in Fig. \ref{wheel}. Thus, the phase spontaneously breaks $Z_2\vee \tilde{Z}_2$ symmetry.}

Furthermore, one could see that the patch symmetry transformation $U_{Z_2}(i,j)$ (see \eqn{TT}) is the same as the string operator that creates a pair of $e$-type excitations at
string ends (see Fig. \ref{wheel}b), and the patch symmetry transformation
$U_{\t Z_2}(i,j)$ (see \eqn{TT}) is the same as the string operator that creates
a pair of $m$-type excitations at string ends (see Fig. \ref{wheel}b). The excitations are only on the boundary of the wheel. This
explains the non-trivial mutual statistics between $Z_2$ and $\t Z_2$  patch
symmetries, since the two string operators in Fig. \ref{wheel}b intersect at
one point.

To summarize, a theory with non-invertible gravitational anomaly has emergent
symmetries (\ie the categorical symmetry), which come from the conservation of
topological excitations in one-higher-dimension bulk.  Thus the categorical
symmetry is fully characterized by the bulk topological order.  Part of the
categorical symmetry must be broken in any gapped phase, and the symmetric
phase must be gapless.  There is a gapless phase that respects the full
categorical symmetry.

\subsection{How categorical symmetry determines gapless state }

Eqn. (\ref{tIsing}) is a $m$-condensed boundary of 2+1D $Z_2$ topological
order. Its partition function has four-components. For such $m$-condensed
boundary (with $|J|<B$), the four-component partition function is given by
(after shifting the ground state energy density to zero)\cite{JW190513279}
\begin{align}
\label{Z2SB}
\begin{pmatrix}
 Z_\one\\
 Z_e\\
 Z_m\\
 Z_f\\
\end{pmatrix} 
&=
\begin{pmatrix}
1\\
0\\
1\\
0\\
\end{pmatrix} .
\end{align}
Here 
\begin{align}
 Z_\one = \Tr_{U_{Z_2} =1} \ee^{-\bt H^\text{Is}_P}=1,
\end{align}
in the large $\bt,L$ limit and in $U_{Z_2}=\prod_i X_i =1$ sector.
Also
\begin{align}
 Z_e = \Tr_{U_{Z_2} =-1} \ee^{-\bt H^\text{Is}_P}=0,
\end{align}
in the large $\bt,L$ limit and in $U_{Z_2}=\prod_i Z_i =-1$ sector.
Similarly
\begin{align}
 Z_m = \Tr_{U_{Z_2} =1} \ee^{-\bt H^\text{Is}_A}=1,
\nonumber\\
 Z_f = \Tr_{U_{Z_2} =-1} \ee^{-\bt H^\text{Is}_A}=0,
\end{align}
where $H^\text{Is}_A$ is the model \eqn{tIsing} with an ``anti-periodic
boundary condition'':
\begin{align}
\label{tIsingA}
H^\text{Is}_A = - B \sum_{i=1}^L  X_i 
- J \sum_{i=1}^{L-1} Z_i Z_{i+1}
+ J Z_L Z_{1}
,
\end{align}
\ie the $e$-type particle moving around the ring see a $\pi$-flux.

The above partition functions describe a $Z_2$ symmetric gapped state.  $Z_e=0$
implies the excitation carrying $Z_2$ charge to have a finite energy gap.
$Z_m=1$ implies that the $Z_2$ symmetry is not spontaneously broken, since the
$Z_2$-symmetry twist has no effect on the ground state.  Also $Z_m=1$ means that
the excitation carrying $Z_2$ flux has no energy gap.  It also means that the
patch symmetry operator $U_{Z_2}(i,j)$ (creating a pair of $Z_2$ flux
excitations) have a non-zero average, \ie the $\t Z_2$ symmetry is
spontaneously broken.

When $J>B$, we obtain the second gapped boundary: 
\begin{align}
\label{tZ2SB}
\begin{pmatrix}
 Z_\one(\tau,\bar \tau)\\
 Z_e(\tau,\bar \tau)\\
 Z_m(\tau,\bar \tau)\\
 Z_f(\tau,\bar \tau)\\
\end{pmatrix} 
&=
\begin{pmatrix}
1\\
1\\
0\\
0\\
\end{pmatrix} .
\end{align}
This corresponds to the $Z_2$ symmetry breaking phase of the Ising model, which
is also $\t Z_2$ symmetric.  We see that, indeed, the critical point of Ising
transition, plus its two neighboring gapped states, can be described by a
gapless edge, and its neighbors, of $2+1$D $Z_2$ topological order.

When $J=B$, the boundary effective theory is gappless.  From $H^\text{Is}_P$
and $H^\text{Is}_A$, we can obtain the gapless partition functions
\cite{JW190513279,CZ190312334,KZ170501087,KZ190504924,KZ191201760}:
\begin{align}
\label{Z2cri}
\begin{pmatrix}
 Z_\one(\tau,\bar \tau)\\
 Z_e(\tau,\bar \tau)\\
 Z_m(\tau,\bar \tau)\\
 Z_f(\tau,\bar \tau)\\
\end{pmatrix} 
&=
\begin{pmatrix}
|\chi^\text{Is}_0(\tau)|^2+|\chi^\text{Is}_\frac12(\tau)|^2\\
|\chi^\text{Is}_{\frac1{16}}(\tau)|^2\\
|\chi^\text{Is}_{\frac1{16}}(\tau)|^2\\
\chi^\text{Is}_0(\tau) \bar \chi^\text{Is}_\frac12(\tau) +\chi^\text{Is}_\frac12(\tau) \bar \chi^\text{Is}_0(\tau) \\
\end{pmatrix} ,
\end{align}
where $\chi^\text{Is}_i(\tau)$ are characters of Ising CFT. This corresponds to
the critical point at the $Z_2$ symmetry breaking transition.  

Since $Z_m$ has the $Z_2$ symmetry twist, $Z_m\neq 0$ implies that the  $Z_2$
symmetry is not spontaneously broken.  Also $Z_e$ has the $\t Z_2$ symmetry
twist. Thus $Z_e\neq 0$ implies that the  $\t Z_2$ symmetry is not
spontaneously broken.  This is why we say the gapless critical point to have
full $Z_2 \vee \t Z_2$ categorical symmetry.

To see how the patch symmetry transformations for the $Z_2 \vee \t Z_2$
categorical symmetry act in the symmetric gapless point, we note that
the patch symmetry operators have the following form at the
symmetric gapless point
\begin{align}
 U_{Z_2}(i,j) &= \t X_{i-\frac12} \t X_{j+\frac12} \sim \mu(i)\mu(j), 
\nonumber\\
 U_{\t Z_2}(i,j) &= Z_i Z_j \sim \si(i)\si(j).
\end{align}
Here $\si$ and $\mu$ are two primary fields with scaling dimensions
$(\frac1{16},\frac1{16})$ and $(\frac1{16},\frac1{16})$ of non-chiral Ising
CFT. The operator product expansion with the fermion primary field $\psi$ in
the  Ising CFT is given by\cite{CFT12}
\begin{align}
\psi \sigma\sim \mu\,, \ \ \  \bar{\psi} \mu \sim \si,\, \ \ \ \sigma \mu\sim \psi+\bar\psi\,.
\end{align}
The monodromy between $\mu$ and $\sigma$ is $-1$. This reflects that
$U_{Z_2}(i,j)$ and $U_{\tilde{Z}_2}(i',j')$ have mutually $\pi$-statistics.

Therefore,
the averages of the
 patch symmetry operators (\ie the
$Z_2$ and $\t Z_2$ order parameters)
have a form
\begin{align}
\begin{split}
 \<U_{Z_2}(x,y)\> &= \<\mu(x)\mu(y)\> \sim |x-y|^{-1/4},\\
 \<U_{\t Z_2}(x,y)\> &= \<\si(x)\si(y)\> \sim |x-y|^{-1/4}.
 \end{split}
\end{align}
They vanish when $|x-y|\to \infty$.  Thus the Ising critical point has the full
$Z_2 \vee \t Z_2$ categorical symmetry.

That in the modular invariant partition function in the Ising CFT there is only one excitation with scaling dimension $\left(\frac{1}{16},\frac{1}{16}\right)$ is consistent with the fact that in this low energy theory, only either $U_{Z_2}(x,y)$ or $U_{\tilde{Z}_2}(x,y)$ is the correlator of the local excitation, while the other is the patch symmetry operator acting on the local excitation.

The above partition functions are essentially determined by the $Z_2 \vee
\t Z_2$ categorical symmetry.  Remenber that the $Z_2 \vee \t Z_2$
categorical symmetry is characterized by 2+1D $Z_2$ topological order, which in
turn is characterized by the following $S,T$ matrices
\begin{align}
   T^{Z_2\vee \t Z_2}=
\begin{pmatrix}
    1&0&0&0\\
    0&1&0&0\\
    0&0&1&0\\
    0&0&0&-1
  \end{pmatrix} ,\
  S^{Z_2\vee \t Z_2}&= \begin{pmatrix}
    \frac12 &\frac12 &\frac12 &\frac12 \\
    \frac12 &\frac12 &-\frac12 &-\frac12 \\
    \frac12 &-\frac12 &\frac12 &-\frac12 \\
    \frac12 &-\frac12 &-\frac12 &\frac12 
  \end{pmatrix}.
\end{align}
Through $S,T$ matrices, the categorical symmetry can determine the partition
functions for low energy fixed points via the following
relation\cite{JW190513279}
\begin{align}
\label{STZ}
Z_i(\tau+1)=T_{ij}^{Z_2\vee \t Z_2} Z_j(\tau),\ \ \ Z_i(-1/\tau)=S_{ij}^{Z_2\vee \t Z_2} Z_j(\tau) .
\end{align}

For the gapped states in a system with $Z_2 \vee \t Z_2$ categorical
symmetry, the partition functions $Z_i$ are $\tau$ independent positive
integers with $Z_\one =1$.  We find that \eqn{Z2SB} and \eqn{tZ2SB} are the
only two gapped solutions of \eqn{STZ}.  This confirms the result in
\Ref{L190309028}: gapped states must partially break categorical symmetry
spontaneously.

Eqn. (\ref{Z2cri}) is a gapless solution of \eqn{STZ} that has the full $Z_2
\vee \t Z_2$ categorical symmetry.  Eqn. (\ref{STZ}) also has other
solutions with the full $Z_2 \vee \t Z_2$ categorical symmetry, such as
\begin{align}
\label{CFT54}
 \begin{pmatrix}
Z_\one \\ Z_e \\ Z_m \\ Z_f \\
\end{pmatrix}=
\begin{pmatrix}
|\chi^{5,4}_0|^2 +  |\chi^{5,4}_\frac{1}{10}|^2 +  |\chi^{5,4}_\frac{3}{5}|^2 +  |\chi^{5,4}_\frac{3}{2}|^2 \\
|\chi^{5,4}_\frac{7}{16}|^2 +  |\chi^{5,4}_\frac{3}{80}|^2 \\
|\chi^{5,4}_\frac{7}{16}|^2 +  |\chi^{5,4}_\frac{3}{80}|^2 \\
\chi^{5,4}_0 \bar\chi^{5,4}_\frac{3}{2} +  \chi^{5,4}_\frac{1}{10} \bar\chi^{5,4}_\frac{3}{5} +  \chi^{5,4}_\frac{3}{5} \bar\chi^{5,4}_\frac{1}{10} +  \chi^{5,4}_\frac{3}{2} \bar\chi^{5,4}_0 \\
\end{pmatrix},
\end{align}
where $\chi^\text{5,4}_h(\tau)$ are characters of $(5,4)$ minimal model CFT
(with cenral charge $c=\frac{7}{10}$). 

We see that categorical symmetry can largely determine the gapless states where
the categorical symmetry is not spontaneously broken, but not uniquely.
However, the  gapless states with larger heat capacity may have additional
emergent categorical symmetry.  Therefore, we consider minimal gapless states
(\ie with minimal central charge $c$) with the full categorical symmetry.  For
$Z_2 \vee \t Z_2$ categorical symmetry, there is only one minimal gapless
state \eqn{Z2cri}.  For categorical symmetry characterized by 2+1D $S_3=Z_3
\rtimes Z_2$ gauge theory there is also only one minimal gapless state
(see Table \ref{S3FusionRules})\cite{JW190513279}
\begin{align}
 Z_{\one} &=  |\chi^{6,5}_{0}|^2 +  |\chi^{6,5}_{3}|^2 +  |\chi^{6,5}_{\frac{2}{5}}|^2 +  |\chi^{6,5}_{\frac{7}{5}}|^2 
 \nonumber \\ 
Z_{a^1} &=  \chi^{6,5}_{0} \bar\chi^{6,5}_{3} +  \chi^{6,5}_{3} \bar\chi^{6,5}_{0} +  \chi^{6,5}_{\frac{2}{5}} \bar\chi^{6,5}_{\frac{7}{5}} +  \chi^{6,5}_{\frac{7}{5}} \bar\chi^{6,5}_{\frac{2}{5}} 
 \nonumber \\ 
Z_{a^2} &=  |\chi^{6,5}_{\frac{2}{3}}|^2 +  |\chi^{6,5}_{\frac{1}{15}}|^2  
\nonumber \\ 
Z_{b} &=  |\chi^{6,5}_{\frac{2}{3}}|^2 +  |\chi^{6,5}_{\frac{1}{15}}|^2 
\\ 
Z_{ b^1} &=  \chi^{6,5}_{0} \bar\chi^{6,5}_{\frac{2}{3}} +  \chi^{6,5}_{3} \bar\chi^{6,5}_{\frac{2}{3}} +  \chi^{6,5}_{\frac{2}{5}} \bar\chi^{6,5}_{\frac{1}{15}} +  \chi^{6,5}_{\frac{7}{5}} \bar\chi^{6,5}_{\frac{1}{15}} 
 \nonumber \\ 
Z_{ b^2} &=  \chi^{6,5}_{\frac{2}{3}} \bar\chi^{6,5}_{0} +  \chi^{6,5}_{\frac{2}{3}} \bar\chi^{6,5}_{3} +  \chi^{6,5}_{\frac{1}{15}} \bar\chi^{6,5}_{\frac{2}{5}} +  \chi^{6,5}_{\frac{1}{15}} \bar\chi^{6,5}_{\frac{7}{5}}  
 \nonumber \\ 
Z_{c} &=  |\chi^{6,5}_{\frac{1}{8}}|^2 +  |\chi^{6,5}_{\frac{13}{8}}|^2 +  |\chi^{6,5}_{\frac{1}{40}}|^2 +  |\chi^{6,5}_{\frac{21}{40}}|^2 
 \nonumber \\ 
Z_{ c^1} &=  \chi^{6,5}_{\frac{1}{8}} \bar\chi^{6,5}_{\frac{13}{8}} +  \chi^{6,5}_{\frac{13}{8}} \bar\chi^{6,5}_{\frac{1}{8}} +  \chi^{6,5}_{\frac{1}{40}} \bar\chi^{6,5}_{\frac{21}{40}} +  \chi^{6,5}_{\frac{21}{40}} \bar\chi^{6,5}_{\frac{1}{40}}  ,
\nonumber 
\end{align}
where $\chi^\text{6,5}_h(\tau)$ are characters of $(6,5)$ minimal model (with
central charge $c=\frac{4}{5}$).

The above examples strongly suggest that categorical symmetry characterized by
$2+1$D $Z_2$ topological order allows us to determine one or a few of minimal
gapless states via \eqn{STZ}. This points to a direction that \emph{gapless
states are largely (might even uniquely) determined by categorical symmetries,
\ie by topological order in one higher dimension}.

\section{$Z_2$ symmetry and $Z_2^{(1)}$ 1-symmetry in 2+1D Ising model}\label{Z2modelsaly}

In two dimensions, it is well known that the critical point for $Z_2$ symmetry
breaking transition has the $Z_2$ symmetry.  In this section, we show that the
critical point also has a 1-symmetry.  

Let us consider the following two models: $Z_2$ Ising model and $Z_2$ gauge
model on the square lattice.  We will demonstrate that the $Z_2$ Ising model
restricted to $Z_2$ even (chargeless) sector is exactly dual to the lattice
$Z_2$-link model in the limit where the $Z_2$ vortex has infinity gap.

The $Z_2$ Ising model is given by
\begin{align}
\label{Ising2d}
H=-J\sum_l Z_{i_1(l)}Z_{i_2(l)} - B \sum_i X_i,
\end{align}
where $\sum_l$ sums over all links, $\sum_i$ sums over all vertices, $i_1(l)$
and $i_2(l)$ are two vertices connected by the link $l$.  The lattice $Z_2$
gauge model is given by
\begin{align}
\label{gauge2d}
\tilde{H}=-J\sum_{l} \tilde{Z}_{l}
-B\sum_{i} \prod_{{l} \supset i}\tilde{X}_{l}
+U\sum_{s} (1-\prod_{{l}\in s}\tilde{Z}_{l}),
\end{align}
where  $\sum_{i}$ sums over all sites, $\sum_{s}$ sums over all squares,
$\prod_{l \supset i}$ is a product over all the four links that contain vertex
$i$, and $\prod_{l \in s}$ is a product over all the four link on the boundary
of square $s$.  We consider the limit $U\rightarrow +\infty$.

The two models can be mapped into each other via the map that preserves the
operator algebra
\begin{align}
X_i\rightarrow \prod_{l\supset i}\tilde{X}_{l},~~~Z_{i_1(l)}Z_{i_2(l)}\rightarrow \tilde{Z}_{l} .
\end{align}
We will refer such a map as the ``duality'' map, or ``gauging'' in a looser sense, from a pure matter theory to a pure gauge theory. 

The Ising model $H$ has a global $Z_2$ symmetry generated by 
\begin{align}
U_{Z_2}= \prod_i X_i.  
\end{align}
After duality, it is mapped into an identity operator $1$.  The lattice
$Z_2$ gauge model has a $Z_2^{(1)}$ $1$-symmetry generated by the Wilson-line
operators
along any closed path $C$
\begin{align}
U_{Z_2^{(1)}} (C)= \prod_{l\in C}\t Z_{l}.
\label{Z21generator}
\end{align}
It corresponds to an identity operator in the Ising model under the duality map. 

Next we compare the low energy sub-Hilbert space of the two models. Assume the space to be a torus with $N=L\times L$ vertices.  The Hilbert space
of the Ising model has a dimension $2^{N}$.  The subspace of $Z_2$ symmetric
states, $\cV_\text{Ising}^\text{symm}$,  has a dimension $2^{N-1}$.  The
Hilbert space of the $Z_2$ gauge model has  a dimension $2^{2N}$. In
$U\rightarrow+\infty$ limit, the low energy subspace has a dimension $2^{N}\cdot 2$, The
extra factor $2$ is due to the operator identity $\prod_{s} \prod_{l\in
s}\tilde{Z}_{l}=1$.  In the low  energy sub-Hilbert space, we have $\prod_{l\in
s}\tilde{Z}_{l}=1$ and
\begin{align}
 U_{Z_2^{(1)}} (C)=  U_{Z_2^{(1)}} (C'),
\end{align}
if $C$ can be deformed into $C'$.  Now we consider the subspace,
$\cV_\text{gauge}^\text{symm}$, of the low  energy Hilbert space where
$U_{Z_2^{(1)}}(S^1_x)= U_{Z_2^{(1)}}(S^1_y)=1$, where $S^1_x$ and $S^1_y$ are
the two non-contractible loops wrapping around the system in $x$- and
$y$-directions.  $\cV_\text{gauge}^\text{symm}$ has a dimension $2^{N-1}$.  $H$
in $\cV_\text{Ising}^\text{symm}$ and $\t H$ in $\cV_\text{gauge}^\text{symm}$
are equivalent via an unitary transformation.  In this sense, the Ising model
\eqn{Ising2d} is exactly dual to the $Z_2$ gauge model \eqn{gauge2d}.

Now let us consider the ground states. Let us assume $J,B>0$.  It is
interesting to note that, for $B \gg J$, the trivial phase of the Ising model
is mapped to the topologically ordered phase (the $Z_2^{(1)}$ 1-symmetry
breaking phase) of the $Z_2$ gauge model, while for $B \ll J$, the $Z_2$
symmetry breaking phase of the Ising model is mapped to the trivial phase (the
symmetric phase of the $Z_2^{(1)}$ 1-symmetry) of the $Z_2$ gauge model.  At
the gapless critical point of the 2+1D $Z_2$ symmetry breaking transition (also
the $Z_2^{(1)}$ 1-symmetry breaking transition), we have the $Z_2\vee
Z_2^{(1)}$ symmetry which is not spontaneously broken.  Therefore, we show
the appearance of 1-symmetry of the ground state at the 2+1D $Z_2$ symmetry
breaking transition.

Now we discuss the charges of the categorical symmetry $Z_2\vee Z_2^{(1)}$. We will find that in the sub-Hilbert space, we can only create charges that the total of them is neutral, as measured by the global symmetry operators, while the charge in a finite region is not neutral and can be measured by patch operators. We may call this kind of charge excitations are neutral charges. 

A single charge of $Z_2$ symmetry is a $e$ particle on a site ($Z_i=-1$ in the Ising model \eqn{Ising2d}). The neutral charge of $Z_2$ symmetry, however, is two $e$ particles, living on two sites, or rather $S^0$.  The charge of the $\tilde{Z}_2^{(1)}$ symmetry is an open $Z_2$ vortex string, living on $D^1$. The neutral charge of the $\tilde{Z_2}^{(1)}$ symmetry is a closed contractible $Z_2$ vortex loop living on $S^1$, let us call it a $s$ string. In the sub-Hilbert space, we can only create neutral charges, excitations on $S^0$ and $S^1$. The operator for a pair of $e$-particles at site $i$ and site $j$ is $Z_iZ_j$. In the $Z_2$ gauge theory, the operator is dual to $U_{Z_2^{(1)}}(C_{ij})=\prod_{l\in C_{ij}}\tilde{Z}_l$, where $C_{ij}$ is any path from $i$-site to $j$-site. It is also the patch (or part) of the generators of the $Z_2$ $1$-symmetry. 

A $s$ string is created by 
\begin{align}
U_{Z_2}\left[(S^1)^\vee\right]=\prod_{l\in (S^1)^{\vee}} \tilde{X}_l, \label{2dsstring}
\end{align} where $(S^1)^\vee$ is a contractible loop on the dual lattice. It corresponds to create $\t
Z_{l}=-1$ along the loop in the $Z_2$ gauge model
\eqn{gauge2d}. In the Ising model, the $s$ string operator is dualed from the patch operator of the $Z_2$ global symmetry, 
\begin{align}
\label{2dZ2patch}
U_{Z_2}(C^\vee)= \prod_{i\in D^2}X_i,
\end{align} 
where $D^2$ is the disk whose boundary is $C^{\vee}$. This charge, when measured by $Z_2^{(1)}$ symmetry generator \eqn{Z21generator}, is neutral. Yet, when measured by part of the generator $U_{Z_2}(C_{ij})$, it has charge $-1$ when there $C^{\vee}$ circles around a single end point of $C_{ij}$, either $i$ or $j$. 

That is to say, the two kinds of patch operators satisfy the following relation,
\begin{align}
U_{Z_2^{(1)}}(C_{ij}) U_{Z_2}(C^\vee)=-U_{Z_2}(C^\vee)U_{Z_2^{(1)}}(C_{ij}),
\end{align}
when only one $e$ particle on either $i$ or $j$-site is circled by the loop excitation along $C^\vee$. This reveals the mutual $\pi$ statistics between the $e$ and $s$ excitation.

In summary, in the sub-Hilbert space, there are states with both conserved $e$ charges and conserved $s$ string, either the Ising model description, or the $Z_2$ gauge theory description. They are created by patch operators. And one kind of patch operator can measure the existence of the other. In other words, the charge of $Z_2$ and that of $Z_2^{(1)}$ have mutual statistics. The $Z_2\vee Z_2^{(1)}$ symmetry (or more precisely, the
$Z_2\vee Z_2^{(1)}$ categorical symmetry) is the conservation of $e$
particles and $s$ strings, together with the mutual statistics.

The above theory with sub-Hilbert space also describes the boundary of the 3+1D $Z_2$ gauge theory with $Z_2$-charge
$e$ and $Z_2$-vortex string $s$,  where $e$ particles and $s$ strings have low energies only at the boundary. In 3+1D $Z_2$ topological gauge theory, the $e$ excitations (living on $S^0$, composed of two points) and $s$ string (living on $(S^1)^\vee$ is created by the following operators defined on the open string $C$ on the lattice and a contractible membrane $(D^2)^\vee$ on the dual lattice,
\begin{align}
 U_{Z_2}(C)=&\prod_{l\in C} \tilde{Z}_l,\\
 U_{Z_2^{(2)}}\left[(D^2)^\vee\right]=& \prod_{l\in (D^2)^\vee} \tilde{X}_l.\label{Z23dmembrane}
\end{align}
The $s$ vortex loop on the boundary of $U_{Z_2^{(2)}}$ brought to the boundary of the $3$d spacial lattice, is the neutral $Z_2^{(1)}$ charge in the $2$d either Ising model \eqn{Ising2d} or $Z_2$-link model \eqn{gauge2d} with a sub-Hilbert space respectively. It is neutral in the sense that the membrane \eqn{Z23dmembrane} creating it commute with any $U_{Z_2}(S^1)$ on a closed string $S^1$.
\begin{figure}[h]
\includegraphics[scale=.7]{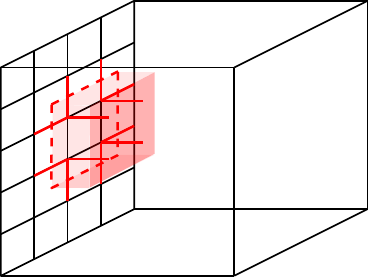}
\caption{A $Z_2^{(1)}$ neutral charge, a $s$ string on the 2d boundary (red dashed loop) is created by a $Z_2$ membrane operator \eqn{Z23dmembrane} in the 3d bulk (red surface). It is only neutral when on the boundary. If we translate this membrane to the bulk, there are $U_{Z_2}(S^1)$ operators in the bulk that anticommute the membrane operator, justifying it as the $Z_2$ vortex topological excitation.}
\label{Z2flux3d}
\end{figure}

Furthermore, the 3+1D bulk has a highly non-trivial effect on the boundary:  in a gapped
ground state of the 2+1D model \eq{Ising2d}, one and only one of the $Z_2$ and
$Z_2^{(1)}$ symmetries must be spontaneously broken. Pictorially  speaking, this is because the two topological excitations with mutual statistics cannot condense simultaneously. Thus, a ground state of
model \eq{Ising2d} with the full $Z_2\vee Z_2^{(1)}$ categorical symmetry
must be gapless.  This is a consequence of 2+1D non-invertible gravitational
anomaly \cite{JW190513279} and the $Z_2\vee Z_2^{(1)}$ categorical symmetry
characterized by 3+1D $Z_2$ topological order (\ie $Z_2$ gauge theory).  

\section{The model with anomalous symmetry as the boundary of topological order
in one higher dimension}
\label{BB}

Through the examples above, we have shown that a model with a finite
symmetry $G$, when restricted in the symmetric sector, can be viewed as the
boundary of $G$-gauge theory in one higher dimension.  The conservation (the
fusion rules) of the point-like gauge charge, and codimension-2 gauge flux give
rise to the symmetry and algebraic higher symmetry (whose combination becomes
the so-called categorical symmetry) of the $G$-symmetric model, The categorical
symmetry is not spontaneously broken at the critical point of the symmetry
breaking transition (see Section \ref{ahsymmG} for more details).

We know that a $G$-gauge theory can be twisted and becomes Dijkgraaf-Witten
theory \cite{DW9093}.  We will show that boundary of such twisted $G$-gauge
theory has an anomalous $G$-symmetry.  This implies that a system with
anomalous $G$-symmetry\cite{H8035,W1313} also has an algebraic higher symmetry.
The combination of the two symmetries corresponds to the categorical symmetry
described by the twisted $G$-gauge theory in one higher dimension.  Such a
system has a gapless state, where the categorical symmetry is not spontaneously
broken.  Also, a state with the unbroken categorical symmetry must be gapless.
And the gapped states of the system must spontaneously break the anomalous
$G$-symmetry.

\subsection{Boundary of double-semion model}

In this section, we will study the boundary of double-semion (DS) model (\ie
the twisted $Z_2$ gauge theory in 2+1D) to illustrate the above result.

A 2+1D double-semion (DS) topological order has four types of excitations
$\one,s,s^*,b$.  Here $s,s^*,f$ are topological excitations $s,s^*$ are
semions with statistics $\pm \ii$, and $b$ is a boson.
They satisfy the following fusion relation
\begin{align}
 s\otimes s =\one, \ \
 s^*\otimes s^* =\one, \ \
 b\otimes b =\one, \ \
 s\otimes s^* =b .
\end{align}
$s$ and $s$ have a mutual $\pi$ statistics and $s$ and $s^*$ have a mutual
boson statistics. As a result, $s$ and $b$ have a mutual $\pi$ statistics.

We consider a gapped boundary from condensing $b$ excitations.  Since $b\otimes b=\one$
and $b$ particles have mod-2 conservation, we assume the $b$ condensation gives
rise to two degenerate ground states, one with $Z_i=1$ and the other with
$Z_i=-1$.  The domain wall between $Z_i=1$ and $Z_i=-1$ regions corresponds a
$s$ particle.

We would like to point out that, on the boundary, although $s$-type particle and
$e$-type particle (in the $Z_2$ gauge theory discussed before) have the same fusion rule $s\otimes
s=\one$ and $e\otimes e=\one$, their fusion $F$-tensor are different
\cite{FNS0428,LW0510}.  In particular, fusing three $s$-type particles into
one $s$-type particle in two different ways differ by a phase $-1$:
\begin{align}
\label{sproc}
 [sss \to \one\one s \to \one s \one ] = (-)
 [sss \to s \one\one  \to \one s \one ].
\end{align}
In contrast, fusing three $e$-type particles (described by $X_i=-1$) into one
$e$-type particle in two different ways have the same phase:
\begin{align}
 [eee \to \one\one e \to \one e \one ] = 
 [eee \to e \one\one  \to \one e \one ]
\end{align}

For the boundary of $Z_2$ topological order, the above two processes of fusing
$e$ particles are induced, respectively, by a pair-annihilation operator
$X^+_iX^+_{i+1}$ and a hopping operator $ X^+_iX^-_{i+1} +X^-_iX^+_{i+1} $,
where
\begin{align}
 X^{\pm} = \frac12 (Y \pm \ii Z).
\end{align}
Indeed, we have
\begin{align}
&\ \ \ \
(X^+_{i-1} X^-_{i}+X^-_{i-1} X^+_{i}) (X^+_{i} X^+_{i+1})|eee\>
\nonumber\\
&=(X^+_{i} X^-_{i+1}+X^-_{i} X^+_{i+1})(X^+_{i-1} X^+_{i})|eee\> .
\end{align}
The pair-annihilation operator $Z^+_iZ^+_{i+1}$ and hopping operator $
Z^+_iZ^-_{i+1} +Z^-_iZ^+_{i+1} $ are allowed local operations, and we
can use them to construct effective boundary Hamiltonian
\begin{align}
\label{tIsingG}
H^\text{Is}_P &= 
-  \sum_{i=1}^L J_1( Z^+_i Z^-_{i+1} +Z^-_i Z^+_{i+1}) 
+J_2(Z^+_i Z^+_{i+1} + h.c.)
\nonumber\\
&\ \ \ \ \ \ \ \ \ \ \
- B \sum_{i=1}^L  Z_i 
,
\end{align}
which describes the boundary of 2+1D $Z_2$ topological order.

For the boundary of DS topological order, the two processes for fusing $s$
particles \eqn{sproc} are also induced by a pair-annihilation operator and a
hopping operator.  Here we choose the hopping operator to be $ X_i - Z_{i-1}X_i
Z_{i+1} $, which shift a domain wall from $i-\frac12$ to $i+\frac12$, or
$i+\frac12$ to $i-\frac12$.  The pair-annihilation or pair-creation operator is
given by $Z_{i-1}( X_i + Z_{i-1}X_i Z_{i+1}) $, which creates or annihilates a
pair of domain walls at $i+\frac12$ and $i-\frac12$.

For three $s$-type particles (the domain walls) at
$i-\frac12,i+\frac12,i+\frac32$, we indeed have
\begin{align}
& 
- ( X_{i+1} - Z_{i}X_{i+1} Z_{i+2}) Z_{i-1}( X_i + Z_{i-1}X_i Z_{i+1}) |sss\>
\nonumber\\
&=
 (X_i - Z_{i-1}X_i Z_{i+1}) Z_{i}( X_{i+1} + Z_{i}X_{i+1} Z_{i+2}) |sss\>, 
\end{align}
where $|sss\>=| \up_{i-1} \down_{i} \up_{i+1} \down_{i+2} \> $.

Now we can construct the boundary effective theory for the $b$ condensed
boundary of DS topological order.  We note that such a boundary contains a
gapped excitation that corresponds to the $s$-type particle.  One might expect
a second boundary excitation corresponding to the $s^*$-type particle.
However, since $b$ is condensed on the boundary, the $s$-type particle and the
$s^*$-type particle are actually equivalent on the boundary.  The simplest
boundary effective lattice Hamiltonian that describes the gapped $s$ particles
has a form
\begin{align}
H = - B \sum_i  Z_iZ_{i+1} , \ \ \ \ B > 0,
\end{align}
which has two degenerate ground states and the
$s$ particles correspond to domain walls.

Using the above allowed local operations $ X_i - Z_{i-1}X_i Z_{i+1} $ and
$Z_{i-1}( X_i + Z_{i-1}X_i Z_{i+1}) $, we can construct a more general boundary
effective theory
\begin{align}
\label{DSb}
 H^\text{DS} 
&=
- B  \sum_{i=1}^L Z_iZ_{i+1} 
- J_1 \sum_{i=1}^{L} (X_i - Z_{i-1}X_i Z_{i+1} ) 
\nonumber\\
&
+ J_2  \sum_{i=1}^{L} Z_{i-1}( X_i + Z_{i-1}X_i Z_{i+1}),
\end{align}
where site-$i$ and site-$(i+L)$ are identified.  

We note that the above Hamiltonian is not invariant under the spin-flip
transformation $\prod_i X_i$.  In fact, it is invariant under a non-on-site
transformation \cite{CLW1141}:
\begin{align}
\label{Xs}
 U_{Z_2} =\prod_i X_i \prod_i s_{i,i+1},
\end{align}
where $s_{ij}$ acts on two spins as 
\begin{align}
\label{sij}
s_{ij}&= |\up\up\>\<\up\up| +|\down\up\>\<\down\up|
-|\up\down\>\<\up\down| +|\down\down\>\<\down\down|
\nonumber\\
&= \frac 12( 1 - Z_i + Z_j + Z_i  Z_j).
\end{align}
The transformation has a simple picture: it flips all the spins and include a
$(-)^{N_{\up\to\down}}$ phase, where $N_{\up\to\down}$ is the number of $\up
\to \down$ domain wall.  We see that the transformation is a $Z_2$
transformation (\ie square to 1).  From Appendix \ref{Z2trns}, the  $Z_2$
transformation has the following action,
\begin{align}
 Z_{i} &\leftrightarrow  - Z_{i} ,
\nonumber\\
 X_{i} &\leftrightarrow  - Z_{i-1}  X_{i}  Z_{i+1} .
\end{align}
We find that \eqn{DSb} is invariant under the $Z_2$ transformation.

From the above discussion, we see that the different fusion properties lead to
different local operators.  The boundary effective theories for $Z_2$
topological order and for the double semion topological order are different.
In particular, the boundary effective theory for $Z_2$ topological order has an
on-site $Z_2$ symmetry, while the boundary effective theory for the double
semion topological order has a non-on-site $Z_2$ symmetry.  The non-on-site
$Z_2$ symmetry $U_{Z_2}$ implies that the model \eq{DSb} cannot have a gapped
$Z_2$ symmetric ground state \cite{CLW1141}.

\subsection{$\t Z_2$ dual symmetry}

We have seen that a 1+1D lattice model \eq{DSb} with an anomalous $Z_2$
symmetry (non-on-site symmetry\cite{CLW1141,W1313}) can be viewed as a boundary
of twisted 2+1D $Z_2$ gauge theory (\ie DS topological order).  The anomalous
$Z_2$ symmetry comes from the mod-2 conserved $b$ particles.  The mod-2
conserved $s$ particles will give rise to another symmetry, which will be
referred as dual $\t Z_2$ symmetry.  In other words, we claim that the  model
\eq{DSb} has both the $Z_2$ symmetry and the $\t Z_2$ symmetry.

To see the $\t Z_2$ symmetry explicitly, we do a dual transformation on the
model \eq{DSb}:
\begin{align}
 Z_iZ_{i+1} &\to  \t Z_{i+\frac12},
\nonumber\\
 X_i  &\to \t X_{i-\frac12} \t X_{i+\frac12}, 
\nonumber\\
 Z_i  &\to \prod_{j\leq i} \t Z_{j-\frac12}. 
\end{align}
We find
\begin{align}
&\ \ \ \
 X_{i} - Z_{i-1} X_{i} Z_{i+1}  
 =X_{i} + Z_{i-1}Z_i X_{i}Z_i Z_{i+1}  
\nonumber\\
&\to
\t X_{i-\frac12}\t X_{i+\frac12} 
+\t Z_{i-\frac12} \t X_{i-\frac12} \t X_{i+\frac12} \t Z_{i+\frac12}
\nonumber\\
&= \t X_{i-\frac12}\t X_{i+\frac12}+\t Y_{i-\frac12}\t Y_{i+\frac12},
\end{align}
\begin{align}
&\ \ \ \
 X_{i} + Z_{i-1} X_{i} Z_{i+1}  
 =X_{i} - Z_{i-1}Z_i X_{i}Z_i Z_{i+1}  
\nonumber\\
&\to
\t X_{i-\frac12}\t X_{i+\frac12} 
-\t Z_{i-\frac12} \t X_{i-\frac12} \t X_{i+\frac12} \t Z_{i+\frac12}
\nonumber\\
&= \t X_{i-\frac12}\t X_{i+\frac12}-\t Y_{i-\frac12}\t Y_{i+\frac12}.
\end{align}
The duality transformation changes the Hamiltonian \eq{DSb} into:
\begin{align}
 &
 \t H^\text{DS}
= 
 +  J_2 \sum_i \prod_{j< i} \t Z_{j-\frac12} (\t X_{i-\frac12}\t X_{i+\frac12}-\t Y_{i-\frac12}\t Y_{i+\frac12})
\nonumber\\
&\ \ \ \ - B  \sum_i   \t Z_{i+\frac12} 
- J_1 \sum_i (\t X_{i-\frac12}\t X_{i+\frac12}+\t Y_{i-\frac12}\t Y_{i+\frac12}) .
\end{align}
We see that the dual $\t Z_2$ symmetry is generated by
\begin{align}
 U_{\t Z_2} = \prod_i \t Z_{i+\frac12}.
\end{align}
This way, we obtain the explicit expression of the dual $\t Z_2$ symmetry.  The
on-site $\t Z_2$ symmetry $U_{\t Z_2}$ implies that the model \eq{DSb} can have
a gapped $\t Z_2$ symmetric ground state, which correspond to a $Z_2$ symmetry
breaking state.

In the dual model, $\t Z_{i+\frac12}=1$ describes a site with no semion $s$,
while $\t Z_{i+\frac12}=-1$ describes a site occupied with a semion $s$.  The
term $\t X_{i-\frac12}\t X_{i+\frac12}+\t Y_{i-\frac12}\t Y_{i+\frac12}$ is the
hopping term for the $s$ particle, while the term $ \prod_{j< i} \t
Z_{j-\frac12} (\t X_{i-\frac12}\t X_{i+\frac12}-\t Y_{i-\frac12}\t
Y_{i+\frac12})$ creates a pair of $s$ particles.

\section{Appearance of algebraic higher symmetry at the symmetry breaking transition
for general finite symmetry}
\label{ahsymmG}

In the previous section, we show the categorical symmetry in 1+1D and 2+1D
models with a local degrees of freedom taking values in $Z_2$. In this
section, we generalize the discussion to any $(n+1)$D dimensional lattice models
with local degrees of freedoms taking values in any finite group $G$. Same as
above, we discuss the lattice model in terms of two descriptions, generalizing
the Ising model and the $Z_2$-link model to the $G$-matter model and the $G$-link model. A major distinction is that when $G$ is non-Abelian, the $0$-symmetry in the $G$-link model is a global symmetry that is not reduced to specifying boundary conditions. We will show the emergence of categorical symmetry at and off the
critical point of Landau symmetry breaking transition in these models. 

\subsection{A duality point of view}

\begin{table*}[tb]
\setlength\extrarowheight{4pt}
\setlength{\tabcolsep}{6pt}
\centering
\begin{tabular}{c ? c|c|c|c|c|c|c|c}
\Xhline{4\arrayrulewidth}
$d,s$ & $1,0$ & $1,0$ & $2,0$ & $2,0$ & $2,\frac13$ & $2,-\frac13$ & $3,0$ & $3,\frac12$\\
\hline
$\otimes$ & $\bm 1$ &  $a^1$  & $a^2$ &  $b$  &  $b^1$  &  $b^2$  & $c$  & $c^1$    \\
\Xhline{2.5\arrayrulewidth}
$ \bm 1 $ & $\bm 1$ & $a^1$ & $a^2$   & $b$  & $b^1$  &  $b^2$          & $c$  & $c^1$    \\
$a^1$ & $a^1$ & $\bm 1$ & $a^2$	  & $b$  &$b^1$  & $b^2$      &  $c^1$  & $c$  \\
$a^2$ & $a^2$ & $a^2$   & $\bm 1\oplus a^1\oplus a^2$      & $b^1\oplus b^2$    & $b\oplus b^2$ & $b\oplus b^1$      & $c\oplus c^1$  & $c\oplus c^1$\\
$b$  & $b$  & $b$ & $b^1\oplus b^2$     & $\bm 1\oplus a^1\oplus b$ & $b^2\oplus a^2$  & $b^1\oplus a^2$     & $c\oplus c^1$   & $c\oplus c^1$  \\
$b^1$  & $b^1$   & $b^1$ & $b\oplus b^2$      & $b^2\oplus a^2$ & $\bm 1\oplus a^1\oplus b^1$  & $b\oplus a^2$      & $c\oplus c^1$ & $c\oplus c^1$ \\
$b^2$  & $b^2$  & $b^2$  & $b\oplus b^1$    & $b^1\oplus a^2$ & $b\oplus a^2$  & $\bm 1\oplus a^1\oplus b^2$      & $c\oplus c^1$  & $c\oplus c^1$ \\
$c$ & $c$ & $c^1$ & $c\oplus c^1$     & $c\oplus c^1$   & $c\oplus c^1$  & $c\oplus c^1$     & $\bm 1\oplus a^2\oplus b\oplus b^1\oplus b^2$ & $a^1 \oplus a^2\oplus b\oplus b^1\oplus b^2$  \\
$c^1$ & $c^1$ & $c$  & $c\oplus c^1$     & $c\oplus c^1$ & $c\oplus c^1$  & $c\oplus c^1$     & $a^1 \oplus a^2\oplus b\oplus b^1\oplus b^2$  & $\bm 1\oplus a^2\oplus b\oplus b^1\oplus b^2$ \\
\Xhline{4\arrayrulewidth}
\end{tabular}
\caption{The point-like excitations and their fusion rules in 2+1D $S_3$
topological order. Here $b$ and $c$ correspond to pure $S_3$ flux excitations,
$a^1$ and $a^2$ pure $S_3$ charge excitations, $\bm 1$ the trivial excitation,
while  $b^1$, $b^2$, and $c^1$ are charge-flux composites.  $d,s$ are the
quantum dimension and the topological spin of an excitation.  }
\label{S3FusionRules} 
\end{table*}

We consider two lattice models defined on the triangulation of $n$-dimensional
space.  The vertices of the triangulation are labeled by $i$, the links labeled
by $ij$, \etc.

In the first model, we may call it $G$-matter model, the physical degrees of
freedom live on the vertices and are labeled by group elements $g$ of a finite
group $G$.  The many-body Hilbert space is spanned in the following local basis
\begin{align}
|\{g_i\}\>, \ \ \ g_i \in G.
\end{align}
The Hamiltonian is given by
\begin{align}
\label{lmG}
 H_1 = - J \sum_{ij} \del(g_ig_j^{-1}) - B \sum_i \sum_{h \in G} L_h(i),
\end{align}

where 
\begin{align}
 \del(g) =
\begin{cases}
 1, & \text{ if } g = 1 \\
 0, & \text{ otherwise } \\
\end{cases}
.
\end{align}
Also, the operator $L_h(i)$ is given by
\begin{align}
 L_h(i) |g_1,\cdots,g_i,\cdots,g_N\>= |g_1,\cdots,hg_i,\cdots,g_N\>.
\end{align}
The Hamiltonian $H_1$ has an on-site  $G$ 0-symmetry
\begin{align}
\label{Uh}
 U_h H_1 = H_1 U_h, \ \ \ 
U_h =\prod_i L_h(i).
\end{align}
We see that when $J \gg B$, $H_1$ is in the symmetry breaking phase, and when
$J \ll B$, $H_1$ is in the symmetric phase.

Our second bosonic lattice model, which we may call the $G$-link model, has
degrees of freedom living on the links.  On an oriented link $ij$ pointing from
$i$-site to $j$ site, the degrees of freedom are labeled by $g_{ij} \in G$. The
many-body Hilbert space has the following local basis
\begin{align}
|\{g_{ij}\}\>, \ \ \ g_{ij} \in G.
\end{align} Here, $g_{ij}$'s on links with
opposite orientations satisfy 
\begin{align}
 g_{ij}=g_{ji}^{-1}.
\end{align}
The second model is related to the first model.  A state
$|g_1,\cdots,g_i,\cdots,g_N\>$ in the first model is mapped to a state
$|\cdots,g_{ij},\cdots\>$ in the second model where $g_{ij}=g_ig_j^{-1}$.

This connection allows us to design the Hamiltonian of the second model as
\begin{align}
\label{lgt}
 H_2 = &- J \sum_{ij} \del(g_{ij}) - B \sum_i \sum_{h \in G} Q_h(i) \nonumber\\
 &-U \sum_{ijk} \del(g_{ij}g_{jk}g_{ik}^{-1}), 
\end{align}
where the star term $Q_h(i)$ acts on all the links that connect to the vertex
$i$:
\begin{align}\label{Qh}
&\ \ \ \
 Q_h(i) |\cdots,g_{ij} , g_{ki},g_{jk},\cdots\> 
\nonumber\\
& = |\cdots,hg_{ij} ,g_{ki}h^{-1}, g_{jk},\cdots\>,
\end{align}
and the plaquette term acts as a projection to zero-flux configurations,
The second model has an algebraic $(n-1)$-symmetry, denoted as $G^{(n-1)}$
\cite{KZ200514178}, 
\begin{align}\label{Wq}
 W_q(S^1) H_2 = H_2 W_q(S^1)  ,~~~
W_q(S^1) = \Tr \prod_{ij \in S^1} R_q(g_{ij}),
\end{align}
for any loop $S^1$ formed by links, where $R_q$ is an irreducible
representation of $G$.  We see that the algebraic $(n-1)$-symmetry $G^{(n-1)}$
is generated by the Wilson loop operators $W_q(S^1)$, for all loops $S^1$ and
all irreducible representations $q$.  We note that the algebraic $0$-symmetry
$G^{(0)}$ is different from the usual 0-symmetry characterized by a group $G$,
when $G$ is non-Abelian. But when $G$ is Abelian the  algebraic $0$-symmetry
$G^{(0)}$ happen to be the usual 0-symmetry $G$.  Also, for Abelian $G$,
$G^{(n)}$ is a $n$-symmetry described by a higher group.  But, for non-Abelian
$G$, $G^{(n)}$ is an algebraic $n$-symmetry beyond higher group.

The Hamiltonian $H_2$ has the  algebraic $(n-1)$-symmetry, because $Q_h(i)$ term
in the Hamiltonian can be viewed as a ``gauge'' transformation and the Wilson
loop operator $W_q(S^1)$ is gauge invariant, and hence 
\begin{align} 
W_q(S^1) Q_h(i) = Q_h(i) W_q(S^1).  
\end{align} 
$W_q(S^1)$ commutes with other terms in $H_2$ since they are all diagonal in
the $|\{g_{ij}\}\>$ basis. 

In the limit $|B| \ll J \ll U$, the ground state of $H_2$ is a trivial product
state
\begin{align}
\label{gij1}
|\{g_{ij}=1\}\>, 
\end{align} 
which is symmetric under the  algebraic $(n-1)$ symmetry $G^{(n-1)}$. In the
other limit $|J| \ll B \ll U$, the ground state of $H_2$ is a topologically
ordered state (described by the $G$-gauge theory), breaking the algebraic
$(n-1)$ symmetry $G^{(n-1)}$ spontaneously.

What is the global $G$ symmetry operator in the first model \eqn{Uh} mapped to? It is mapped to 
a global $0$-symmetry operator,
\begin{align}
\calU_h H_2=H_2\,\calU_h,\ \ \ \calU_h=\prod_i Q_h (i)\,. \label{innG}
\end{align} 
In particular, when the model has periodic boundary condition, this $0$-symmetry acts as $\calU_h |g_{ij}\rangle = |hg_{ij}h^{-1}\rangle$. Thus the global symmetry is an inner automorphism of $G$, denoted as $\Inn (G)$. When the centralizer of $G$ is trivial, $\Inn (G)\cong G$. 

Only when $G$ is Abelian, the global symmetry action in (\ref{innG}) reduces to claiming the boundary conditions or the twisted sectors of the model. For example, when $G=Z_2$ and $d=1$, it reduces to $U_{\tilde{Z}_2}$ in (\ref{HDW}). 

Furthermore, the symmetry generators of the algebraic $(n-1)$-symmetry $G^{(n-1)}$
and the $0$-symmetry $\Inn (G)$ commute,
\begin{align}
 W_q(S^1)\, \calU_h=\calU_h W_q (S^1)\,.
 \label{symcommute}
\end{align}

In the limit $U\to +\infty$, the low energy part of $H_2$ can be mapped to
$H_1$ via the following duality and inverse duality map,
\begin{align}
 g_{ij} &= g_ig_j^{-1},
&
\left(g_{i_0}\right)^{-1}g_i & =   \left(g_{i_0j} g_{jk} \cdots g_{li}\right)^{-1} ,
\end{align}
where $i_0$ is a fixed base point. 
Note that to map a configuration $g_i$
to a configuration $g_{ij}$, we need to pick a base point $i_0$ and a value
$g_{i_0}$.  Therefore, the above map is a $|G|$-to-one map. It maps the
following $|G|$ configurations of $H_1$ (label by $h\in G$), $|\{ hg_i \}\>$,
into the same configuration configuration of $H_2$, $|\{g_{ij}\}\>$. Thus the
spectrum of $H_1$ formed by $G$ invariant states, $|\Psi\> =U_h |\Psi\>$, is
identical to the low energy spectrum of $H_2$ below $U$.  $H_1$ and $H_2$ have
the same $G$-symmetric low energy dynamics.  In particular they have the same
phase transition and critical point.

The $G$-symmetry breaking phase of $H_1$ corresponds to the trivial phase of
$H_2$ (which is the symmetric phase of the algebraic $(n-1)$-symmetry
$G^{(n-1)}$) and the $G$-symmetric phase of $H_1$ corresponds to the
topologically ordered phase of $H_2$ (which is the symmetry breaking phase of
the algebraic $(n-1)$-symmetry $G^{(n-1)}$) \cite{GW14125148,KZ200514178}.  Now we
see that the critical point at the symmetry breaking transition neighbors a
phase with $G$ 0-symmetry and a phase with algebraic $(n-1)$-symmetry.
Heuristically, the emergent symmetry at the critical point is the same or
larger than the neighboring gapped phases. Thus the critical point has both the
$G$ 0-symmetry and the algebraic $(n-1)$-symmetry $G^{(n-1)}$.  In other words,
the  critical point has a categorical symmetry $G\vee G^{(n-1)}$ which is
the combination of the $G$ 0-symmetry and the algebraic $(n-1)$-symmetry
$G^{(n-1)}$. 

\subsection{Patch symmetry operators}
\label{ps2}

Now let us discuss the charges of the categorical symmetry in the model given
by the previous two descriptions. Just as the case $G=Z_2$ discussed before, we
can only create \emph{neutral charges}, by patch operators. The $0$-symmetry
patch operator creates conserved charges of $(n-1)$-symmetry, part of the
conserved charges can be measured by the $(n-1)$-symmetry patch operator, and
vice versa.

We start with the simple case that $n=1$. For a generic group, one set of patch
operators, from site $i_1$ to site $i_2$, acting on a state
$|\{g_{ij}\}\rangle$, are 
\begin{align}
W_{q,\al\bt}(i_1, i_2)= \left(\prod_{i=i_1}^{i_2}R_q (g_{ij})\right)_{\al\bt}\,,
\label{n-1patch}
\end{align}
where $\al,\bt$ runs from $1$ to $n_R$, the dimension of the irreducible representation $R_q$ of $G$. The other set of patch operators are 
\begin{align}
\calU_h (i_1, i_2)=\prod_{i=i_1}^{i_2} Q_h(i)\,.
\end{align}

 They satisfy the following commutation relations with the ordering $i_1<i_2<i_3<i_4$, and a simplified notation $W_{q}(i_1,i_3)=W_{q,13},\,\calU_h (i_2,i_4)=\calU_{h,24}$, (with the subscript $\al,\bt$ of $W_q(l_1,l_2)$ suppressed),
\begin{align}
\begin{split}
\Tr \left[W_{q,13}\;\calU_{h,24}\,(W_{q,13})^\dagger\right]=&\, \chi^q (h^{-1}) \,\calU_{h,24}\,,\\
\Tr \left[(W_{q,13})^\dagger\;\calU_{q,24}\,W_{q,13}\right]=&\, \chi^q (h) \,\calU_{h,24}\,,\\
\Tr \left[(W_{q,24})\;\calU_{h,13}\,(W_{q,24})^\dagger\right]=&\, \chi^q (h) \,\calU_{h,13}\,,\\
\Tr \left[(W_{q,24})^\dagger\;\calU_{h,13}\,W_{q,24}\right]=&\, \chi^q (h^{-1}) \,\calU_{h,13}\,.
\end{split}
\label{eq:patchop}
\end{align}
where $\chi^q(h)$ is the character of $h$ in the $q$ representation. The character represents that the $0$-symmetry charge and the algebraic $(n-1)$-symmetry charge are mutually non-local.

More generally, for any $n$, the patch operator that creates the neutral charge
for dual $(n-1)$-symmetry $G^{(n-1)}$ is defined on a $n$-dimensional patch
(disk), $D^n$, and is given by the product of star terms,\footnote{In the low energy sub-Hilbert space symmetric under the $\Inn (G)$ symmetry, $\calU_g=1$. It follows that the patch operator is defined up to a conjugacy class of a representative $h\in G$, $\calU_{g^{-1}}\calU_h (D^n) \calU_g= \calU_{ghg^{-1}}(D^n)$.}
\begin{align} 
\calU_h
(D^n)= \prod_{i\in D^n} Q_h(i).  
\label{0patch} 
\end{align} 
The $G^{(n-1)}$ neutral charge  lives on the boundary of $D^n$, denoted as
$(S^{n-1})^\vee$, living on the dual lattice. Let us call it $s$. In
particular, when $G$ is Abelian, $\calU_h (D^n)$ acts trivially inside $D^n$. 
That is $\calU_h$ is in fact defined on the boundary of $D^n$,\footnote{Note that in the
case $G$ is Abelian, we can take $U_h[(S^{n-1})^\vee]$ as the generator of a $1$-symmetry.}\footnote{In the case that $G$ is Abelian, 
$\sum_{h\in C_a}\calU_h$, where the sum is over a
conjugacy class of $a\in G$, has codimension-2, relative to the spacetime
dimension. They are the Gukov-Witten operators\cite{Gukov:2013zka}.}  

\begin{align}
\calU_h[(S^{n-1})^\vee]=&\prod_{ij\in (S^{n-1})^\vee}T_h(ij)
,\\
T_h
(ij)|g_{ij}\rangle=&\begin{cases}
|hg_{ij}\rangle & i \in D^n \\
|g_{ij}h^{-1}\rangle & j \in D^n \\
\end{cases}.\label{linkop}
\end{align}
For $G=Z_2$ and $d=2$, we recover the $s$ string operator \eqn{2dsstring}.

The other patch operator that creates conserved charges for the 0-symmetry $G$
is  $W_q (C_{ij})$ defined on any open string $C_{ij}$. The $0$-symmetry
charges are at the end points $i$ and $j$ sites of the open string. Let us call them $e$ particles.

These charges can be thought of as the topological excitations in
$(n+1)$D topological
order\cite{KW1458,LW170404221,LW180108530,GJ190509566,J200306663}. The $e$ particle corresponds to the point-like topological excitations, and the $s$ corresponds to the other topological excitations on the closed $(S^{n-1})^\vee$ surface.
For example, when $d=2$, and $G=Z_2$, $\calU_h\left[ (S^1)^\vee\right]$  is the
closed $Z_2$ vortex string operator. This conserved charge of
$\tilde{Z}_2^{(1)}$ can be thought of as coming from the topological string
excitation in $3+1$-dimensional $Z_2$ topological field theory, as discussed in
section \ref{Z2modelsaly} and illustrated in Fig. \ref{Z2flux3d}.

\subsection{An example of algebraic $1$-symmetry $S_3^{(1)}$ in $2+1$D theories}

The simplest example where the algebraic symmetry is beyond a higher symmetry
is in the $(n+1)$D model \eq{lgt} with $d=2$ and $G=S_3$.  Here, $S_3=\langle
s,r|s^3=r^2=1, rsr=s^2\rangle$ is the permutation group on $3$ elements.  The
topologically ordered phase of \eq{lgt} is described by $S_3$ gauge theory.
There are $8$ types of anyonic excitations in the model. Their fusion rules are shown in Table \ref{S3FusionRules}.

The model \eq{lgt} has an algebraic 1-symmetry, which denoted as $S_3^{(1)}$.
The generators are two Wilson line operators (see \eqn{Wq}), labeled by the two
non-trivial irreducible representations $a^1$ and $a^2$ of $S_3$.  The end of
Wilson line operators create anyons $a^1$ and $a^2$ whose fusion is descrined
in Table  \ref{S3FusionRules}.  The product of Wilson line operators is given
by the fusion of irreducible representations (\ie the fusion of the anyons
$a^1$ and $a^2$):
\begin{align}
\begin{split}
&W^{a^1}(S^1)W^{a^1}(S^1)=\one, \\
& W^{a^2}(S^1)W^{a^2}(S^1)=\one+ W^{a^1}(S^1)+ W^{a^2}(S^1), \\
&W^{a^1}(S^1)W^{a^2}(S^1)=W^{a^2}(S^1).
\end{split}
\end{align}
The product of two $ W^{a^2}(S^1)$'s reveals that the symmetry is an algebraic
$(n-1)$-symmetry beyond higher group.  In general, if $G$ is non-Abelian,
$G^{(n-1)}$ is an algebraic $(n-1)$-symmetry beyond higher group.

\subsection{A holographic point of view}
\label{catG}

\begin{figure}[t]
\includegraphics[scale=0.9]{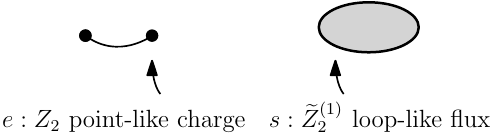}
\caption{
The conservation (the fusion rule) of the $Z_2$ point-like charge and $\t
Z^{(1)}_2$ loop-like flux in the 3+1D $Z_2$ gauge theory give rise to the
categorical symmetry of the 2+1D lattice model $H_1$ \eq{lmG}.  The mod-2
conservation of $Z_2$ charge $e$ gives rise to the $Z_2$ 0-symmetry. The mod-2
conservation of $Z_2^{(1)}$ flux $s$ gives rise to the $Z_2^{(1)}$ 1-symmetry.
$e$ and $s$ has a mutual $\pi$ statistics between them.
}
\label{3Dcatsym}
\end{figure}

Let us start with the lattice  model $H_1$ \eq{lmG} with a finite 0-symmetry
$G$.  We would like to study the $G$-symmetry within the restricted symmetric
sub-Hilbert space.  In the symmetric sub-Hilbert space, the $G$-symmetry
transformation \eq{Uh} will be trivial.  It appears that we cannot see the
$G$-symmetry.  But we can see the $G$-symmetry via point-like excitations in a
finite region, which carry non-trivial representations of $G$.  The non-trivial
fusion of $G$-representations (in particular, the fusion channel of nontrivial
representations into a trivial representation) signifies the $G$ 0-symmetry in
the symmetric sub-Hilbert space.  Thus, restricting to the symmetric
sub-Hilbert space forces us to view the $G$-symmetry via the fusion category of
the symmetry charges.  This is the categorical view of
symmetry.\cite{LW160205946}

The lattice  model $H_1$ \eq{lmG}, when restricted to the symmetric sub-Hilbert
space, has a gravitational anomaly. This is because the symmetric sub-Hilbert
space $\cV_\text{symm}$ does not have a tensor product decomposition $
\cV_\text{symm} \neq \otimes_i \cV_i$, in terms of the local  Hilbert space
$\cV_i$ on each site.  This suggests that the fusion category of the symmetry
charges is anomalous,\cite{W1313,KW1458} \ie the  fusion category can only be
realized at a boundary of a topological order in one higher dimension.  Indeed,
the model $H_1$ \eq{lmG}, when restricted to the symmetric sub-Hilbert space,
can be viewed as a boundary of $G$-gauge theory in one higher dimension, where
a simple example is discussed in Section \ref{IsTO}.\cite{JW190513279}  The
gauge charges in the $G$-gauge theory also carry representations of $G$.  The
non-trivial fusion of $G$-representations give rise to the $G$ 0-symmetry both
in the bulk 
and at the boundary. This is how the finite 0-symmetry $G$ in the model $H_1$
\eq{lmG} appears via the $G$-gauge theory in one higher dimension.

But the $G$-gauge theory also has other excitations (such as the gauge flux --
codimension-2 excitations), which also fuse in a non-trivial way and give rise
to additional symmetry to the lattice model $H_1$ \eq{lmG}.  So the complete
symmetry of the lattice model $H_1$ \eq{lmG} is given by the non-trivial fusion
of all the excitations (see Fig. \ref{3Dcatsym}). Such a complete symmetry is
called categorical symmetry of the lattice model $H_1$ \eq{lmG} (when
restricted to the symmetric sub-Hilbert space).  The categorical symmetry is
fully characterized by the $G$-gauge theory in one higher dimension. The data
in one higher dimension includes gauge charges, gauge fluxes, their fusion
rules, and the mutually non-local property. The set of data realizes on the
boundary as the global symmetry charges, the global algebraic higher symmetry
charges, their fusion as well as their mutual statistics. 
This is the holographic understanding of the categorical symmetry.  Compare to
our patch-operator understanding of the categorical symmetry discussed in
Sections \ref{ps1} and \ref{ps2}, holographic view reveals the essence of the
categorical symmetry more clearly.

Let us rephase the above holographic point of view using a categorical language
(for details see \Ref{KZ200514178} and Appendix \ref{cat}). The braiding and
fusion of the particles carrying $G$ representations is described by the fusion
$n$-category $n\Rep(G)$.  Every fusion higher category can be mapped into a
braided fusion higher category by a $Z_1$ functor, called $Z_1$ center in this
paper (see \eqn{Z1}, and, for a physical description, see for example
\Ref{KZ200514178}). The $Z_1$ center of the fusion $n$-category $n\Rep(G)$ is
denoted as $Z_1(n\Rep(G))$, which is a braided fusion $n$-category.  In fact,
$Z_1(n\Rep(G))$ describes the excitations in the $G$-gauge theory in one higher
dimension (\ie in $(d+2)$-dimensional spacetime) and is denoted as $\eG^{d}_G$.
In other words, $\eG^{d}_G=Z_1(n\Rep(G))$\cite{Freed:2018cec,KZ200514178}.  

Therefore, for a system with symmetry described by a fusion $n$-category
$n\Rep(G)$ (which is nothing but the $G$ 0-symmetry), to find its categorical
symmetry is to find the $Z_1$ center of $n\Rep(G)$: $Z_1(n\Rep(G))$.
$Z_1(n\Rep(G))$ describes the excitations in the $G$-gauge theory in one higher
dimension.  This is the holographic point of view of the categorical symmetry.

We stress that the lattice model $H_1$ \eq{lmG} (when restricted to the
symmetric sub-Hilbert space) has the full  categorical symmetry, but its ground
states may spontaneously break part of the categorical symmetry.  Those
different ground states correspond to different boundaries of the $G$-gauge
theory.  Since a gapped boundary of $G$-gauge theory always comes from
condensation of gauge charges, or gauge flux, or some combinations of them, a
gapped boundary always spontaneously breaks some part of the categorical
symmetry.  Therefore, the gapped ground states of $H_1$ always spontaneously
break some part of the categorical symmetry.  Because the gauge charge and
gauge flux have non-trivial mutual statistics between them, we cannot condense
all gauge charges and gauge fluxes simultaneously.  Therefore, any ground
states of the lattice model $H_1$ \eq{lmG} cannot break the categorical
symmetry completely.

The $G$-gauge theory has a gapless boundary if none of the gauge charges and
gauge flux is condensed. Such a boundary does not break the categorical
symmetry. Thus the lattice model $H_1$ \eq{lmG} has a gapless ground state
where the categorical symmetry is not spontaneously broken.  This gapless state
should correspond to the critical point of the Landau $G$-symmetry breaking
transition. The above discussions also apply to the dual model $H_2$ \eq{lgt}.

\section{The emergence of algebraic higher symmetry}

\label{eAHS}

We have seen that a $G$ 0-symmetry in $n$-dimensional space can be fully
characterized by a fusion $n$-category $n\Rep(G)$, describing the fusion of the
charge objects of the $G$-symmetry.  In fact, the charge objects described by
$n\Rep(G)$ are nothing but the excitations on top of a product state with the
$G$-symmetry.  To be precise, $n\Rep(G)$ describes the types of the
excitations, which are the equivalence classes under the $G$-symmetry
preserving deformations (\ie two excitations are equivalent if they can deform
into each other smoothly without breaking the symmetry).  This is why
$n\Rep(G)$ depends on the symmetry $G$, despite it describes excitations in a
trivial product state.

Similarly, the algebraic $(n-1)$-symmetry $G^{(n-1)}$ in the lattice model
\eqn{lgt} is fully characterized by a fusion $n$-category $n\Vec_G$, describing
the fusion of the charge objects of the $G^{(n-1)}$-symmetry.  Again, $n\Vec_G$
describe the types of the excitations on top of a product state with the
$G^{(n-1)}$-symmetry. Now types are the equivalence classes under the
$G^{(n-1)}$-symmetry preserving deformations.

This result can be generalized. Consider a lattice model with an algebraic
higher symmetry in $n$-dimensional space, which has a symmetric product state
as its ground state (such as \eqn{gij1}).  The types of the excitations on top
of the product state is described by a fusion $n$-category $\cR$, and the
algebraic higher symmetry is fully characterized by $\cR$.  So we will refer an
algebraic higher symmetry as $\cR$.  

We would like to remark that the fusion higher category $\cR$ describe excitations on
top of a product state is a special class of fusion higher category , called
the \emph{local fusion higher category}.  Actually, describing types of
excitations on top of a symmetric product state is the defining property of
local fusion higher category.  We believe that \emph{local fusion
$n$-categories classify algebraic higher symmetries in $n$-dimensional
space}.\cite{KZ200514178}

Now, let us consider a lattice theory or a field theory in $n$-dimensional
space, whose low energy excitations happen to be described by a local fusion
$n$-category $\cR$. 
If we ignore all the high energy excitations and pretend $\cR$ are only
excitations, then we can pretend $\cR$ to be the excitations in a product state
with the algebraic higher symmetry $\cR$.  In this way, we say that the theory
has an emergent algebraic higher symmetry $\cR$, and we can regard the system
to be in a trivial $\cR$-symmetric phase.

Let us elaborate with some examples of emergent algebraic higher symmetries.  The
first is the model with a finite 0-symmetry $G$ in $n$-dimensional space, which
we now discuss using the point of view of local fusion higher
category. If the ground state of the model is a product state with $G$
symmetry, then the excitations will be point-like and are labeled by the
irreducible representations of $G$.  Those excitations are described by a local
fusion $n$-category $n\Rep(G)$.  Thus the $G$ 0-symmetry can also be denoted as
$n\Rep(G)$ symmetry.  If the model is in the spontaneous symmetry breaking
phase, the ground states will be degenerate and are labeled by the ground
elements $g\in G$.  The excitations will be $(n-1)$-dimensional domain walls
between different degenerate ground states.  Those  domain wall excitations are
labeled by pairs $(g_1,g_2)$ if the domain wall connect the ground state $g_1$
and the ground state $g_2$.  Under a symmetry transformation $g\in G$, the
domain wall transform as $(g_1,g_2) \to (g g_1 , g g_1)$. We say $(g_1,g_2)$
and $(g g_1 , g g_1)$ are equivalent.  The equivalent classes of domain walls
(\ie symmetrized domain walls) are labeled by a single group element $h =
g_1^{-1} g_2$.  Those excitations are described by a fusion $n$-category
$n\Vec_G$.  It turns out that $n\Vec_G$ is also a local fusion
$n$-category.\cite{KZ200514178} The algebraic higher symmetry $n\Vec_G$ is nothing
but the algebraic $(n-1)$-symmetry $G^{(n-1)}$ generated by Wilson loop
operators that we discussed before.  

The second example of 2d lattice model has $S_3=Z_3\rtimes Z_2$ $0$-symmetry.
We have a phase with $S_3$ 0-symmetry.  We have another phase that
spontaneously breaks the $S_3$ 0-symmetry.  This phase has an emergent
$S_3^{(1)}$ algebraic $1$-symmetry.  We also have some other phases that break
different symmetries, and thus have different emergent algebraic higher symmetries.  All
those phases and their emergent algebraic higher symmetries are listed below:
\begin{itemize}
\itemsep0em
\item $S_3$ symmetric phase,  whose point-like charges are
\begin{align}
\cR_{S_3}=2\Rep(S_3)=\{\one, p_1, p_2\}.
\end{align}
\item $Z_2$ (charge conjugation) spontaneous symmetry breaking phase with $Z_3$ symmetry, 
whose point-like and string-like
excitations are
\begin{align}
\cR_{Z_3,Z_2^{(1)}}=\{\one, p,\bar p, s\}.
\end{align}
which include $Z_3$ charges $p,\bar p$ 
(point-like) and $Z_2$ domain wall $s$ (string-like).
\item $Z_3$ spontaneous symmetry breaking phase with $Z_2$ symmetry:
\begin{align}
\cR_{Z_2,Z_3^{(1)}}=\{\one, p, s, \bar s\}.
\end{align}
which include $Z_2$ charge $p$ 
(point-like) and $Z_3$ domain wall $s,\bar s$ (string-like).
\item $S_3$ spontaneous symmetry breaking phase, whose string-like excitations are labeled
group elements 
\begin{align}
\cR_{S_3^{(1)}}=2\Vec_{S_3} = \{s_h| h\in G\}.
\end{align}
\end{itemize}
$\cR_{S_3}=2\Rep(S_3)$ and $\cR_{S_3^{(1)}}=2\Vec_{S_3}$ are local fusion
2-categories, since they describe excitations on top of symmetric product
states, as explicilty shown in Section \ref{ahsymmG}.  They correspond to
algebraic higher symmetries: the 0-symmetry $S_3$ and the algebraic 1-symmetry
$S_3^{(1)}$.  Using the results in \Ref{KZ200514178}, we find that
$\cR_{Z_2,Z_3^{(1)}}$ and $\cR_{Z_3,Z_2^{(1)}}$ are also  local fusion
2-categories, and they also correspond to two algebraic higher symmetries.  The
algebraic higher symmetry $\cR_{Z_2,Z_3^{(1)}}$ contains a 0-symmetry $Z_2$
(the conservation of the $Z_2$ charges) and a 1-symmetry $Z_3^{(1)}$ (the
conservation of the $Z_3$ domain walls).  The algebraic higher symmetry
$\cR_{Z_3,Z_2^{(1)}}$ contains a 0-symmetry $Z_3$ (the conservation of the
$Z_3$ charges) and a 1-symmetry $Z_2^{(1)}$ (the conservation of the $Z_2$
domain walls).

Those four algebraic higher symmetries form two dual pairs:
$(2\Rep(S_3),2\Vec_{S_3})$ and $(\cR_{Z_2,Z_3^{(1)}},\cR_{Z_3,Z_2^{(1)}})$.
Moreover, the $Z_1$ center of all above $\cR$'s is the same $Z_1(\cR)=
\eG^2_{S_3}$, the same category that describes the topological data of $3$d
$S_3$ topological order, which characterizes the $S_3\vee S_3^{(1)}$
categorical symmetry. 

In fact, all the four phases discussed above have the same emergent categorical
symmetry $S_3\vee S_3^{(1)}$.  But in different phases, the categorical
symmetry is spontaneously broken in different ways.  It turns out that, to
understand the emergent algebraic higher symmetry, it is better to understand
the  emergent categorical symmetry first.  Then, the  emergent algebraic higher
symmetry is just the unbroken part of the emergent categorical symmetry.  In
the next section, we use this point of view to understand the emergent
categorical symmetry and emergent algebraic higher symmetry in a more general
setting.

\section{The emergence of categorical symmetry
(and algebraic higher symmetry)}

\label{eCS}

In this section, we consider a lattice theory or a field theory in
$n$-dimensional space, whose low energy excitations are described by a fusion
$n$-category $\cC$.  Some excitations in $\cC$ may correspond to charge objects
of a certain symmetry, and other excitations correspond topological excitations
not associated with symmetry.  Here we ignore all the high energy excitations
and pretend $\cC$ are the only excitations.  Moreover, we use the categorical point
of view of symmetry, \ie we view all the charge objects as topological
excitations, and ignore their symmetry origin.  This is possible since the
symmetry is fully encoded in the fusion of the charge objects.  Now, we would like to
ask: what is the emergent algebraic higher symmetry in a theory with low energy
excitations $\cC$?  First we would like to understand what is the emergent
categorical symmetry in a theory with low energy excitations $\cC$.

Let us consider an example of 2+1D product state with a $Z_2$-symmetry.  The
state has point-like excitations $\cC_{Z_2\text{-sym}}=\{\one, e\}$, where
$\one$ has $Z_2$ charge-0, and $e$ has $Z_2$ charge-1.  Since the
Hamiltonian has $Z_2$ symmetry, $\cC_{Z_2\text{-sym}}=\{\one, e\}$ should give
rise to the $Z_2$ symmetry.  The second example is the $Z_2$ topological order
(described by $Z_2$ gauge theory) without any symmetry.  The topological phase
has point-like excitations $\cC_{Z_2\text{-top}}=\{\one, e, m, f\}$.  Since the
Hamiltonian has no symmetry, $\cC_{Z_2\text{-top}}=\{\one, e, m, f\}$ should
not give rise to any symmetry.

When viewed as two fusion 2-categories describing topological excitations 
in 2d topological orders, why $\cC_{Z_2\text{-sym}}$ give rise to symmetry
while $\cC_{Z_2\text{-top}}$ gives rise to no symmetry?  To see their
difference, here we would like to introduce the notion of gravitational
anomaly. Conventionally, the gravitational anomaly refers to a non-invariance
of the path integral measure under the diffeomorphism transformations of spacetime
manifold.  Here, following \Ref{W1313,KW1458}, we define \emph{gravitational
anomaly} differently, as the obstruction to regularize the theory by a local
lattice bosonic model without symmetry in the same dimension.
We ask whether there exists a local lattice bosonic model without symmetry in
the same dimension, whose \emph{complete} excitations reproduce the fusion category $\cC$.  If
such a lattice model exist, then the fusion category $\cC$ is free of
gravitational anomaly.  If the lattice regularization without symmetry does not
exist, then the fusion category $\cC$ has a gravitational anomaly.  It turns
out that $\cC_{Z_2\text{-top}}$ has no gravitational anomaly, while
$\cC_{Z_2\text{-sym}}$ has a gravitational anomaly.  One may say that
$\cC_{Z_2\text{-sym}}$ can be realize as excitations in a lattice model, and
should be anomaly-free.  However, the lattice regularization of
$\cC_{Z_2\text{-sym}}$ requires a $Z_2$ symmetry.  $\cC_{Z_2\text{-sym}}$ has
no lattice regularization without symmetry, and thus has a gravitational
anomaly.

The above examples reveals a general property: if the excitations are described
by anomaly-free fusion higher category $\cC$, then there is no emergent
symmetry. Here, the emergent symmetries are global symmetries that can be
beyond group-like.\footnote{By ``beyond group-like'', we allow at least the
following two kinds of generalizations, first the multiplication of symmetry
generators is given by an algebra that is not group-like, second the charges of
the symmetry can be mutually non-local.} If $\cC$ is anomalous, then there is
an  emergent symmetry.  \footnote{When the emergent symmetry is a global
symmetry of a (finite) group $G$, the theory is a low energy theory of either a
symmetry protected phase or a spontaneously symmetry breaking phase on a local
lattice bosonic model with a global symmetry $G$. } 
We see
that \emph{emergent symmetry $\sim$ gravitational anomaly}.  So to understand
emergent symmetry, we need to understand gravitational anomaly.

But in the above, we just defined what is ``no gravitational anomaly'' as the
existence of lattice regularization in the same dimension.  We did not define
what is gravitational anomaly.  To define what is gravitational anomaly, we
rely on the following conjecture, the \textbf{holographic principle of
topological order}:\cite{KW1458,KZ150201690,KZ170200673} \emph{The excitations
in $n$-dimensional space described by a fusion $n$-category $\cC$ can always be
realized at a boundary of an anomaly-free topological order (denoted as $\sM$)
in one higher dimensions.  Moreover, the topological order $\sM$ is uniquely
determined by $\cC$, we denote this map from $\cC$ and $\sM$ as
$\sM=\bulk(\cC)$.}
Using the holographic principle, we can rephrase the anomaly-free condition for
a fusion higher category $\cC$ as
\begin{align}
\bulk(\cC)=\one, 
\end{align}
where $\one$ denotes the trivial topological order (\ie a product state with no
symmetry).  This is because if $\cC$ can be realized by a boundary of a product
state in a lattice model in one higher dimension, we can always remove the bulk
product state, and conclude that $\cC$ can be realized by a  lattice model in
the same dimension. Thus the holographic principle can tell us when there
exists a gravitational anomaly. Furthermore, the holographic principle gives a
way to regularize the anomalous theory $\cC$ on a $n+1$ dimensional lattice,
$\cC$ is realized as one low energy phase of a boundary of the lattice model.
Thus the holographic principle tells us what is gravitational anomaly:
\emph{a gravitational anomaly is a topological order in one higher
dimension.}\cite{KW1458,KZ150201690,KZ170200673}

Should we view the bulk topological order $\sM$ (the gravitational anomaly) as
the emergent symmetry?  May be not. The emergent symmetry should be related to
conservation laws encoded by fusion rules, or more precisely a fusion higher
category.  In fact, from the bulk topological order $\sM$ in
$(n+1)$-dimensional space, we can get a fusion $(n+1)$-category $\cM$
describing its excitations.  Since the  bulk topological order $\sM$ is
anomaly-free all the codimension-1 excitations are descendent (\ie formed by
codimension-2 and higher excitations). We can drop the  codimension-1
excitations, which turns the fusion $(n+1)$-category $\cM$ into a braided
fusion $n$-category $\eM$.  This turns the map from $\cC$ to $\sM$, $\sM=\bulk(\cC)$, into a map from $\cC$ to $\eM$:
\begin{align}
\label{Z1}
 \eM=Z_1(\cC).
\end{align}
So we should view the braided fusion $n$-category $\eM$ as the emergent
symmetry $\cC$.  Since $\eM$ describes the excitations in a topological order
in one higher dimension where $\cC$ is realized as a boundary, we see that
$\eM$ is actually the emergent categorical symmetry.

Now, we can say that the fusion 2-category $\cC_{Z_2\text{-top}}=\{\one, e, m,
f\}$ that describes the excitations in the 2+1D $Z_2$ topological order has no
emergent categorical symmetry since $Z_1(\cC_{Z_2\text{-top}})=\{\one\}$.
Similarly, the fusion 2-category $\cC_0=\{\one\}$ that describes the
excitations in the 2+1D product state with no symmetry has no emergent
categorical symmetry since $Z_1(\cC_0)=\{\one\}$.  Since the both phases have
no categorical symmetry.  We can find a lattice model with no symmetry in which
the above two phases can transform into each other via phase transitions. As
there is no global symmetry to be concerned, we just turn off the Hamiltonian
for $Z_2$ topological order and turn on another for the product state. 

On the other hand, the fusion 2-category $\cC_{Z_2\text{-sym}}=\{\one, e\}$
that describes the excitations in the 2+1D $Z_2$-symmetric product state has an
emergent categorical symmetry described by $Z_1(\cC_{Z_2\text{-sym}})$.  In
fact, $Z_1(\cC_{Z_2\text{-sym}})$ is the braided fusion $2$-category (denoted
as $\eG^2_{Z_2}$) describing the excitations in a 3+1D $Z_2$ gauge theory.
Therefore, $\cC_{Z_2\text{-sym}}=\{\one, e\}$ has an emergent categorical
symmetry $\eG^2_{Z_2}=Z_1(\cC_{Z_2\text{-sym}})$.  In this case, we cannot
connect a phase described by the fusion 2-category
$\cC_{Z_2\text{-sym}}=\{\one, e\}$ to a phase described by the fusion
2-category $\cC_0=\{\one\}$.

The above result sounds counter-intuitive, since $\cC_{Z_2\text{-sym}}=\{\one, e\}$ can be
realized by a $Z_2$-symmetric product state and  $\cC_0=\{\one\}$ can be
realized by a product state with no symmetry.  Two product states should be in
the same phase and are connected by phase transitions (actually connected by
zero phase transition).  When we say $Z_2$-symmetric product state, we also
specify the deformation class of the Hamiltonians, which are all required to
have the $Z_2$-symmetry.  In the phase diagram of the $Z_2$-symmetric
Hamiltonians, there is no phase whose excitations are given by
$\cC_0=\{\one\}$, but there is a phase whose excitations are given by
$\cC_{Z_2\text{-sym}}=\{\one, e\}$.  
This is what we mean by ``a phase described
by the fusion 2-category $\cC_{Z_2\text{-sym}}=\{\one, e\}$ is not connected to
a phase described by the fusion 2-category $\cC_0=\{\one\}$''. 
In general, if low energy excitations $\cC$ has non-trivial categorical
symmetry $Z_1(\cC)$, then any gapped phase formed by condensing those low
energy excitations cannot be a trivial phase with excitation $\{\one\}$.
This comes from the knowledge that $Z_1(\cC)\neq Z_1(\{\one\})$.

We see that the emergent categorical symmetry can
constrain the possible phases and phase transitions, just like the usual
symmetry does. This represents one of the most important application of
categorical symmetry (see Section \ref{dual}).

Let us  discuss more examples of emergent categorical symmetry using the 2+1D
$Z_2$ topological order.\cite{RS9173,W9164} The $Z_2$-topological order has a
trivial excitation $\one$ and three non-trivial excitations $e,\ m,\ f$ with
mod 2 conservation.  $e,\ m$ are bosons and $f$ is a fermion, and they fuse as
$e\otimes m =f$.  If the low energy excitations are $\cR=\{\one,e\}$ (and $m,f$
are assume to have very high energies), then $\cR$ is a local fusion
$2$-category $\cR=2\Rep(Z_2)$, which describes an anomaly-free $Z_2$
0-symmetry.  The system also has an emergent categorical symmetry described by
$Z_1(\cR)=Z_1(2\Rep(Z_2))=\eG^2_{Z_2}$, which is the braided fusion category
describing the excitations in $3+1$D $Z_2$-gauge theory.  Such a categorical
symmetry contains a $Z_2$ 0-symmetry and a $Z_2^{(1)}$ 1-symmetry and is
denoted as $Z_2\vee Z_2^{(1)}$.  We may also say that the low energy physics of
$\{\one,e\}$ is controlled by the emergent categorical symmetry $Z_2\vee
Z_2^{(1)}$.  For example, the $e$ excitations may condense and drive the system
to another gapped phase the spontaneously break the $Z_2$ symmetry but has the
$Z_2^{(1)}$ 1-symmetry.  The excitations in the new phase are described by
$\{\one, s\} = 2\Vec_{Z_2}$, where $s$ is a string-like excitation with $Z_2$
fusion $s\otimes s=\one$.  Such a gapped phase is possible since its has the
same categorical symmetry $Z_1(2\Vec_{Z_2}) =Z_1(2\Rep(Z_2))$.  At the
transition between the phases, we have a gapless critical point formed by
$\{\one,e\}$ (or equvalently, formed by $\{\one, s\}$). which has the unbroken
categorical symmetry $ Z_2\vee Z_2^{(1)}$.  It requires that both the $Z_2$
charges $e$ and $Z_2^{(1)}$ charges $s$ are not condensed at the critical
point.

If the low energy excitations are $\cC=\{\one,f\}$, (which is a fusion
2-category denoted as $2\Rep\left(Z_2^f\right)$), then $\cC$ is not a local
fusion higher category.  The emergent categorical symmetry is described by
$Z_1(\cC)=Z_1\left(2\Rep\left(Z_2^f\right)\right)$,  which is the braided
fusion category describing the excitations in a twisted $3+1$D $Z_2^f$-gauge
theory where the $Z_2^f$ charge is a fermion.\cite{LW0316} Such a categorical
symmetry contains a $Z_2^f$ 0-symmetry (the fermion number parity) and a $\t
Z_2^{(1)}$ 1-symmetry, and is denoted as $Z_2^f \vee Z_2^{(1)}$.  The
categorical symmetry controls the low energy dynamics of $\{\one,f\}$, which
can be simulated by the boundary of the twisted $3+1$D $Z_2^f$-gauge theory.
For example, we cannot have a phase that spontaneously breaks the fermionic
$Z_2^f$ symmetry.  Also, the  categorical symmetry $Z_2^f \vee Z_2^{(1)}$
is an example that no corresponding algebraic higher symmetry $\cR$, \ie there
is no local fusion 2-category $\cR$ and satisfies $Z_1(\cR)=Z_1\left(2\Rep\left(Z_2^f\right)\right)$. Physically, this implies that there is no 2+1D bosonic systems, with
or without symmetry, whose gapped state gives rise the excitations described
$\cC=\{\one,f\}$.

In summary, for a system with low energy excitations described by a fusion
category $\cC$, when $\cC$ is a local one, the system has an emergent algebraic
higher symmetry. In general, the largest emergent symmetry is the categorical
symmetry given by $Z_1(\cC)$. Phases that have the same categorical symmetries
are connected through phase transitions.  Since each of them is a phase
spontaneously breaking part of the same categorical symmetry. Starting from the
critical point that has the full category symmetry, the system can break
different part of the category symmetry and drive a transition to those spontaneous symmetry breaking phases.

\subsection{Categorical symmetry, anomaly, and duality}
\label{dual}

Let us consider two field theories\footnote{Here by \emph{field theory}, we
mean that the UV regularization is not specified. In particular, when we say
two phases in two \emph{field theories} are connected by phase transitions, we
mean that there exist UV regularizations (such as lattice models) for each
field theories, and for such regularized field theories, the two phases are
connected by phase transitions.  (There are may be other different UV
regularizations where the two phases are not connected by phase transitions.)}
whose low energy excitations are described by two fusion higher categories
$\cC_1$ and $\cC_2$ respectively.  The two field theories may have different
algebraic higher symmetries, and $\cC_1, \cC_2$ may contain the charge objects
of those algebraic higher symmetries.  We would like to ask the following
questions. Can the two theories be connected through phase transitions?  Can we
use one theory to simulate the other theory?  Are the two theories dual to each
other?  In fact, the above three questions are the same question, since the
following five statements are equivalent:\footnote{See for example\cite{Gattringer:2014nxa}, for the recent development of simulating lattice gauge theories in dual variables with defects included.}
\begin{enumerate}[label=(\arabic*)] \itemsep0em 
\item The two theories describing phases which are connected through phase 
transitions and other phases.
\item  The two theories can simulate each other.  
\item The two theories are dual to each other.  
\item The two theories have the same anomaly.  
\item The two theories have the same categorical symmetry 
$Z_1(\cC_1)=Z_1(\cC_2)$.
\end{enumerate} \Ref{W1313,KW1458} pointed out that anomaly is simply
topological/SPT order in one higher dimension. 
In light of this view of
anomaly, $Z_1(\cC_1)$ and $Z_1(\cC_2)$ simply correspond to the anomalies of
the two theories, and $Z_1(\cC_1)=Z_1(\cC_2)$ is simply the anomaly matching
condition.  We would like to remark that $Z_1(\cC_1)$ and $Z_1(\cC_2)$ are in general
non-invertible anomalies.\cite{JW190513279}

However, to be more precise, we should replace $Z_1(\cC_1)$ and $Z_1(\cC_2)$ in
the above by the bulk topological orders of $\cC_1$ and $\cC_2$, which are
denoted as $\bulk(\cC_1)$ and $\bulk(\cC_2)$ respectively.  The anomaly
matching condition is really $\bulk(\cC_1)=\bulk(\cC_2)$.  Such a condition
implies that $\cC_1$ and $\cC_2$ can be viewed as two boundaries of the same
bulk topological order.  We know that two theories that are boundary theories
of the same bulk topological order can simulate each other. 

We remark that $Z_1(\cC_1)$ describes the excitations in the bulk topological
order $\bulk(\cC_1)$. $Z_1(\cC_1)$ contains a little less information than the
bulk topological order $\bulk(\cC_1)$: two topological orders differ by an
invertible topological order has the same excitations.

We know that symmetry breaking transitions are induced by the condensation of
charge objects of the symmetry, while topological phase transitions are induced
by condensing topological excitations. But in our setup , the charge objects and
topological excitations are treated at equal footing, and thus symmetry
breaking transitions and topological transitions  are treated at equal footing.
So the emergence of categorical symmetry happens at both symmetry breaking
transitions and topological transitions, as well as their mixtures.

As an application, we may consider a 2+1D field theory with a $Z_2$
$0$-symmetry and another 2+1D field theory with a $Z_2^{(1)}$ $1$-symmetry.
Both symmetries give rise to the same categorical symmetry (\ie the same bulk
topological order), which is described by the 3+1D $Z_2$ gauge theory. As a
result, the two theories have the same gravitational anomaly.  Then, the two theories can be
connected by phase transitions, can simulate each other, and are dual to
each other, even if we do not explicitly break the two symmetries (\ie under
the constraint of the two symmetries).

\subsection{An example: Higgs and confinement transition in 3+1D $Z_2$ gauge
theory}\label{3dIsing}

Let us discuss the simple example of 3+1D $Z_2$ topological order to
illustrate the above general results.  In the first case, we choose the low
energy subcategory $\cR_p$ to be the one formed by all the point-like
excitations, \ie the $Z_2$ charges (which is denoted as $3\Rep(Z_2)$).  We
assume all other excitations (such as gauge flux) to have infinite energy.  In
this case, we can focus only on excitations described by $\cR_p=3\Rep(Z_2)$,
and our system can be viewed as a system with $Z_2$ 0-symmetry. Our previous
discussions on $G$-symmetric system will apply.  In particular, we have an
(emergent) categorical symmetry characterized by $Z_1(\cR_p)=\eG^3_{Z_2}$,
which is a $Z_2$-gauge theory in 4+1D. The categorical symmetry  contains $Z_2$
0-symmetry from the fusion of the point-like $Z_2$ charges in the 4+1D
$Z_2$-gauge theory.  The categorical symmetry  also contains $Z_2^{(2)}$
2-symmetry from the fusion of the membrane-like $Z_2$ flux in the 4+1D
$Z_2$-gauge theory.  The condensation of the $Z_2$ charge induces a Higgs
transition from the $3+1D$ $Z_2$ topological order to trivial order.  The
critical point of the Higgs transition has the $Z_2 \vee Z_2^{(2)}$
categorical symmetry.  In fact, such a critical point is the same as the $Z_2$
symmetry breaking critical point discussed before.

In the second case, we choose the low energy  subcategory $\cR_s$ to be the one
formed by the pure string-like excitations, \ie the $Z_2$ flux lines.  We
assume all other excitations (such as gauge charges) to have infinite energy.
After ignoring other excitations, the only excitations are described by
$\cR_s$, and our system can be viewed as a system with $Z_2^{(1)}$ 1-symmetry.
Such a system has a categorical symmetry given by
$Z_1(\cR_s)=\eG^3_{Z_2^{(1)}}$, which is a $Z_2^{(1)}$ 2-gauge theory in
4+1D.\cite{BSm0511710,GP07083051,BH10034485,KT13094721,BM160606639,BM170200868,CT171104186,ZW180809394}
The categorical symmetry  contains $Z_2$ 1-symmetry from the fusion of the
string-like $Z_2^{(1)}$ charges in the 4+1D $Z_2^{(1)}$ 2-gauge theory.  The
categorical symmetry  also contains $\t Z_2^{(1)}$ 1-symmetry. The charge of
$\t Z_2^{(1)}$ symmetry is the string-like $Z_2^{(1)}$ flux in the 4+1D
$Z_2^{(1)}$ 2-gauge theory.  The condensation of the string-like $Z_2^{(1)}$
charges induces a confinement transition from the $Z_2$ topological order to
trivial order.  The critical point of the confinement transition has the
$Z_2^{(1)} \vee \t Z_2^{(1)}$ categorical symmetry.  

Such a critical point with $Z_2^{(1)} \vee \t Z_2^{(1)}$ categorical
symmetry can be realized by a model on a cubic lattice with a spin-1/2 living
on each link.  We label the sites by $i$ and links by $ij$. The Hamiltonian of
our model is given by
\begin{align}
\label{H3d}
 H &= - B \sum_{\<ij\>} Z_{ij} 
- J \sum_{\<ijkl\>} X_{ij} X_{jk}X_{kl}X_{li} 
\nonumber\\
&\ \ \ \
+ U \sum_i\left(1- \prod_{j\text{ next to } i} Z_{ij}\right),
\end{align}
where $\sum_i$ sums over all sites, $\sum_{\<ij\>}$ over all links, and
$\sum_{\<ijkl\>}$ over all squares of the cubic lattice.
We also assume $U\to +\infty$ and $B,J>0$.

When $B=0$ and $J>0$, the above model $H$ has ground state with a $Z_2$
topological order.  When $B>0$ and $J=0$, the model has a trivial product state
$|\{Z_{ij}=1\}\>$ as its ground state and is in the trivial phase (the confined
phase). Changing from $(B=0,J>0)$ to $(B>0,J=0)$ induces a confinement
transition.

The model explicitly has a $Z_2^{(1)}$ 1-symmetry generated by the
following membrane operator
\begin{align}
 C_{Z_2^{(1)}} (\t M^2) = \prod_{\<ij\> \in \t M^2} Z_{ij}.
\end{align} 
The membrane operator $C_{Z_2^{(1)}} (\t M^2)$ acts on a membrane $\t M^2$
in the dual cubic lattice, where the membrane $\t M^2$ is formed by the squares of
the dual lattice. Since the squares of the dual lattice one-to-one correspond
to the links in the original cubic lattice, thus the membrane $\t M^2$ is
formed by the links of the original lattice.  $ \prod_{\<ij\> \in \t M^2}$ is a
product over all the links from $\t M^2$.  Such a $Z_2^{(1)}$
1-symmetry is spontaneously broken in the phase with $Z_2$ topological order,
and is unbroken in the trivial phase (the confined phase).

\begin{figure}[t]
\begin{center}
\includegraphics[scale=0.5]{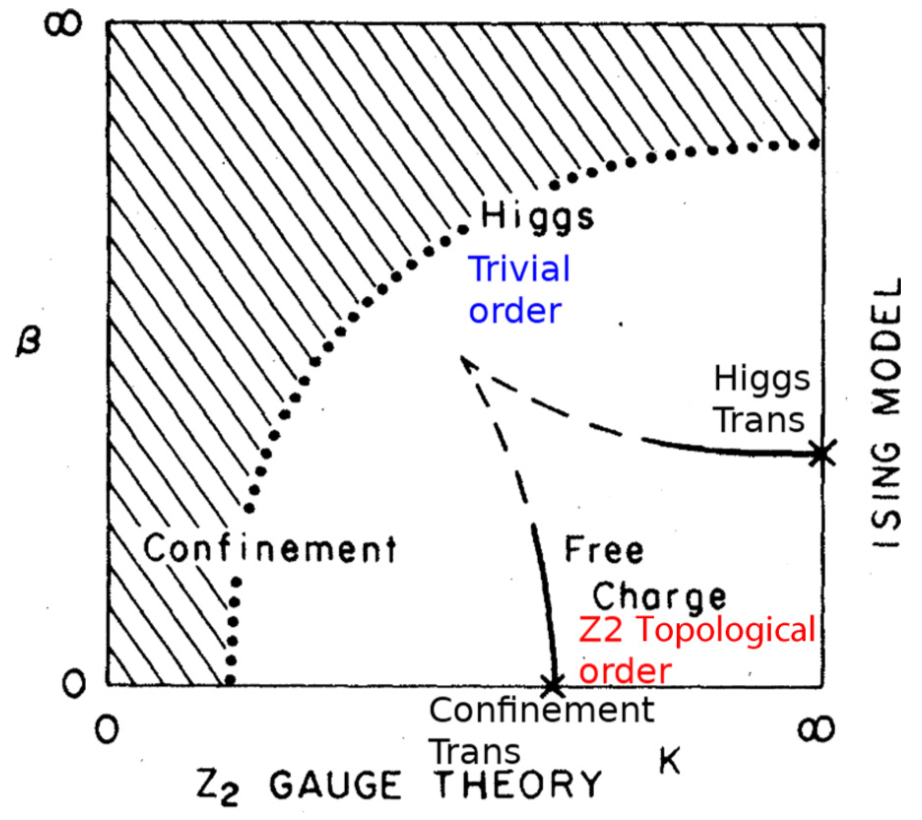} \end{center}
\caption{ 
The phase diagram of \eqn{FS}.
}
\label{FradkinShenker}
\end{figure}

To see the  $\t Z_2^{(1)}$ 1-symmetry explicitly, we need to do a duality
transformation to obtain a new model with a spin-1/2 living on each square face
$\<ijkl\>$ of the cubic lattice. Let $ \t X_{ijkl}, \t Y_{ijkl}, \t Z_{ijkl} $
be the Pauli operators acting on the spin-1/2 on face $\<ijkl\>$.  The duality
mapping is given by
\begin{align}
X_{ij} X_{jk}X_{kl}X_{li} &\to \t X_{ijkl}, &
Z_{ij}  &\to \prod_{k,l\text{ next to }i,j} \t Z_{ijkl}, 
\nonumber \\
\prod_{\<ijkl\> \in c} \t X_{ijkl} &=1.
\end{align}
where $ \prod_{\<ijkl\> \in c}$ is the product over all the six faces of a cube
$c$.  The dual Hamiltonian is given by
\begin{align}
\label{tH3d}
\t H &= - B \sum_{\<ij\>} \prod_{k,l\text{ next to }i,j} \t Z_{ijkl}
- J \sum_{\<ijkl\>} \t X_{ijkl}
\nonumber\\
&\ \ \ \
+ \t U \sum_c\left(1- \prod_{\<ijkl\> \in c} \t X_{ijkl}\right),
\end{align}
where $\t U\to \infty$.

When $B>0$ and $J=0$, the above dual model $\t H$ has ground state with a dual
$\t Z_2$ topological order (which corresponds to the confined phase with the
trivial $Z_2$ topological order in the original model \eq{H3d}).  When $B=0$
and $J>0$, the model has trivial product state $|\{\t X_{ijkl}=1\}\>$ as its
ground state and is in the dual trivial phase (which corresponds to the phase
with the $Z_2$ topological order in the original model \eq{H3d}).  Changing
from $(B>0,J=0)$ to $(B=0,J>0)$ induces a dual confinement transition, while
changing from $(B=0,J>0)$ to $(B>0,J=0)$ induces a confinement transition.

The dual model $\t H$ explicitly has a $\t Z_2^{(1)}$ 1-symmetry generated by
the following membrane operator
\begin{align}
 C_{\t Z_2^{(1)}} (M^2) = \prod_{\<ijkl\> \in M^2} \t X_{ijkl}.
\end{align} 
The membrane operator $C_{\t Z_2^{(1)}} (M^2)$ acts on a membrane $M^2$ which
is formed by the squares of the lattice.  Such a $\t Z_2^{(1)}$ 1-symmetry is
spontaneously broken in the phase with dual $\t Z_2$ topological order (the
confined phase with the trivial $Z_2$ topological order), and is unbroken in
the dual trivial phase (the phase with the $Z_2$ topological order).  \emph{If
the confinement transition from the $\t Z_2$ topological order to the trivial
order is given by a single critical point}, then such a gapless critical state
will have the full  $ Z_2^{(1)} \vee \t Z_2^{(1)}$ categorical symmetry
which is not spontaneously broken.  

It is well known that there are two ways to go from 3+1D $Z_2$
topological order (\ie $Z_2$ gauge theory) to trivial order, either via Higgs
condensation or via confinement. \Ref{FS7982} studied a model defined by the
3+1D path integral for the following action
\begin{align}
\label{FS}
S &= \frac{K}{2} \sum_{\<ijkl\>} \si_{ij} \si_{jk} \si_{kl} \si_{li}
+\frac{\bt}{2} \sum_{ij} \phi_i^\dag  \si_{ij} \phi_j
\nonumber\\
&  \si_{ij}=\pm 1,\ \ \
\phi_{i}=\pm 1.
\end{align}
The model has $Z_2$ topological order when $\bt\sim 0$ and $K\sim +\infty$.
The model has trivial order when $\bt\sim +\infty$ or $K\sim 0$ (see Figure
\ref{FradkinShenker}).

When $K=+\infty$, as we change $\bt$ from 0 to $+\infty$, the model goes
through a second-order Ising transition (\ie Higgs transition) described by a
critical point.  When $\bt =0$, as we change $K$ from 0 to $+\infty$, the model
goes through a first-order confinement transition\cite{CJR7915}.  Let us assume
that we can modify the model to make the  confinement transition to be a
continuous transition described by a critical point.  One may wonder whether
the two critical points for the two transitions are the same or not.  

Our above discussions suggest that the two critical points to be different
since they have different categorical symmetries. One has $ Z_2 \vee Z_2^{(2)}$
categorical symmetry and the other has $ Z_2^{(1)} \vee \t Z_2^{(1)}$
categorical symmetry, and $Z_1(3\Rep (Z_2))$ are different from
$Z_1\left(3\Vec_{Z_2^{(1)}}\right)$.  Thus categorical symmetries deepen our
understanding of the Higgs transition and the confinement transition, as well
as their relation.

\subsection{The emergent maximal categorical symmetry}

We know that the 1+1D critical point of $Z_2$ symmetry breaking transition has the
following modular invariant partition function:
\begin{align}
\label{Z2CFT}
Z(\tau)=|\chi^\text{Is}_0(\tau)|^2+|\chi^\text{Is}_\frac12(\tau)|^2+
|\chi^\text{Is}_{\frac1{16}}(\tau)|^2 .
\end{align}
Such a CFT and its modular invariant partition function has no gravitational
anomaly and can be realized by a 1+1D lattice model.  In other words, such a CFT
can be realized on the boundary of a trivial 2+1D topological order.  Since the
1+1D categorical symmetry is just another name for 1+1D non-invertible
gravitational anomaly, which is another name for 2+1D topological order, we say
the above formulation of the $Z_2$-symmetry-breaking critical point does not
expose any categorical symmetry.

We know that the critical point actually has a $Z_2$ symmetry. To expose the
$Z_2$ symmetry, we consider systems with $Z_2$-symmetry twisted boundary
condition, as well as even-odd $Z_2$ charges.  This gives us a 4-component
partition function as in \eqn{Z2cri}.  The modular covariance of the
4-component partition function implies that the CFT have a non-invertible
gravitational anomaly, which is characterized by the 2+1D $Z_2$ gauge theory.  In other
words, the CFT \eq{Z2cri} has the $Z_2\vee \t Z_2$ categorical symmetry.

We see that the different formulation of the $Z_2$-symmetry-breaking critical
point reveal or expose different emergent symmetries.  The first modular
invariant formulation \eq{Z2CFT} does not reveal any categorical symmetry.  The
second formulation \eq{Z2cri} exposes the $Z_2\vee \t Z_2$ categorical
symmetry.  In particular, it exposes both the $Z_2$ symmetry and the $\t Z_2$
dual symmetry.

Here, we would like to ask whether the $Z_2$-symmetry-breaking critical point has
additional emergent symmetries.  What is the maximal categorical symmetry
of the $Z_2$-symmetry-breaking critical point?  In fact, the $Z_2\vee \t Z_2$
categorical symmetry is not maximal. The $Z_2$-symmetry-breaking critical point
has even larger categorical symmetry.

One way to find large emergent categorical symmetry of a CFT, is to choose a
smaller set of conformal characters as the trivial component of the partition
function. For the example of the $Z_2$-symmetry-breaking critical point,
we may choose the  trivial component of the partition
function to be
\begin{align}
 Z_\one(\tau) = |\chi^\text{Is}_0(\tau)|^2.
\end{align}
Such a partition function is a part of a 9-component modular covariant
partition function
\begin{align}
\label{Z2CFT2}
 Z_{h,h'} = \chi^\text{Is}_h(\tau) \bar \chi^\text{Is}_{h'}(\bar \tau),
\ \ \ h,h' =0,\frac12,\frac1{16},
\end{align}
where $Z_\one(\tau) =|\chi^\text{Is}_0(\tau)|^2 =Z_{0,0}(\tau) $.  The modular
covariance of the 9-component partition function implies that the CFT have a
non-invertible gravitational anomaly, which is characterized by 2+1D
double-Ising $3^B_{1/2}\boxtimes 3^B_{-1/2}$ topological order (using the
notation in \Ref{W150605768} where $n^{B}_c$ represents a bosonic topological other with $n$ superselection sectors and chiral central charge $c$, and $\boxtimes$ is the stacking operation).  In
other words, the CFT \eq{Z2CFT2} has a double-Ising $3^B_{1/2}\boxtimes
3^B_{-1/2}$ categorical symmetry.  In particular, \eqn{Z2CFT2} exposes both the
$Z_2$ symmetry and the $\t Z_2$ dual symmetry, as well as the $Z_2$ symmetry
from the Kramers-Wannier duality.  We believe that the double-Ising categorical
symmetry $3^B_{1/2}\boxtimes 3^B_{-1/2}$ is the maximal emergent categorical
symmetry of the $Z_2$-symmetry-breaking critical point.\cite{MS8916}

There is another way to obtain  the larger emergent categorical symmetry.  We
know that the 1+1D $Z_2$-symmetry-breaking critical point can be viewed as a
boundary of 2+1D $Z_2$ topological order, which has four types of topological
excitations $\one,e,m,f$.  This point of view reveals the emergent $Z_2\vee \t
Z_2$ categorical symmetry of the critical point.  We can fine-tune the 2+1D
$Z_2$ topological order to make it to have a $e,m$ exchange symmetry.  We can
then gauge such a $e,m$ exchange symmetry.\cite{Barkeshli19,HL160607816}  The
resulting 2+1D topological order is the double-Ising topological order
$3^B_{1/2}\boxtimes 3^B_{-1/2}$.  This corresponds to the large double-Ising
$3^B_{1/2}\boxtimes 3^B_{-1/2}$ categorical symmetry. 

Next, let us consider another example of $c=\bar c= \frac 7{10}$ CFT.  The
4-component partition function \eq{CFT54} reveal the emergent $Z_2\vee \t Z_2$
categorical symmetry of the CFT.  However, the  $c=\bar c= \frac 7{10}$ CFT has
a larger emergent categorical symmetry, as exposed by the following
36-component partition function
\begin{align}
\label{CFT54M}
\begin{split}
 Z_{h,h'} =& \chi^{5,4}_h(\tau) \bar \chi^{5,4}_{h'}(\bar \tau),\\
h,h' =&\,0,\frac32,\frac7{16},\frac1{10},\frac35,\frac3{80}.
\end{split}
\end{align}
Such a maximal emergent categorical symmetry is characterized by the 2+1D
topological order $6^B_{7/10}\boxtimes 6^B_{-7/10}$ (again using the notation
in \Ref{W150605768}).  In fact, $6^B_{-7/10}=2^B_{14/5}\boxtimes 3^B_{-7/2}$,
where $2^B_{14/5}$ is the Fibonacci topological order, and $3^B_{-7/2}$ is a
twisted Ising topological order, whose $\sigma$ anyon has a topological spin
$e^{-\ii 2\pi \frac{7}{16}}$.

The above two examples suggest that a CFT can have a maximal categorical
symmetry.  We hope that the maximal categorical symmetry carries a lot of
information about the corresponding CFT.  It may be possible that we can use
the maximal categorical symmetry to characterize or even classify the CFTs.
Thus maximal categorical symmetry represents a research direction that may
allow us to develop a systematic theory of gapless CFTs, via mapping class
group representations for various spacetime topologies.  We also note that the
maximal categorical symmetry of a 1+1D CFT has a form $\cC\boxtimes \bar \cC$,
where $\cC$ is a 2+1D topological order and $\bar\cC$ is the time-reversal
conjugate of $\cC$.  This structure was referred to as naturality principle of
maximally extended CFT in \Ref{MS8916}. In fact, modular tensor category -- the
mathematical description of bosonic topological orders, has been used to
classify all rational modular invariant conformal field theories
\cite{FSh0110133,FSh0204148,FSh0306164,FSh0403157,FSh0412290,FSh0503194,RSm0512076,KR08073356,K11073649,KR13101875}
Such classification has also been interpretated physically in terms of surface
operators of $2+1$D topological order \cite{KS10120911}. 


\section{Summary}
\label{sum}

\begin{figure}[t]
\includegraphics[scale=0.55]{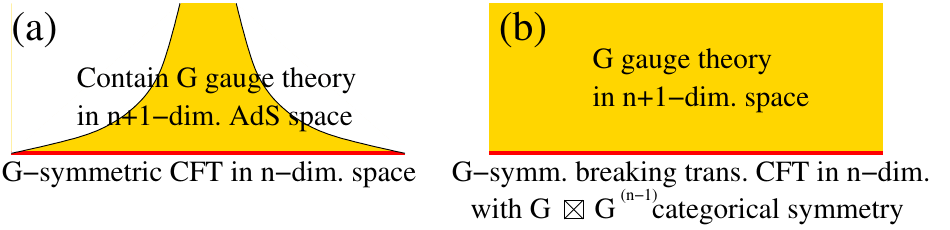}
\caption{
(a) A CFT with 0-symmetry $G$ is described by a AdS/CFT dual
that contains a $G$ gauge theory in the AdS space in one higher dimension.
(b) The CFT at the transition of spontaneously $G$-symmetry breaking
has a categorical symmetry $\eG^n_G = G\vee G^{(n-1)}$
described by excitations in a $G$-gauge theory in one higher dimension.
}
\label{AdSCFTgauge}
\end{figure}

\begin{figure}[t]
\includegraphics[scale=0.6]{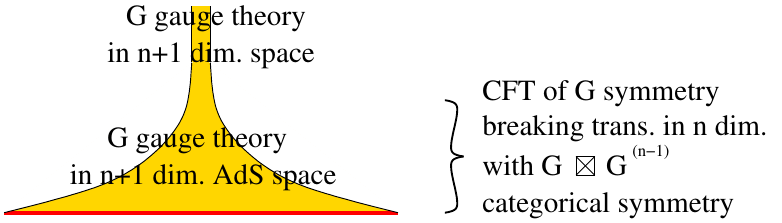}
\caption{
A combination of the holographic picture for AdS/CFT duality and the
holographic picture for emergent categorical symmetry in the CFT for
the $G$-symmetry breaking transition.  } \label{AdSCFTcatsymm}
\end{figure}

In this paper, we show that a quantum \emph{system} in $n$-dimensional space
with a 0-symmetry $G$ actually has a larger symmetry, which includes both the
0-symmetry $G$ and an algebraic $(n-1)$-symmetry, denoted as $G^{(n-1)}$, whose
transformations are given by Wilson loop operators.  In fact, the $G$-symmetric
quantum system actually has a larger categorical symmetry characterized by a
braided fusion $n$-category $\eG^n_G$.  We also denote the categorical symmetry
$\eG^n_G$ as $G\vee G^{(n-1)}$, since it includes both the 0-symmetry $G$
and the algebraic $(n-1)$-symmetry $G^{(n-1)}$.

We find that any gapped \emph{state} in a \emph{system} with a categorical
symmetry must partially (and only partially) break the categorical symmetry
spontaneously.  However, for the gapless critical state at the transition of
the spontaneous $G$-symmetry breaking, the categorical symmetry $G\vee
G^{(n-1)}$ is not broken.  In particular, the critical state has both the
0-symmetry $G$ and the algebraic $(n-1)$-symmetry $G^{(n-1)}$.

It was proposed that a CFT in $n$-dimensional space with 0-symmetry $G$ has a
AdS/CFT dual that contain a $G$-gauge theory in the $(n+1)$d AdS space (see
Fig.  \ref{AdSCFTgauge}a).\cite{M9831,Wh9802150,KWh9905104,HO181005338} If the
AdS/CFT dual contains only a $G$-gauge theory (with the fluctuations of both
gauge charges and gauge flux) and gravity in the $(n+1)$d AdS space (see Fig.
\ref{AdSCFTgauge}a), then the corresponding CFT in $n$-dimensional space must
be a particular one with 0-symmetry $G$.  But there are many CFT's with
0-symmetry $G$. Which is the right one?  The result in this paper (see Fig.
\ref{AdSCFTgauge}b) suggests that: It is the CFT at the transition of the
spontaneous $G$-symmetry breaking that is dual to a theory that contains only a
$G$-gauge theory and gravity in the $(n+1)$d AdS space. This is because the specific CFT has the categorical symmetry $G\vee G^{(n-1)}$, whose excitations matches that of the dual theory in the AdS bulk.  In other words, the whole Fig.
\ref{AdSCFTgauge}a corresponds to the CFT boundary in Fig.  \ref{AdSCFTgauge}b.
This is demonstrated in Fig. \ref{AdSCFTcatsymm}.  The categorical symmetry of
a CFT can help us to select the AdS/CFT dual of the CFT, which is an important
application of the categorical symmetry.

We would like to mention that the close relation between topological order and its
canonical gapless CFT boundary has been discussed extensively in
\Ref{KZ170501087,KZ190504924,KZ191201760}, for the case of 2+1D topological
order and 1+1D gapless boundary.  A notion of topological Wick-rotation is
introduced to describe how a 2+1D topological order can determine a canonical
1+1D CFT boundary.

~

~

We would like to thank Liang Kong, Tian Lan, Shu-heng Shao, Hao Zheng for many
helpful discussions.   This research was partially supported by NSF DMS-1664412.
This work was also partially supported by the Simons Collaboration on
Ultra-Quantum Matter, which is a grant from the Simons Foundation (651440).

\appendix 
\allowdisplaybreaks

\section{List of terminologies} \label{cat}

Here, we explain some mathematical terminologies used in this paper, at the
level of physical rigorousness.  An ordinary \textbf{category} has
\textbf{objects} and \textbf{morphisms} (also called 1-morphisms), and it is also called 1-category.  A
2-category generalizes this by also including 2-morphisms between the
1-morphisms. Continuing this up to $n$-morphisms between $(n-1)$-morphisms
gives an \textbf{$n$-category}. 

In condensed matter physics, a \textbf{$n$-category} can correspond to a
collection of topological orders in $n$D spacetime.  The topological orders in
the collection correspond to the simple objects in the category.  The composite
objects correspond to degenerate states where several different topological
orders happen to have the same energy.  The gapped domain walls between
different topological orders correspond to 1-morphisms between different simple
objects.  The gapped domain walls within the same topological order correspond
to 1-morphisms between the same simple object.  The domain walls (1-morphisms)
can also have domain walls with one lower dimension between them, which
correspond to 2-morphisms, \etc.  In general, a $m$-morphism that connects the
trivial $m-1$ morphism to itself corresponds to a codimension-$m$ excitation.
A $m$-morphism that connects a non-trivial $m-1$ morphism to itself corresponds
to a domain wall on the codimension-$m-1$ excitation.  A $m$-morphism that
connects two $m-1$ morphisms corresponds to a domain wall between the two
codimension-$m-1$ excitations.  A $n$-morphism corresponds to an ``instanton''
in spacetime (\ie an insertion of a local operator in spacetime).  A
$(n-1)$-morphism corresponds a point-like excitation.

Two topological orders can be stacked to form the third topological order.
Under the stacking operation, topological orders form a monoid (which is
similar to a group, but no need for an inverse operation) \cite{KW1458}.  If we
include the stacking operation in the $n$-category, the $n$-category will
become a \textbf{fusion $n$-category}.  Thus a collection of topological orders
in $n$-dimensional spacetime is actually described by a monoidal $n$-category.

The excitations in a single topological order in $n$D spacetime is described by
a fusion $(n-1)$-category.  Now the objects in the fusion $(n-1)$-category
correspond codimension-1 excitations (domain walls).  The 1-morphisms
correspond codimension-2 excitations (domain walls on domain walls), \etc.

An example of fusion 1-category is $\Vec$.  The objects in $\Vec$ are the
vector spaces.  The simple objects are the (equivalent classes of)
1-dimensional vector spaces and composite objects are multi-dimensional vector
spaces.  There is only one simple object.  We see that composite objects are
direct sums of simple objects.  The morphisms (the 1-morphisms) correspond to
the linear operators acting on the vector spaces.  The tensor product of the
vector spaces defines the fusion of the objects.  We see that 1-dimensional
vector space is the unit of the tensor product, and hence the simple object is
the unit of the fusion. We also call such a fusion unit as the trivial object.
The 1-morphisms that connect the simple object to itself is proportional to the
one-by-one identity operator, and thus are also trivial.  So we refer fusion
1-category $\Vec$ as \textbf{trivial category}.  We also have \textbf{trivial
higher category}, which has only one simple non-descendant morphism (which are
not condensation of other
excitations\cite{GJ190509566,J200306663,KZ200308898,KZ200514178}) at every
level.  We denote the trivial higher category as $n\Vec$.

Another example of fusion $n$-category is $\Rep(G)$.  The $(n-1)$-morphisms in
$\Rep(G)$ are the representations of the group $G$ which correspond to
point-like excitations in $n$-dimensional space.  The tensor product of the
$G$-representations defines the fusion of the $(n-1)$-morphisms.  The simple
$(n-1)$-morphisms are the (equivalent classes of) irreducible representations
and composite objects are the reducible representations.  The reducible
representations are direct sums of irreducible representations, and thus
composite morphisms are direct sums of simple morphisms.  The $n$-morphisms
correspond the symmetric operators acting one on group representations.  

All the 1-morphisms, 2-morphisms on the trivial object in a fusion $n$-category
form a \textbf{braided fusion $(n-1)$-category}.  It can be used to describe
codimension-2 and higher excitations in an anomaly-free topological order
(after dropping the codimension-1 excitations which are always descendent for
anomaly-free topological orders).

We also need a notion of \textbf{local fusion $n$-category}: A fusion
$n$-category $\cF$ is \textbf{local} if we can add morphisms in a
consistent way, such that all the resulting simple morphisms are isomorphic to
the trivial one.  Physically, the process of ``adding morphisms'' corresponds
to explicit breaking of the (algebraic higher) symmetry.  This is because,
$\cF$ only has morphisms that correspond to symmetric operators.  Adding
morphisms means include morphisms that correspond to symmetry breaking
operators.  If after breaking all the symmetry, $\cF$ becomes a trivial product
phase of bosons or fermions, then $\cF$ is a local fusion $n$-category.

\section{Proof of equivalence of the 1d (projected) $Z_2$ minimally coupled model and the (projected) Ising model}\label{Z2equivmodels}

We are going to prove that the (projected) minimally coupled model \eqn{HZ2Z2}, the (projected) Ising model \eqn{HIs} and the (projected) dual Ising model \eqn{HDW} are all equivalent. In this subsection, we always consider the projected sub-Hilbert space respectively.

The strategy is to show that in the minimally coupled model, the Hamiltonian is built from the logical operators within a stabilized subspace. 
We begin with $2N$ physical qubits on $2N$ sites. We take those states stabilized by $Z_{2i-1}Z_{2i}Z_{2i+1}$, for $i=1,\cdots, N$. This is to constraint ourselves to states $Z_{2i-1}Z_{2i}Z_{2i+1}=1$.

We first show that the minimally coupled model is equivalent to the Ising model. The proof uses the stabilized code formalism in quantum information (QI). 

For simplicity, we switch the notation in the minimal coupled model to that in QI, $X_i\rightarrow X_{2i-1}, \tilde{X}_{i+\frac{1}{2}}\rightarrow X_{2i}$, and similarly for the Pauli $Z$'s. In this way, the low energy constraints in the limit $U\rightarrow \infty$ are mapped to the stabilizers. And the Hamiltonian becomes
\begin{align}
H=-B\sum_i X_{2i}X_{2i+1}X_{2i+2}-J\sum_i Z_{2i},
\end{align}
acting on the $2N$ number of physical qubits. And 
there are then in total $2^N$ logical qubits.

First, we show 
\begin{align}
\mathcal{X}_{i}=X_{2i}X_{2i+1}X_{2i+2},~~~\mathcal{Z}_{i}=Z_{2i+1},
\label{logicPauli}
\end{align}
are the Pauli-$X$ and Pauli-$Z$ operators on the logical qubits. (And $\mathcal{X}_N=X_{2N}X_1X_2, \mathcal{Z}_N=Z_1$.) Obviously, they satisfy the algebra of the Pauli matrices, and each of them commute with the stabilizers. Since there are $N$ Pauli-$Z$ operators we find, their eigenvalues labels are in one-to-one correspond to the states in the stabilized space. Therefore, these logical Pauli operators $\{\mathcal{X}_i, \mathcal{Z}_{i}|i=1,\cdots, N\}$ form a complete basis for all operators acting on the logic qubits. 

Then we find that $\mathcal{Z}_{i}\mathcal{Z}_{i+1}=Z_{2i+2}$ (and $\mathcal{Z}_N\mathcal{Z}_1=Z_2$) are the same as a pair of Pauli-$Z$ operators on the logical qubits. It follows from (\ref{logicPauli}) that $\mathcal{Z}_i\mathcal{Z}_{i+1}=Z_{2i+1}Z_{2i+3}(Z_{2i+1}Z_{2i+2}Z_{2i+3})$, where it is a stabilizer that we multiply on in the bracket. The model\footnote{This is not a Hamiltonian built of stabilizers, but built from logical Pauli operators within the stabilized subspace.} in terms of logical operators becomes
\begin{align}
H=-B\sum_i \mathcal{X}_i -J\sum_i \mathcal{Z}_{i-1}\mathcal{Z}_i.
\end{align}

Finally, the global symmetry operator $\prod_i X_{2i+1}$ in the minimally coupled model also maps to $\prod_i \mathcal{X}_i$, the global $Z_2$ symmetry operator on the logical qubits. We may, of course, add this operator initially to the set of stabilizers and the analysis remains the same but with $2^{N-1}$ logical qubits.

There is a caveat in this proof. If different models can be mapped to the same  Hamiltonian on the same set of logical qubits defined in a stabilized space, they are indeed the same at the algebraic level. However, we need to further check if they have the same ``locality property''. Even though it is hard to define the property generically, for the one-dimensional models we study here, they all have the properties that neighboring logical Pauli operators come from neighboring local terms composed of physical Pauli operators. 

\section{Non-on-site $Z_2$ symmetry transformations} \label{Z2trns}

The non-on-site $Z_2$ symmetry transformation $U=\prod_i X_{i} \prod_i
s_{i-\frac12,i+\frac12}$ (where $s_{ij}$ is given in \eqn{sij}) transforms
$X_i$ in the following way:
\begin{align}
&\ \ \ \  
\left(\prod_j X_j \prod_j s_{j,j+1}\right)
X_i
\left(\prod_j X_j \prod_j s_{j,j+1}\right)
\nonumber\\
&=
\frac { 1 -Z_{i-1} +Z_i +Z_{i-1} Z_i}2
\frac {1 -Z_i +Z_{i+1} +Z_i Z_{i+1}}2 
X_i 
\nonumber\\ 
&\ \ \ \
\frac { 1 -Z_{i-1} +Z_i +Z_{i-1} Z_i}2
\frac {1 -Z_i +Z_{i+1} +Z_i Z_{i+1}}2 
\nonumber\\
& = 
\frac { 1 -Z_{i-1} +Z_i +Z_{i-1} Z_i}2
\frac {1 -Z_i +Z_{i+1} +Z_i Z_{i+1}}2 
\nonumber\\
&\ \ \ \
\frac { 1 -Z_{i-1} -Z_i -Z_{i-1} Z_i}2
\frac {1 +Z_i +Z_{i+1} -Z_i Z_{i+1}}2 
X_i 
\nonumber\\ 
& = 
\Big(\frac { 1 -Z_{i-1} +Z_i +Z_{i-1} Z_i}2
\frac { 1 -Z_{i-1} -Z_i -Z_{i-1} Z_i}2 \Big)
\nonumber\\
&\ \ \ \
\Big(\frac {1 -Z_i +Z_{i+1} +Z_i Z_{i+1}}2 
\frac {1 +Z_i +Z_{i+1} -Z_i Z_{i+1}}2 \Big)
X_i 
\nonumber\\ 
& = 
\frac { (1 -Z_{i-1}) -(1 +Z_{i-1}) }2
\frac { (1 +Z_{i+1}) -(1 - Z_{i+1})}2 
X_i 
\nonumber\\ 
& =  -
Z_{i-1}
X_i 
Z_{i+1}.
\end{align}
In other words
\begin{align}
X_i \leftrightarrow  - Z_{i-1} X_i Z_{i+1}.
\end{align}

\bibliography{../../bib/all,../../bib/allnew,../../bib/publst,./local}

\begin{thebibliography}{92}%
\makeatletter
\providecommand \@ifxundefined [1]{%
 \@ifx{#1\undefined}
}%
\providecommand \@ifnum [1]{%
 \ifnum #1\expandafter \@firstoftwo
 \else \expandafter \@secondoftwo
 \fi
}%
\providecommand \@ifx [1]{%
 \ifx #1\expandafter \@firstoftwo
 \else \expandafter \@secondoftwo
 \fi
}%
\providecommand \natexlab [1]{#1}%
\providecommand \enquote  [1]{``#1''}%
\providecommand \bibnamefont  [1]{#1}%
\providecommand \bibfnamefont [1]{#1}%
\providecommand \citenamefont [1]{#1}%
\providecommand \href@noop [0]{\@secondoftwo}%
\providecommand \href [0]{\begingroup \@sanitize@url \@href}%
\providecommand \@href[1]{\@@startlink{#1}\@@href}%
\providecommand \@@href[1]{\endgroup#1\@@endlink}%
\providecommand \@sanitize@url [0]{\catcode `\\12\catcode `\$12\catcode
  `\&12\catcode `\#12\catcode `\^12\catcode `\_12\catcode `\%12\relax}%
\providecommand \@@startlink[1]{}%
\providecommand \@@endlink[0]{}%
\providecommand \url  [0]{\begingroup\@sanitize@url \@url }%
\providecommand \@url [1]{\endgroup\@href {#1}{\urlprefix }}%
\providecommand \urlprefix  [0]{URL }%
\providecommand \Eprint [0]{\href }%
\providecommand \doibase [0]{http://dx.doi.org/}%
\providecommand \selectlanguage [0]{\@gobble}%
\providecommand \bibinfo  [0]{\@secondoftwo}%
\providecommand \bibfield  [0]{\@secondoftwo}%
\providecommand \translation [1]{[#1]}%
\providecommand \BibitemOpen [0]{}%
\providecommand \bibitemStop [0]{}%
\providecommand \bibitemNoStop [0]{.\EOS\space}%
\providecommand \EOS [0]{\spacefactor3000\relax}%
\providecommand \BibitemShut  [1]{\csname bibitem#1\endcsname}%
\let\auto@bib@innerbib\@empty
\bibitem [{\citenamefont {Landau}(1937{\natexlab{a}})}]{L3726}%
  \BibitemOpen
  \bibfield  {author} {\bibinfo {author} {\bibfnamefont {L.~D.}\ \bibnamefont
  {Landau}},\ }\href@noop {} {\bibfield  {journal} {\bibinfo  {journal} {Phys.
  Z. Sowjetunion}\ }\textbf {\bibinfo {volume} {11}},\ \bibinfo {pages} {26}
  (\bibinfo {year} {1937}{\natexlab{a}})}\BibitemShut {NoStop}%
\bibitem [{\citenamefont {Landau}(1937{\natexlab{b}})}]{L3745}%
  \BibitemOpen
  \bibfield  {author} {\bibinfo {author} {\bibfnamefont {L.~D.}\ \bibnamefont
  {Landau}},\ }\href@noop {} {\bibfield  {journal} {\bibinfo  {journal} {Phys.
  Z. Sowjetunion}\ }\textbf {\bibinfo {volume} {11}},\ \bibinfo {pages} {545}
  (\bibinfo {year} {1937}{\natexlab{b}})}\BibitemShut {NoStop}%
\bibitem [{\citenamefont {{Ginsparg}}(1991)}]{Ght9108028}%
  \BibitemOpen
  \bibfield  {author} {\bibinfo {author} {\bibfnamefont {P.}~\bibnamefont
  {{Ginsparg}}},\ }\href@noop {} {\  (\bibinfo {year} {1991})},\ \Eprint
  {http://arxiv.org/abs/hep-th/9108028} {hep-th/9108028} \BibitemShut {NoStop}%
\bibitem [{\citenamefont {Di~Francesco}\ \emph {et~al.}(1997)\citenamefont
  {Di~Francesco}, \citenamefont {Mathieu},\ and\ \citenamefont
  {S??n??chal}}]{CFT12}%
  \BibitemOpen
  \bibfield  {author} {\bibinfo {author} {\bibfnamefont {P.}~\bibnamefont
  {Di~Francesco}}, \bibinfo {author} {\bibfnamefont {P.}~\bibnamefont
  {Mathieu}}, \ and\ \bibinfo {author} {\bibfnamefont {D.}~\bibnamefont
  {S??n??chal}},\ }\href {\doibase 10.1007/978-1-4612-2256-9} {\emph {\bibinfo
  {title} {Conformal Field Theory}}}\ (\bibinfo  {publisher} {Springer New
  York},\ \bibinfo {year} {1997})\BibitemShut {NoStop}%
\bibitem [{\citenamefont {Lan}\ \emph {et~al.}(2018)\citenamefont {Lan},
  \citenamefont {Kong},\ and\ \citenamefont {Wen}}]{LW170404221}%
  \BibitemOpen
  \bibfield  {author} {\bibinfo {author} {\bibfnamefont {T.}~\bibnamefont
  {Lan}}, \bibinfo {author} {\bibfnamefont {L.}~\bibnamefont {Kong}}, \ and\
  \bibinfo {author} {\bibfnamefont {X.-G.}\ \bibnamefont {Wen}},\ }\href
  {\doibase 10.1103/physrevx.8.021074} {\bibfield  {journal} {\bibinfo
  {journal} {Phys. Rev. X}\ }\textbf {\bibinfo {volume} {8}},\ \bibinfo {pages}
  {021074} (\bibinfo {year} {2018})},\ \Eprint
  {http://arxiv.org/abs/1704.04221} {arXiv:1704.04221} \BibitemShut {NoStop}%
\bibitem [{\citenamefont {Lan}\ and\ \citenamefont {Wen}(2019)}]{LW180108530}%
  \BibitemOpen
  \bibfield  {author} {\bibinfo {author} {\bibfnamefont {T.}~\bibnamefont
  {Lan}}\ and\ \bibinfo {author} {\bibfnamefont {X.-G.}\ \bibnamefont {Wen}},\
  }\href {\doibase 10.1103/physrevx.9.021005} {\bibfield  {journal} {\bibinfo
  {journal} {Phys. Rev. X}\ }\textbf {\bibinfo {volume} {9}},\ \bibinfo {pages}
  {021005} (\bibinfo {year} {2019})},\ \Eprint
  {http://arxiv.org/abs/1801.08530} {arXiv:1801.08530} \BibitemShut {NoStop}%
\bibitem [{\citenamefont {{Kong}}\ \emph
  {et~al.}(2020{\natexlab{a}})\citenamefont {{Kong}}, \citenamefont {{Lan}},
  \citenamefont {{Wen}}, \citenamefont {{Zhang}},\ and\ \citenamefont
  {{Zheng}}}]{KZ200308898}%
  \BibitemOpen
  \bibfield  {author} {\bibinfo {author} {\bibfnamefont {L.}~\bibnamefont
  {{Kong}}}, \bibinfo {author} {\bibfnamefont {T.}~\bibnamefont {{Lan}}},
  \bibinfo {author} {\bibfnamefont {X.-G.}\ \bibnamefont {{Wen}}}, \bibinfo
  {author} {\bibfnamefont {Z.-H.}\ \bibnamefont {{Zhang}}}, \ and\ \bibinfo
  {author} {\bibfnamefont {H.}~\bibnamefont {{Zheng}}},\ }\href {\doibase
  doi.org/10.1007/JHEP09(2020)093} {\bibfield  {journal} {\bibinfo  {journal}
  {J. High Energ. Phys.}\ }\textbf {\bibinfo {volume} {2020}},\ \bibinfo
  {pages} {93} (\bibinfo {year} {2020}{\natexlab{a}})},\ \Eprint
  {http://arxiv.org/abs/2003.08898} {arXiv:2003.08898} \BibitemShut {NoStop}%
\bibitem [{\citenamefont {{Kong}}\ \emph
  {et~al.}(2020{\natexlab{b}})\citenamefont {{Kong}}, \citenamefont {{Lan}},
  \citenamefont {{Wen}}, \citenamefont {{Zhang}},\ and\ \citenamefont
  {{Zheng}}}]{KZ200514178}%
  \BibitemOpen
  \bibfield  {author} {\bibinfo {author} {\bibfnamefont {L.}~\bibnamefont
  {{Kong}}}, \bibinfo {author} {\bibfnamefont {T.}~\bibnamefont {{Lan}}},
  \bibinfo {author} {\bibfnamefont {X.-G.}\ \bibnamefont {{Wen}}}, \bibinfo
  {author} {\bibfnamefont {Z.-H.}\ \bibnamefont {{Zhang}}}, \ and\ \bibinfo
  {author} {\bibfnamefont {H.}~\bibnamefont {{Zheng}}},\ }\href@noop {} {\
  (\bibinfo {year} {2020}{\natexlab{b}})},\ \Eprint
  {http://arxiv.org/abs/2005.14178} {arXiv:2005.14178} \BibitemShut {NoStop}%
\bibitem [{\citenamefont {Gaiotto}\ \emph {et~al.}(2015)\citenamefont
  {Gaiotto}, \citenamefont {Kapustin}, \citenamefont {Seiberg},\ and\
  \citenamefont {Willett}}]{GW14125148}%
  \BibitemOpen
  \bibfield  {author} {\bibinfo {author} {\bibfnamefont {D.}~\bibnamefont
  {Gaiotto}}, \bibinfo {author} {\bibfnamefont {A.}~\bibnamefont {Kapustin}},
  \bibinfo {author} {\bibfnamefont {N.}~\bibnamefont {Seiberg}}, \ and\
  \bibinfo {author} {\bibfnamefont {B.}~\bibnamefont {Willett}},\ }\href
  {\doibase 10.1007/jhep02(2015)172} {\bibfield  {journal} {\bibinfo  {journal}
  {J. High Energ. Phys.}\ }\textbf {\bibinfo {volume} {2015}},\ \bibinfo
  {pages} {172} (\bibinfo {year} {2015})},\ \Eprint
  {http://arxiv.org/abs/1412.5148} {arXiv:1412.5148} \BibitemShut {NoStop}%
\bibitem [{\citenamefont {Fr\"ohlich}\ \emph {et~al.}(2007)\citenamefont
  {Fr\"ohlich}, \citenamefont {Fuchs}, \citenamefont {Runkel},\ and\
  \citenamefont {Schweigert}}]{FSh0607247}%
  \BibitemOpen
  \bibfield  {author} {\bibinfo {author} {\bibfnamefont {J.}~\bibnamefont
  {Fr\"ohlich}}, \bibinfo {author} {\bibfnamefont {J.}~\bibnamefont {Fuchs}},
  \bibinfo {author} {\bibfnamefont {I.}~\bibnamefont {Runkel}}, \ and\ \bibinfo
  {author} {\bibfnamefont {C.}~\bibnamefont {Schweigert}},\ }\href@noop {}
  {\bibfield  {journal} {\bibinfo  {journal} {Nucl. Phys. B}\ }\textbf
  {\bibinfo {volume} {763}},\ \bibinfo {pages} {354} (\bibinfo {year}
  {2007})},\ \Eprint {http://arxiv.org/abs/hep-th/0607247}
  {arXiv:hep-th/0607247} \BibitemShut {NoStop}%
\bibitem [{\citenamefont {Davydov}\ \emph {et~al.}(2011)\citenamefont
  {Davydov}, \citenamefont {Kong},\ and\ \citenamefont {Runkel}}]{DR10044725}%
  \BibitemOpen
  \bibfield  {author} {\bibinfo {author} {\bibfnamefont {A.}~\bibnamefont
  {Davydov}}, \bibinfo {author} {\bibfnamefont {L.}~\bibnamefont {Kong}}, \
  and\ \bibinfo {author} {\bibfnamefont {I.}~\bibnamefont {Runkel}},\
  }\href@noop {} {\bibfield  {journal} {\bibinfo  {journal} {Adv. Theor. Math.
  Phys.}\ }\textbf {\bibinfo {volume} {15}},\ \bibinfo {pages} {43} (\bibinfo
  {year} {2011})},\ \Eprint {http://arxiv.org/abs/1004.4725} {arXiv:1004.4725}
  \BibitemShut {NoStop}%
\bibitem [{\citenamefont {Chang}\ \emph {et~al.}(2019)\citenamefont {Chang},
  \citenamefont {Lin}, \citenamefont {Shao}, \citenamefont {Wang},\ and\
  \citenamefont {Yin}}]{CY180204445}%
  \BibitemOpen
  \bibfield  {author} {\bibinfo {author} {\bibfnamefont {C.-M.}\ \bibnamefont
  {Chang}}, \bibinfo {author} {\bibfnamefont {Y.-H.}\ \bibnamefont {Lin}},
  \bibinfo {author} {\bibfnamefont {S.-H.}\ \bibnamefont {Shao}}, \bibinfo
  {author} {\bibfnamefont {Y.}~\bibnamefont {Wang}}, \ and\ \bibinfo {author}
  {\bibfnamefont {X.}~\bibnamefont {Yin}},\ }\href {\doibase
  10.1007/jhep01(2019)026} {\bibfield  {journal} {\bibinfo  {journal} {J. High
  Energ. Phys.}\ }\textbf {\bibinfo {volume} {2019}},\ \bibinfo {pages} {26}
  (\bibinfo {year} {2019})},\ \Eprint {http://arxiv.org/abs/1802.04445}
  {arXiv:1802.04445} \BibitemShut {NoStop}%
\bibitem [{\citenamefont {{Thorngren}}\ and\ \citenamefont
  {{Wang}}(2019)}]{TW191202817}%
  \BibitemOpen
  \bibfield  {author} {\bibinfo {author} {\bibfnamefont {R.}~\bibnamefont
  {{Thorngren}}}\ and\ \bibinfo {author} {\bibfnamefont {Y.}~\bibnamefont
  {{Wang}}},\ }\href@noop {} {\  (\bibinfo {year} {2019})},\ \Eprint
  {http://arxiv.org/abs/1912.02817} {arXiv:1912.02817} \BibitemShut {NoStop}%
\bibitem [{\citenamefont {Kong}\ \emph {et~al.}(2019)\citenamefont {Kong},
  \citenamefont {Yuan},\ and\ \citenamefont {Zheng}}]{KZ191213168}%
  \BibitemOpen
  \bibfield  {author} {\bibinfo {author} {\bibfnamefont {L.}~\bibnamefont
  {Kong}}, \bibinfo {author} {\bibfnamefont {W.}~\bibnamefont {Yuan}}, \ and\
  \bibinfo {author} {\bibfnamefont {H.}~\bibnamefont {Zheng}},\ }\href@noop {}
  {\  (\bibinfo {year} {2019})},\ \Eprint {http://arxiv.org/abs/1912.13168}
  {arXiv:1912.13168} \BibitemShut {NoStop}%
\bibitem [{\citenamefont {{Kapustin}}\ and\ \citenamefont
  {{Thorngren}}(2013)}]{KT13094721}%
  \BibitemOpen
  \bibfield  {author} {\bibinfo {author} {\bibfnamefont {A.}~\bibnamefont
  {{Kapustin}}}\ and\ \bibinfo {author} {\bibfnamefont {R.}~\bibnamefont
  {{Thorngren}}},\ }\href@noop {} {\  (\bibinfo {year} {2013})},\ \Eprint
  {http://arxiv.org/abs/1309.4721} {arXiv:1309.4721} \BibitemShut {NoStop}%
\bibitem [{\citenamefont {{Thorngren}}\ and\ \citenamefont {{von
  Keyserlingk}}(2015)}]{TK151102929}%
  \BibitemOpen
  \bibfield  {author} {\bibinfo {author} {\bibfnamefont {R.}~\bibnamefont
  {{Thorngren}}}\ and\ \bibinfo {author} {\bibfnamefont {C.}~\bibnamefont {{von
  Keyserlingk}}},\ }\href@noop {} {\  (\bibinfo {year} {2015})},\ \Eprint
  {http://arxiv.org/abs/1511.02929} {arXiv:1511.02929} \BibitemShut {NoStop}%
\bibitem [{\citenamefont {Ibieta-Jimenez}\ \emph {et~al.}(2020)\citenamefont
  {Ibieta-Jimenez}, \citenamefont {Petrucci}, \citenamefont {Xavier},\ and\
  \citenamefont {Teotonio-Sobrinho}}]{CT171104186}%
  \BibitemOpen
  \bibfield  {author} {\bibinfo {author} {\bibfnamefont {J.}~\bibnamefont
  {Ibieta-Jimenez}}, \bibinfo {author} {\bibfnamefont {M.}~\bibnamefont
  {Petrucci}}, \bibinfo {author} {\bibfnamefont {L.~Q.}\ \bibnamefont
  {Xavier}}, \ and\ \bibinfo {author} {\bibfnamefont {P.}~\bibnamefont
  {Teotonio-Sobrinho}},\ }\href {\doibase 10.1007/jhep03(2020)167} {\bibfield
  {journal} {\bibinfo  {journal} {J. High Energ. Phys.}\ }\textbf {\bibinfo
  {volume} {2020}} (\bibinfo {year} {2020}),\ 10.1007/jhep03(2020)167},\
  \Eprint {http://arxiv.org/abs/1711.04186} {arXiv:1711.04186} \BibitemShut
  {NoStop}%
\bibitem [{\citenamefont {Kobayashi}\ \emph {et~al.}(2019)\citenamefont
  {Kobayashi}, \citenamefont {Shiozaki}, \citenamefont {Kikuchi},\ and\
  \citenamefont {Ryu}}]{KR180505367}%
  \BibitemOpen
  \bibfield  {author} {\bibinfo {author} {\bibfnamefont {R.}~\bibnamefont
  {Kobayashi}}, \bibinfo {author} {\bibfnamefont {K.}~\bibnamefont {Shiozaki}},
  \bibinfo {author} {\bibfnamefont {Y.}~\bibnamefont {Kikuchi}}, \ and\
  \bibinfo {author} {\bibfnamefont {S.}~\bibnamefont {Ryu}},\ }\href {\doibase
  10.1103/physrevb.99.014402} {\bibfield  {journal} {\bibinfo  {journal} {Phys.
  Rev. B}\ }\textbf {\bibinfo {volume} {99}},\ \bibinfo {pages} {014402}
  (\bibinfo {year} {2019})},\ \Eprint {http://arxiv.org/abs/1805.05367}
  {arXiv:1805.05367} \BibitemShut {NoStop}%
\bibitem [{\citenamefont {{Lake}}(2018)}]{L180207747}%
  \BibitemOpen
  \bibfield  {author} {\bibinfo {author} {\bibfnamefont {E.}~\bibnamefont
  {{Lake}}},\ }\href@noop {} {\  (\bibinfo {year} {2018})},\ \Eprint
  {http://arxiv.org/abs/1802.07747} {arXiv:1802.07747} \BibitemShut {NoStop}%
\bibitem [{\citenamefont {Wen}(2019)}]{W181202517}%
  \BibitemOpen
  \bibfield  {author} {\bibinfo {author} {\bibfnamefont {X.-G.}\ \bibnamefont
  {Wen}},\ }\href {\doibase 10.1103/physrevb.99.205139} {\bibfield  {journal}
  {\bibinfo  {journal} {Phys. Rev. B}\ }\textbf {\bibinfo {volume} {99}},\
  \bibinfo {pages} {205139} (\bibinfo {year} {2019})},\ \Eprint
  {http://arxiv.org/abs/1812.02517} {arXiv:1812.02517} \BibitemShut {NoStop}%
\bibitem [{\citenamefont {Wan}\ and\ \citenamefont {Wang}(2019)}]{WW181211955}%
  \BibitemOpen
  \bibfield  {author} {\bibinfo {author} {\bibfnamefont {Z.}~\bibnamefont
  {Wan}}\ and\ \bibinfo {author} {\bibfnamefont {J.}~\bibnamefont {Wang}},\
  }\href {\doibase 10.1103/physrevd.99.065013} {\bibfield  {journal} {\bibinfo
  {journal} {Phys. Rev. D}\ }\textbf {\bibinfo {volume} {99}},\ \bibinfo
  {pages} {065013} (\bibinfo {year} {2019})},\ \Eprint
  {http://arxiv.org/abs/1812.11955} {arXiv:1812.11955} \BibitemShut {NoStop}%
\bibitem [{\citenamefont {{Wan}}\ and\ \citenamefont
  {{Wang}}(2019)}]{WW181211967}%
  \BibitemOpen
  \bibfield  {author} {\bibinfo {author} {\bibfnamefont {Z.}~\bibnamefont
  {{Wan}}}\ and\ \bibinfo {author} {\bibfnamefont {J.}~\bibnamefont {{Wang}}},\
  }\href@noop {} {\bibfield  {journal} {\bibinfo  {journal} {Annals of
  Mathematical Sciences and Applications}\ }\textbf {\bibinfo {volume} {4}},\
  \bibinfo {pages} {107} (\bibinfo {year} {2019})},\ \Eprint
  {http://arxiv.org/abs/1812.11967} {arXiv:1812.11967} \BibitemShut {NoStop}%
\bibitem [{\citenamefont {Kitaev}(2003)}]{K032}%
  \BibitemOpen
  \bibfield  {author} {\bibinfo {author} {\bibfnamefont {A.}~\bibnamefont
  {Kitaev}},\ }\href {\doibase 10.1016/s0003-4916(02)00018-0} {\bibfield
  {journal} {\bibinfo  {journal} {Ann. Phys.}\ }\textbf {\bibinfo {volume}
  {303}},\ \bibinfo {pages} {2} (\bibinfo {year} {2003})}\BibitemShut {NoStop}%
\bibitem [{\citenamefont {Wen}(2003)}]{W0303}%
  \BibitemOpen
  \bibfield  {author} {\bibinfo {author} {\bibfnamefont {X.-G.}\ \bibnamefont
  {Wen}},\ }\href {\doibase 10.1103/physrevlett.90.016803} {\bibfield
  {journal} {\bibinfo  {journal} {Phys. Rev. Lett.}\ }\textbf {\bibinfo
  {volume} {90}},\ \bibinfo {pages} {016803} (\bibinfo {year} {2003})},\
  \Eprint {http://arxiv.org/abs/quant-ph/0205004} {arXiv:quant-ph/0205004}
  \BibitemShut {NoStop}%
\bibitem [{\citenamefont {Levin}\ and\ \citenamefont {Wen}(2003)}]{LW0316}%
  \BibitemOpen
  \bibfield  {author} {\bibinfo {author} {\bibfnamefont {M.}~\bibnamefont
  {Levin}}\ and\ \bibinfo {author} {\bibfnamefont {X.-G.}\ \bibnamefont
  {Wen}},\ }\href {\doibase 10.1103/physrevb.67.245316} {\bibfield  {journal}
  {\bibinfo  {journal} {Phys. Rev. B}\ }\textbf {\bibinfo {volume} {67}},\
  \bibinfo {pages} {245316} (\bibinfo {year} {2003})},\ \Eprint
  {http://arxiv.org/abs/cond-mat/0302460} {arXiv:cond-mat/0302460} \BibitemShut
  {NoStop}%
\bibitem [{\citenamefont {Nussinov}\ and\ \citenamefont
  {Ortiz}(2009{\natexlab{a}})}]{NOc0605316}%
  \BibitemOpen
  \bibfield  {author} {\bibinfo {author} {\bibfnamefont {Z.}~\bibnamefont
  {Nussinov}}\ and\ \bibinfo {author} {\bibfnamefont {G.}~\bibnamefont
  {Ortiz}},\ }\href {\doibase 10.1073/pnas.0803726105} {\bibfield  {journal}
  {\bibinfo  {journal} {Proceedings of the National Academy of Sciences}\
  }\textbf {\bibinfo {volume} {106}},\ \bibinfo {pages} {16944} (\bibinfo
  {year} {2009}{\natexlab{a}})},\ \Eprint
  {http://arxiv.org/abs/cond-mat/0605316} {arXiv:cond-mat/0605316} \BibitemShut
  {NoStop}%
\bibitem [{\citenamefont {Nussinov}\ and\ \citenamefont
  {Ortiz}(2009{\natexlab{b}})}]{NOc0702377}%
  \BibitemOpen
  \bibfield  {author} {\bibinfo {author} {\bibfnamefont {Z.}~\bibnamefont
  {Nussinov}}\ and\ \bibinfo {author} {\bibfnamefont {G.}~\bibnamefont
  {Ortiz}},\ }\href {\doibase 10.1016/j.aop.2008.11.002} {\bibfield  {journal}
  {\bibinfo  {journal} {Ann. Phys.}\ }\textbf {\bibinfo {volume} {324}},\
  \bibinfo {pages} {977} (\bibinfo {year} {2009}{\natexlab{b}})},\ \Eprint
  {http://arxiv.org/abs/cond-mat/0702377} {arXiv:cond-mat/0702377} \BibitemShut
  {NoStop}%
\bibitem [{\citenamefont {Yoshida}(2011)}]{Y10074601}%
  \BibitemOpen
  \bibfield  {author} {\bibinfo {author} {\bibfnamefont {B.}~\bibnamefont
  {Yoshida}},\ }\href {\doibase 10.1016/j.aop.2010.10.009} {\bibfield
  {journal} {\bibinfo  {journal} {Ann. Phys.}\ }\textbf {\bibinfo {volume}
  {326}},\ \bibinfo {pages} {15} (\bibinfo {year} {2011})},\ \Eprint
  {http://arxiv.org/abs/1007.4601} {arXiv:1007.4601} \BibitemShut {NoStop}%
\bibitem [{\citenamefont {Bomb\'in}(2014)}]{B11072707}%
  \BibitemOpen
  \bibfield  {author} {\bibinfo {author} {\bibfnamefont {H.}~\bibnamefont
  {Bomb\'in}},\ }\href {\doibase 10.1007/s00220-014-1893-4} {\bibfield
  {journal} {\bibinfo  {journal} {Commun. Math. Phys.}\ }\textbf {\bibinfo
  {volume} {327}},\ \bibinfo {pages} {387} (\bibinfo {year} {2014})},\ \Eprint
  {http://arxiv.org/abs/1107.2707} {arXiv:1107.2707} \BibitemShut {NoStop}%
\bibitem [{\citenamefont {Kong}\ and\ \citenamefont {Wen}(2014)}]{KW1458}%
  \BibitemOpen
  \bibfield  {author} {\bibinfo {author} {\bibfnamefont {L.}~\bibnamefont
  {Kong}}\ and\ \bibinfo {author} {\bibfnamefont {X.-G.}\ \bibnamefont {Wen}},\
  }\href@noop {} {\  (\bibinfo {year} {2014})},\ \Eprint
  {http://arxiv.org/abs/1405.5858} {arXiv:1405.5858} \BibitemShut {NoStop}%
\bibitem [{\citenamefont {Ji}\ and\ \citenamefont {Wen}(2019)}]{JW190513279}%
  \BibitemOpen
  \bibfield  {author} {\bibinfo {author} {\bibfnamefont {W.}~\bibnamefont
  {Ji}}\ and\ \bibinfo {author} {\bibfnamefont {X.-G.}\ \bibnamefont {Wen}},\
  }\href {\doibase 10.1103/PhysRevResearch.1.033054} {\bibfield  {journal}
  {\bibinfo  {journal} {Phys. Rev. Research}\ }\textbf {\bibinfo {volume}
  {1}},\ \bibinfo {pages} {033054} (\bibinfo {year} {2019})},\ \Eprint
  {http://arxiv.org/abs/1905.13279} {arXiv:1905.13279} \BibitemShut {NoStop}%
\bibitem [{\citenamefont {Maldacena}(1998)}]{M9831}%
  \BibitemOpen
  \bibfield  {author} {\bibinfo {author} {\bibfnamefont {J.}~\bibnamefont
  {Maldacena}},\ }\href {\doibase 10.4310/atmp.1998.v2.n2.a1} {\bibfield
  {journal} {\bibinfo  {journal} {Adv. Theor. Math. Phys.}\ }\textbf {\bibinfo
  {volume} {2}},\ \bibinfo {pages} {231} (\bibinfo {year} {1998})},\ \Eprint
  {http://arxiv.org/abs/hep-th/9711200} {hep-th/9711200} \BibitemShut {NoStop}%
\bibitem [{\citenamefont {{Witten}}(1998)}]{Wh9802150}%
  \BibitemOpen
  \bibfield  {author} {\bibinfo {author} {\bibfnamefont {E.}~\bibnamefont
  {{Witten}}},\ }\href@noop {} {\bibfield  {journal} {\bibinfo  {journal}
  {Advances in Theoretical and Mathematical Physics}\ }\textbf {\bibinfo
  {volume} {2}},\ \bibinfo {pages} {253} (\bibinfo {year} {1998})},\ \Eprint
  {http://arxiv.org/abs/hep-th/9802150} {arXiv:hep-th/9802150} \BibitemShut
  {NoStop}%
\bibitem [{\citenamefont {{Klebanov}}\ and\ \citenamefont
  {{Witten}}(1999)}]{KWh9905104}%
  \BibitemOpen
  \bibfield  {author} {\bibinfo {author} {\bibfnamefont {I.~R.}\ \bibnamefont
  {{Klebanov}}}\ and\ \bibinfo {author} {\bibfnamefont {E.}~\bibnamefont
  {{Witten}}},\ }\href {\doibase 10.1016/S0550-3213(99)00387-9} {\bibfield
  {journal} {\bibinfo  {journal} {Nuclear Physics B}\ }\textbf {\bibinfo
  {volume} {556}},\ \bibinfo {pages} {89} (\bibinfo {year} {1999})},\ \Eprint
  {http://arxiv.org/abs/hep-th/9905104} {arXiv:hep-th/9905104} \BibitemShut
  {NoStop}%
\bibitem [{\citenamefont {{Harlow}}\ and\ \citenamefont
  {{Ooguri}}(2018)}]{HO181005338}%
  \BibitemOpen
  \bibfield  {author} {\bibinfo {author} {\bibfnamefont {D.}~\bibnamefont
  {{Harlow}}}\ and\ \bibinfo {author} {\bibfnamefont {H.}~\bibnamefont
  {{Ooguri}}},\ }\href@noop {} {\  (\bibinfo {year} {2018})},\ \Eprint
  {http://arxiv.org/abs/1810.05338} {arXiv:1810.05338} \BibitemShut {NoStop}%
\bibitem [{\citenamefont {Belavin}\ \emph {et~al.}(1984)\citenamefont
  {Belavin}, \citenamefont {Polyakov},\ and\ \citenamefont
  {Zamolodchikov}}]{BPZ8433}%
  \BibitemOpen
  \bibfield  {author} {\bibinfo {author} {\bibfnamefont {A.}~\bibnamefont
  {Belavin}}, \bibinfo {author} {\bibfnamefont {A.}~\bibnamefont {Polyakov}}, \
  and\ \bibinfo {author} {\bibfnamefont {A.}~\bibnamefont {Zamolodchikov}},\
  }\href {\doibase 10.1016/0550-3213(84)90052-x} {\bibfield  {journal}
  {\bibinfo  {journal} {Nucl. Phys. B}\ }\textbf {\bibinfo {volume} {241}},\
  \bibinfo {pages} {333} (\bibinfo {year} {1984})}\BibitemShut {NoStop}%
\bibitem [{\citenamefont {Zamolodchikov}\ and\ \citenamefont
  {Fateev}(1985)}]{ZF8515}%
  \BibitemOpen
  \bibfield  {author} {\bibinfo {author} {\bibfnamefont {A.}~\bibnamefont
  {Zamolodchikov}}\ and\ \bibinfo {author} {\bibfnamefont {V.}~\bibnamefont
  {Fateev}},\ }\href@noop {} {\bibfield  {journal} {\bibinfo  {journal} {Sov.
  Phys. JETP}\ }\textbf {\bibinfo {volume} {62}},\ \bibinfo {pages} {215}
  (\bibinfo {year} {1985})}\BibitemShut {NoStop}%
\bibitem [{\citenamefont {{Verresen}}\ \emph {et~al.}(2019)\citenamefont
  {{Verresen}}, \citenamefont {{Thorngren}}, \citenamefont {{Jones}},\ and\
  \citenamefont {{Pollmann}}}]{VP190506969}%
  \BibitemOpen
  \bibfield  {author} {\bibinfo {author} {\bibfnamefont {R.}~\bibnamefont
  {{Verresen}}}, \bibinfo {author} {\bibfnamefont {R.}~\bibnamefont
  {{Thorngren}}}, \bibinfo {author} {\bibfnamefont {N.~G.}\ \bibnamefont
  {{Jones}}}, \ and\ \bibinfo {author} {\bibfnamefont {F.}~\bibnamefont
  {{Pollmann}}},\ }\href@noop {} {\  (\bibinfo {year} {2019})},\ \Eprint
  {http://arxiv.org/abs/1905.06969} {arXiv:1905.06969} \BibitemShut {NoStop}%
\bibitem [{\citenamefont {Ji}\ \emph {et~al.}(2020)\citenamefont {Ji},
  \citenamefont {Shao},\ and\ \citenamefont {Wen}}]{JW190901425}%
  \BibitemOpen
  \bibfield  {author} {\bibinfo {author} {\bibfnamefont {W.}~\bibnamefont
  {Ji}}, \bibinfo {author} {\bibfnamefont {S.-H.}\ \bibnamefont {Shao}}, \ and\
  \bibinfo {author} {\bibfnamefont {X.-G.}\ \bibnamefont {Wen}},\ }\href
  {\doibase 10.1103/PhysRevResearch.2.033317} {\bibfield  {journal} {\bibinfo
  {journal} {Phys. Rev. Research}\ }\textbf {\bibinfo {volume} {2}},\ \bibinfo
  {pages} {033317} (\bibinfo {year} {2020})},\ \Eprint
  {http://arxiv.org/abs/1909.01425} {arXiv:1909.01425} \BibitemShut {NoStop}%
\bibitem [{\citenamefont {Levin}(2020)}]{L190309028}%
  \BibitemOpen
  \bibfield  {author} {\bibinfo {author} {\bibfnamefont {M.}~\bibnamefont
  {Levin}},\ }\href {\doibase 10.1007/s00220-020-03802-4} {\bibfield  {journal}
  {\bibinfo  {journal} {Commun. Math. Phys.}\ }\textbf {\bibinfo {volume}
  {378}},\ \bibinfo {pages} {1081} (\bibinfo {year} {2020})},\ \Eprint
  {http://arxiv.org/abs/1903.09028} {arXiv:1903.09028} \BibitemShut {NoStop}%
\bibitem [{\citenamefont {Harlow}\ and\ \citenamefont
  {Ooguri}(2019)}]{Harlow:2018jwu}%
  \BibitemOpen
  \bibfield  {author} {\bibinfo {author} {\bibfnamefont {D.}~\bibnamefont
  {Harlow}}\ and\ \bibinfo {author} {\bibfnamefont {H.}~\bibnamefont
  {Ooguri}},\ }\href {\doibase 10.1103/PhysRevLett.122.191601} {\bibfield
  {journal} {\bibinfo  {journal} {Phys. Rev. Lett.}\ }\textbf {\bibinfo
  {volume} {122}},\ \bibinfo {pages} {191601} (\bibinfo {year} {2019})},\
  \Eprint {http://arxiv.org/abs/1810.05337} {arXiv:1810.05337 [hep-th]}
  \BibitemShut {NoStop}%
\bibitem [{\citenamefont {Doplicher}(1982)}]{Doplicher:1982cv}%
  \BibitemOpen
  \bibfield  {author} {\bibinfo {author} {\bibfnamefont {S.}~\bibnamefont
  {Doplicher}},\ }\href {\doibase 10.1007/BF02029134} {\bibfield  {journal}
  {\bibinfo  {journal} {Commun. Math. Phys.}\ }\textbf {\bibinfo {volume}
  {85}},\ \bibinfo {pages} {73} (\bibinfo {year} {1982})}\BibitemShut {NoStop}%
\bibitem [{\citenamefont {Doplicher}\ and\ \citenamefont
  {Longo}(1983)}]{Doplicher:1983if}%
  \BibitemOpen
  \bibfield  {author} {\bibinfo {author} {\bibfnamefont {S.}~\bibnamefont
  {Doplicher}}\ and\ \bibinfo {author} {\bibfnamefont {R.}~\bibnamefont
  {Longo}},\ }\href {\doibase 10.1007/BF01213216} {\bibfield  {journal}
  {\bibinfo  {journal} {Commun. Math. Phys.}\ }\textbf {\bibinfo {volume}
  {88}},\ \bibinfo {pages} {399} (\bibinfo {year} {1983})}\BibitemShut
  {NoStop}%
\bibitem [{\citenamefont {Buchholz}\ \emph {et~al.}(1986)\citenamefont
  {Buchholz}, \citenamefont {Doplicher},\ and\ \citenamefont
  {Longo}}]{Buchholz:1985ii}%
  \BibitemOpen
  \bibfield  {author} {\bibinfo {author} {\bibfnamefont {D.}~\bibnamefont
  {Buchholz}}, \bibinfo {author} {\bibfnamefont {S.}~\bibnamefont {Doplicher}},
  \ and\ \bibinfo {author} {\bibfnamefont {R.}~\bibnamefont {Longo}},\ }\href
  {\doibase 10.1016/0003-4916(86)90086-2} {\bibfield  {journal} {\bibinfo
  {journal} {Annals Phys.}\ }\textbf {\bibinfo {volume} {170}},\ \bibinfo
  {pages} {1} (\bibinfo {year} {1986})}\BibitemShut {NoStop}%
\bibitem [{\citenamefont {Verstraete}\ and\ \citenamefont
  {Cirac}(2004)}]{VC0466}%
  \BibitemOpen
  \bibfield  {author} {\bibinfo {author} {\bibfnamefont {F.}~\bibnamefont
  {Verstraete}}\ and\ \bibinfo {author} {\bibfnamefont {J.}~\bibnamefont
  {Cirac}},\ }\href@noop {} {\  (\bibinfo {year} {2004})},\ \Eprint
  {http://arxiv.org/abs/cond-mat/0407066} {cond-mat/0407066} \BibitemShut
  {NoStop}%
\bibitem [{\citenamefont {Vidal}(2007)}]{V0705}%
  \BibitemOpen
  \bibfield  {author} {\bibinfo {author} {\bibfnamefont {G.}~\bibnamefont
  {Vidal}},\ }\href {\doibase 10.1103/physrevlett.99.220405} {\bibfield
  {journal} {\bibinfo  {journal} {Phys. Rev. Lett.}\ }\textbf {\bibinfo
  {volume} {99}},\ \bibinfo {pages} {220405} (\bibinfo {year} {2007})},\
  \Eprint {http://arxiv.org/abs/cond-mat/0512165} {arXiv:cond-mat/0512165}
  \BibitemShut {NoStop}%
\bibitem [{\citenamefont {Levin}\ and\ \citenamefont {Nave}(2007)}]{LN0701}%
  \BibitemOpen
  \bibfield  {author} {\bibinfo {author} {\bibfnamefont {M.}~\bibnamefont
  {Levin}}\ and\ \bibinfo {author} {\bibfnamefont {C.~P.}\ \bibnamefont
  {Nave}},\ }\href {\doibase 10.1103/physrevlett.99.120601} {\bibfield
  {journal} {\bibinfo  {journal} {Phys. Rev. Lett.}\ }\textbf {\bibinfo
  {volume} {99}},\ \bibinfo {pages} {120601} (\bibinfo {year}
  {2007})}\BibitemShut {NoStop}%
\bibitem [{\citenamefont {Gu}\ and\ \citenamefont {Wen}(2009)}]{GW0931}%
  \BibitemOpen
  \bibfield  {author} {\bibinfo {author} {\bibfnamefont {Z.-C.}\ \bibnamefont
  {Gu}}\ and\ \bibinfo {author} {\bibfnamefont {X.-G.}\ \bibnamefont {Wen}},\
  }\href {\doibase 10.1103/PhysRevB.80.155131} {\bibfield  {journal} {\bibinfo
  {journal} {Phys. Rev. B}\ }\textbf {\bibinfo {volume} {80}},\ \bibinfo
  {pages} {155131} (\bibinfo {year} {2009})},\ \Eprint
  {http://arxiv.org/abs/0903.1069} {arXiv:0903.1069} \BibitemShut {NoStop}%
\bibitem [{\citenamefont {Chen}\ \emph {et~al.}(2012)\citenamefont {Chen},
  \citenamefont {Gu}, \citenamefont {Liu},\ and\ \citenamefont
  {Wen}}]{CGL1204}%
  \BibitemOpen
  \bibfield  {author} {\bibinfo {author} {\bibfnamefont {X.}~\bibnamefont
  {Chen}}, \bibinfo {author} {\bibfnamefont {Z.-C.}\ \bibnamefont {Gu}},
  \bibinfo {author} {\bibfnamefont {Z.-X.}\ \bibnamefont {Liu}}, \ and\
  \bibinfo {author} {\bibfnamefont {X.-G.}\ \bibnamefont {Wen}},\ }\href
  {\doibase 10.1126/science.1227224} {\bibfield  {journal} {\bibinfo  {journal}
  {Science}\ }\textbf {\bibinfo {volume} {338}},\ \bibinfo {pages} {1604}
  (\bibinfo {year} {2012})},\ \Eprint {http://arxiv.org/abs/1301.0861}
  {arXiv:1301.0861} \BibitemShut {NoStop}%
\bibitem [{\citenamefont {Chen}\ \emph {et~al.}(2020)\citenamefont {Chen},
  \citenamefont {Jian}, \citenamefont {Kong}, \citenamefont {You},\ and\
  \citenamefont {Zheng}}]{CZ190312334}%
  \BibitemOpen
  \bibfield  {author} {\bibinfo {author} {\bibfnamefont {W.-Q.}\ \bibnamefont
  {Chen}}, \bibinfo {author} {\bibfnamefont {C.-M.}\ \bibnamefont {Jian}},
  \bibinfo {author} {\bibfnamefont {L.}~\bibnamefont {Kong}}, \bibinfo {author}
  {\bibfnamefont {Y.-Z.}\ \bibnamefont {You}}, \ and\ \bibinfo {author}
  {\bibfnamefont {H.}~\bibnamefont {Zheng}},\ }\href {\doibase
  10.1103/PhysRevB.102.045139} {\bibfield  {journal} {\bibinfo  {journal}
  {Phys. Rev. B}\ }\textbf {\bibinfo {volume} {102}},\ \bibinfo {pages}
  {045139} (\bibinfo {year} {2020})},\ \Eprint
  {http://arxiv.org/abs/1903.12334} {arXiv:1903.12334} \BibitemShut {NoStop}%
\bibitem [{\citenamefont {Kong}\ and\ \citenamefont
  {Zheng}(2018)}]{KZ170501087}%
  \BibitemOpen
  \bibfield  {author} {\bibinfo {author} {\bibfnamefont {L.}~\bibnamefont
  {Kong}}\ and\ \bibinfo {author} {\bibfnamefont {H.}~\bibnamefont {Zheng}},\
  }\href {\doibase 10.1016/j.nuclphysb.2017.12.007} {\bibfield  {journal}
  {\bibinfo  {journal} {Nucl. Phys. B}\ }\textbf {\bibinfo {volume} {927}},\
  \bibinfo {pages} {140} (\bibinfo {year} {2018})},\ \Eprint
  {http://arxiv.org/abs/1705.01087} {arXiv:1705.01087} \BibitemShut {NoStop}%
\bibitem [{\citenamefont {{Kong}}\ and\ \citenamefont
  {{Zheng}}(2020)}]{KZ190504924}%
  \BibitemOpen
  \bibfield  {author} {\bibinfo {author} {\bibfnamefont {L.}~\bibnamefont
  {{Kong}}}\ and\ \bibinfo {author} {\bibfnamefont {H.}~\bibnamefont
  {{Zheng}}},\ }\href {\doibase 10.1007/JHEP02(2020)150} {\bibfield  {journal}
  {\bibinfo  {journal} {J. High Energ. Phys.}\ }\textbf {\bibinfo {volume}
  {2020}},\ \bibinfo {pages} {150} (\bibinfo {year} {2020})},\ \Eprint
  {http://arxiv.org/abs/1905.04924} {arXiv:1905.04924} \BibitemShut {NoStop}%
\bibitem [{\citenamefont {{Kong}}\ and\ \citenamefont
  {{Zheng}}(2019)}]{KZ191201760}%
  \BibitemOpen
  \bibfield  {author} {\bibinfo {author} {\bibfnamefont {L.}~\bibnamefont
  {{Kong}}}\ and\ \bibinfo {author} {\bibfnamefont {H.}~\bibnamefont
  {{Zheng}}},\ }\href@noop {} {\  (\bibinfo {year} {2019})},\ \Eprint
  {http://arxiv.org/abs/1912.01760} {arXiv:1912.01760} \BibitemShut {NoStop}%
\bibitem [{\citenamefont {Dijkgraaf}\ and\ \citenamefont
  {Witten}(1990)}]{DW9093}%
  \BibitemOpen
  \bibfield  {author} {\bibinfo {author} {\bibfnamefont {R.}~\bibnamefont
  {Dijkgraaf}}\ and\ \bibinfo {author} {\bibfnamefont {E.}~\bibnamefont
  {Witten}},\ }\href {\doibase 10.1007/bf02096988} {\bibfield  {journal}
  {\bibinfo  {journal} {Commun.Math. Phys.}\ }\textbf {\bibinfo {volume}
  {129}},\ \bibinfo {pages} {393} (\bibinfo {year} {1990})}\BibitemShut
  {NoStop}%
\bibitem [{\citenamefont {'t~Hooft}(1980)}]{H8035}%
  \BibitemOpen
  \bibfield  {author} {\bibinfo {author} {\bibfnamefont {G.}~\bibnamefont
  {'t~Hooft}},\ }\href@noop {} {\bibfield  {journal} {\bibinfo  {journal} {NATO
  Adv. Study Inst. Ser. B Phys.}\ }\textbf {\bibinfo {volume} {59}},\ \bibinfo
  {pages} {135} (\bibinfo {year} {1980})}\BibitemShut {NoStop}%
\bibitem [{\citenamefont {Wen}(2013)}]{W1313}%
  \BibitemOpen
  \bibfield  {author} {\bibinfo {author} {\bibfnamefont {X.-G.}\ \bibnamefont
  {Wen}},\ }\href {\doibase 10.1103/physrevd.88.045013} {\bibfield  {journal}
  {\bibinfo  {journal} {Phys. Rev. D}\ }\textbf {\bibinfo {volume} {88}},\
  \bibinfo {pages} {045013} (\bibinfo {year} {2013})},\ \Eprint
  {http://arxiv.org/abs/1303.1803} {arXiv:1303.1803} \BibitemShut {NoStop}%
\bibitem [{\citenamefont {Freedman}\ \emph {et~al.}(2004)\citenamefont
  {Freedman}, \citenamefont {Nayak}, \citenamefont {Shtengel}, \citenamefont
  {Walker},\ and\ \citenamefont {Wang}}]{FNS0428}%
  \BibitemOpen
  \bibfield  {author} {\bibinfo {author} {\bibfnamefont {M.}~\bibnamefont
  {Freedman}}, \bibinfo {author} {\bibfnamefont {C.}~\bibnamefont {Nayak}},
  \bibinfo {author} {\bibfnamefont {K.}~\bibnamefont {Shtengel}}, \bibinfo
  {author} {\bibfnamefont {K.}~\bibnamefont {Walker}}, \ and\ \bibinfo {author}
  {\bibfnamefont {Z.}~\bibnamefont {Wang}},\ }\href {\doibase
  10.1016/j.aop.2004.01.006} {\bibfield  {journal} {\bibinfo  {journal} {Ann.
  Phys.}\ }\textbf {\bibinfo {volume} {310}},\ \bibinfo {pages} {428} (\bibinfo
  {year} {2004})},\ \Eprint {http://arxiv.org/abs/cond-mat/0307511}
  {arXiv:cond-mat/0307511} \BibitemShut {NoStop}%
\bibitem [{\citenamefont {Levin}\ and\ \citenamefont {Wen}(2005)}]{LW0510}%
  \BibitemOpen
  \bibfield  {author} {\bibinfo {author} {\bibfnamefont {M.~A.}\ \bibnamefont
  {Levin}}\ and\ \bibinfo {author} {\bibfnamefont {X.-G.}\ \bibnamefont
  {Wen}},\ }\href {\doibase 10.1103/physrevb.71.045110} {\bibfield  {journal}
  {\bibinfo  {journal} {Phys. Rev. B}\ }\textbf {\bibinfo {volume} {71}},\
  \bibinfo {pages} {045110} (\bibinfo {year} {2005})},\ \Eprint
  {http://arxiv.org/abs/cond-mat/0404617} {arXiv:cond-mat/0404617} \BibitemShut
  {NoStop}%
\bibitem [{\citenamefont {Chen}\ \emph {et~al.}(2011)\citenamefont {Chen},
  \citenamefont {Liu},\ and\ \citenamefont {Wen}}]{CLW1141}%
  \BibitemOpen
  \bibfield  {author} {\bibinfo {author} {\bibfnamefont {X.}~\bibnamefont
  {Chen}}, \bibinfo {author} {\bibfnamefont {Z.-X.}\ \bibnamefont {Liu}}, \
  and\ \bibinfo {author} {\bibfnamefont {X.-G.}\ \bibnamefont {Wen}},\ }\href
  {\doibase 10.1103/physrevb.84.235141} {\bibfield  {journal} {\bibinfo
  {journal} {Phys. Rev. B}\ }\textbf {\bibinfo {volume} {84}},\ \bibinfo
  {pages} {235141} (\bibinfo {year} {2011})},\ \Eprint
  {http://arxiv.org/abs/1106.4752} {arXiv:1106.4752} \BibitemShut {NoStop}%
\bibitem [{\citenamefont {Gukov}\ and\ \citenamefont
  {Kapustin}(2013)}]{Gukov:2013zka}%
  \BibitemOpen
  \bibfield  {author} {\bibinfo {author} {\bibfnamefont {S.}~\bibnamefont
  {Gukov}}\ and\ \bibinfo {author} {\bibfnamefont {A.}~\bibnamefont
  {Kapustin}},\ }\href@noop {} {\  (\bibinfo {year} {2013})},\ \Eprint
  {http://arxiv.org/abs/1307.4793} {arXiv:1307.4793 [hep-th]} \BibitemShut
  {NoStop}%
\bibitem [{\citenamefont {Gaiotto}\ and\ \citenamefont
  {Johnson-Freyd}(2019)}]{GJ190509566}%
  \BibitemOpen
  \bibfield  {author} {\bibinfo {author} {\bibfnamefont {D.}~\bibnamefont
  {Gaiotto}}\ and\ \bibinfo {author} {\bibfnamefont {T.}~\bibnamefont
  {Johnson-Freyd}},\ }\href {\doibase 10.1007/jhep05(2019)007} {\bibfield
  {journal} {\bibinfo  {journal} {J. High Energ. Phys.}\ }\textbf {\bibinfo
  {volume} {2019}},\ \bibinfo {pages} {7} (\bibinfo {year} {2019})},\ \Eprint
  {http://arxiv.org/abs/1905.09566} {arXiv:1905.09566} \BibitemShut {NoStop}%
\bibitem [{\citenamefont {{Johnson-Freyd}}(2020)}]{J200306663}%
  \BibitemOpen
  \bibfield  {author} {\bibinfo {author} {\bibfnamefont {T.}~\bibnamefont
  {{Johnson-Freyd}}},\ }\href@noop {} {\  (\bibinfo {year} {2020})},\ \Eprint
  {http://arxiv.org/abs/2003.06663} {arXiv:2003.06663} \BibitemShut {NoStop}%
\bibitem [{\citenamefont {Lan}\ \emph {et~al.}(2017)\citenamefont {Lan},
  \citenamefont {Kong},\ and\ \citenamefont {Wen}}]{LW160205946}%
  \BibitemOpen
  \bibfield  {author} {\bibinfo {author} {\bibfnamefont {T.}~\bibnamefont
  {Lan}}, \bibinfo {author} {\bibfnamefont {L.}~\bibnamefont {Kong}}, \ and\
  \bibinfo {author} {\bibfnamefont {X.-G.}\ \bibnamefont {Wen}},\ }\href
  {\doibase 10.1103/physrevb.95.235140} {\bibfield  {journal} {\bibinfo
  {journal} {Phys. Rev. B}\ }\textbf {\bibinfo {volume} {95}},\ \bibinfo
  {pages} {235140} (\bibinfo {year} {2017})},\ \Eprint
  {http://arxiv.org/abs/1602.05946} {arXiv:1602.05946} \BibitemShut {NoStop}%
\bibitem [{\citenamefont {Freed}\ and\ \citenamefont
  {Teleman}(2018)}]{Freed:2018cec}%
  \BibitemOpen
  \bibfield  {author} {\bibinfo {author} {\bibfnamefont {D.~S.}\ \bibnamefont
  {Freed}}\ and\ \bibinfo {author} {\bibfnamefont {C.}~\bibnamefont
  {Teleman}},\ }\href@noop {} {\  (\bibinfo {year} {2018})},\ \Eprint
  {http://arxiv.org/abs/1806.00008} {arXiv:1806.00008 [math.AT]} \BibitemShut
  {NoStop}%
\bibitem [{\citenamefont {Kong}\ \emph {et~al.}(2015)\citenamefont {Kong},
  \citenamefont {Wen},\ and\ \citenamefont {Zheng}}]{KZ150201690}%
  \BibitemOpen
  \bibfield  {author} {\bibinfo {author} {\bibfnamefont {L.}~\bibnamefont
  {Kong}}, \bibinfo {author} {\bibfnamefont {X.-G.}\ \bibnamefont {Wen}}, \
  and\ \bibinfo {author} {\bibfnamefont {H.}~\bibnamefont {Zheng}},\
  }\href@noop {} {\  (\bibinfo {year} {2015})},\ \Eprint
  {http://arxiv.org/abs/1502.01690} {arXiv:1502.01690} \BibitemShut {NoStop}%
\bibitem [{\citenamefont {Kong}\ \emph {et~al.}(2017)\citenamefont {Kong},
  \citenamefont {Wen},\ and\ \citenamefont {Zheng}}]{KZ170200673}%
  \BibitemOpen
  \bibfield  {author} {\bibinfo {author} {\bibfnamefont {L.}~\bibnamefont
  {Kong}}, \bibinfo {author} {\bibfnamefont {X.-G.}\ \bibnamefont {Wen}}, \
  and\ \bibinfo {author} {\bibfnamefont {H.}~\bibnamefont {Zheng}},\ }\href
  {\doibase 10.1016/j.nuclphysb.2017.06.023} {\bibfield  {journal} {\bibinfo
  {journal} {Nucl. Phys. B}\ }\textbf {\bibinfo {volume} {922}},\ \bibinfo
  {pages} {62} (\bibinfo {year} {2017})},\ \Eprint
  {http://arxiv.org/abs/1702.00673} {arXiv:1702.00673} \BibitemShut {NoStop}%
\bibitem [{\citenamefont {Read}\ and\ \citenamefont {Sachdev}(1991)}]{RS9173}%
  \BibitemOpen
  \bibfield  {author} {\bibinfo {author} {\bibfnamefont {N.}~\bibnamefont
  {Read}}\ and\ \bibinfo {author} {\bibfnamefont {S.}~\bibnamefont {Sachdev}},\
  }\href@noop {} {\bibfield  {journal} {\bibinfo  {journal} {Phys. Rev. Lett.}\
  }\textbf {\bibinfo {volume} {66}},\ \bibinfo {pages} {1773} (\bibinfo {year}
  {1991})}\BibitemShut {NoStop}%
\bibitem [{\citenamefont {Wen}(1991)}]{W9164}%
  \BibitemOpen
  \bibfield  {author} {\bibinfo {author} {\bibfnamefont {X.-G.}\ \bibnamefont
  {Wen}},\ }\href {\doibase 10.1103/physrevb.44.2664} {\bibfield  {journal}
  {\bibinfo  {journal} {Phys. Rev. B}\ }\textbf {\bibinfo {volume} {44}},\
  \bibinfo {pages} {2664} (\bibinfo {year} {1991})}\BibitemShut {NoStop}%
\bibitem [{\citenamefont {Gattringer}(2014)}]{Gattringer:2014nxa}%
  \BibitemOpen
  \bibfield  {author} {\bibinfo {author} {\bibfnamefont {C.}~\bibnamefont
  {Gattringer}},\ }\href {\doibase 10.22323/1.187.0002} {\bibfield  {journal}
  {\bibinfo  {journal} {PoS}\ }\textbf {\bibinfo {volume} {LATTICE2013}},\
  \bibinfo {pages} {002} (\bibinfo {year} {2014})},\ \Eprint
  {http://arxiv.org/abs/1401.7788} {arXiv:1401.7788 [hep-lat]} \BibitemShut
  {NoStop}%
\bibitem [{\citenamefont {{Baez}}\ and\ \citenamefont
  {{Schreiber}}(2007)}]{BSm0511710}%
  \BibitemOpen
  \bibfield  {author} {\bibinfo {author} {\bibfnamefont {J.~C.}\ \bibnamefont
  {{Baez}}}\ and\ \bibinfo {author} {\bibfnamefont {U.}~\bibnamefont
  {{Schreiber}}},\ }\href@noop {} {\bibfield  {journal} {\bibinfo  {journal}
  {Contemp. Math. 431, AMS, Providence, Rhode Island}\ ,\ \bibinfo {pages} {7}}
  (\bibinfo {year} {2007})},\ \Eprint {http://arxiv.org/abs/math/0511710}
  {math/0511710} \BibitemShut {NoStop}%
\bibitem [{\citenamefont {Girelli}\ \emph {et~al.}(2008)\citenamefont
  {Girelli}, \citenamefont {Pfeiffer},\ and\ \citenamefont
  {Popescu}}]{GP07083051}%
  \BibitemOpen
  \bibfield  {author} {\bibinfo {author} {\bibfnamefont {F.}~\bibnamefont
  {Girelli}}, \bibinfo {author} {\bibfnamefont {H.}~\bibnamefont {Pfeiffer}}, \
  and\ \bibinfo {author} {\bibfnamefont {E.~M.}\ \bibnamefont {Popescu}},\
  }\href {\doibase 10.1063/1.2888764} {\bibfield  {journal} {\bibinfo
  {journal} {J. Math. Phys.}\ }\textbf {\bibinfo {volume} {49}},\ \bibinfo
  {pages} {032503} (\bibinfo {year} {2008})},\ \Eprint
  {http://arxiv.org/abs/0708.3051} {arXiv:0708.3051} \BibitemShut {NoStop}%
\bibitem [{\citenamefont {Baez}\ and\ \citenamefont
  {Huerta}(2010)}]{BH10034485}%
  \BibitemOpen
  \bibfield  {author} {\bibinfo {author} {\bibfnamefont {J.~C.}\ \bibnamefont
  {Baez}}\ and\ \bibinfo {author} {\bibfnamefont {J.}~\bibnamefont {Huerta}},\
  }\href {\doibase 10.1007/s10714-010-1070-9} {\bibfield  {journal} {\bibinfo
  {journal} {Gen Relativ Gravit}\ }\textbf {\bibinfo {volume} {43}},\ \bibinfo
  {pages} {2335} (\bibinfo {year} {2010})},\ \Eprint
  {http://arxiv.org/abs/1003.4485} {arXiv:1003.4485} \BibitemShut {NoStop}%
\bibitem [{\citenamefont {Bullivant}\ \emph
  {et~al.}(2017{\natexlab{a}})\citenamefont {Bullivant}, \citenamefont
  {Calcada}, \citenamefont {K\'ad\'ar}, \citenamefont {Martin},\ and\
  \citenamefont {Martins}}]{BM160606639}%
  \BibitemOpen
  \bibfield  {author} {\bibinfo {author} {\bibfnamefont {A.}~\bibnamefont
  {Bullivant}}, \bibinfo {author} {\bibfnamefont {M.}~\bibnamefont {Calcada}},
  \bibinfo {author} {\bibfnamefont {Z.}~\bibnamefont {K\'ad\'ar}}, \bibinfo
  {author} {\bibfnamefont {P.}~\bibnamefont {Martin}}, \ and\ \bibinfo {author}
  {\bibfnamefont {J.~a.~F.}\ \bibnamefont {Martins}},\ }\href {\doibase
  10.1103/physrevb.95.155118} {\bibfield  {journal} {\bibinfo  {journal} {Phys.
  Rev. B}\ }\textbf {\bibinfo {volume} {95}},\ \bibinfo {pages} {155118}
  (\bibinfo {year} {2017}{\natexlab{a}})},\ \Eprint
  {http://arxiv.org/abs/1606.06639} {arXiv:1606.06639} \BibitemShut {NoStop}%
\bibitem [{\citenamefont {Bullivant}\ \emph
  {et~al.}(2017{\natexlab{b}})\citenamefont {Bullivant}, \citenamefont
  {Calcada}, \citenamefont {K\'ad\'ar}, \citenamefont {Martin},\ and\
  \citenamefont {Martins}}]{BM170200868}%
  \BibitemOpen
  \bibfield  {author} {\bibinfo {author} {\bibfnamefont {A.}~\bibnamefont
  {Bullivant}}, \bibinfo {author} {\bibfnamefont {M.}~\bibnamefont {Calcada}},
  \bibinfo {author} {\bibfnamefont {Z.}~\bibnamefont {K\'ad\'ar}}, \bibinfo
  {author} {\bibfnamefont {P.}~\bibnamefont {Martin}}, \ and\ \bibinfo {author}
  {\bibfnamefont {J.~F.}\ \bibnamefont {Martins}},\ }\href {\doibase
  10.1142/S0129055X20500117} {\bibfield  {journal} {\bibinfo  {journal}
  {Reviews in Mathematical Physics}\ } (\bibinfo {year} {2017}{\natexlab{b}}),\
  10.1142/S0129055X20500117},\ \Eprint {http://arxiv.org/abs/1702.00868}
  {arXiv:1702.00868} \BibitemShut {NoStop}%
\bibitem [{\citenamefont {Zhu}\ \emph {et~al.}(2019)\citenamefont {Zhu},
  \citenamefont {Lan},\ and\ \citenamefont {Wen}}]{ZW180809394}%
  \BibitemOpen
  \bibfield  {author} {\bibinfo {author} {\bibfnamefont {C.}~\bibnamefont
  {Zhu}}, \bibinfo {author} {\bibfnamefont {T.}~\bibnamefont {Lan}}, \ and\
  \bibinfo {author} {\bibfnamefont {X.-G.}\ \bibnamefont {Wen}},\ }\href
  {\doibase 10.1103/physrevb.100.045105} {\bibfield  {journal} {\bibinfo
  {journal} {Phys. Rev. B}\ }\textbf {\bibinfo {volume} {100}},\ \bibinfo
  {pages} {045105} (\bibinfo {year} {2019})},\ \Eprint
  {http://arxiv.org/abs/1808.09394} {arXiv:1808.09394} \BibitemShut {NoStop}%
\bibitem [{\citenamefont {Fradkin}\ and\ \citenamefont
  {Shenker}(1979)}]{FS7982}%
  \BibitemOpen
  \bibfield  {author} {\bibinfo {author} {\bibfnamefont {E.}~\bibnamefont
  {Fradkin}}\ and\ \bibinfo {author} {\bibfnamefont {S.~H.}\ \bibnamefont
  {Shenker}},\ }\href {\doibase 10.1103/physrevd.19.3682} {\bibfield  {journal}
  {\bibinfo  {journal} {Phys. Rev. D}\ }\textbf {\bibinfo {volume} {19}},\
  \bibinfo {pages} {3682} (\bibinfo {year} {1979})}\BibitemShut {NoStop}%
\bibitem [{\citenamefont {Creutz}\ \emph {et~al.}(1979)\citenamefont {Creutz},
  \citenamefont {Jacobs},\ and\ \citenamefont {Rebbi}}]{CJR7915}%
  \BibitemOpen
  \bibfield  {author} {\bibinfo {author} {\bibfnamefont {M.}~\bibnamefont
  {Creutz}}, \bibinfo {author} {\bibfnamefont {L.}~\bibnamefont {Jacobs}}, \
  and\ \bibinfo {author} {\bibfnamefont {C.}~\bibnamefont {Rebbi}},\ }\href
  {\doibase 10.1103/PhysRevD.20.1915} {\bibfield  {journal} {\bibinfo
  {journal} {Phys. Rev. D}\ }\textbf {\bibinfo {volume} {20}},\ \bibinfo
  {pages} {1915} (\bibinfo {year} {1979})}\BibitemShut {NoStop}%
\bibitem [{\citenamefont {Wen}(2015)}]{W150605768}%
  \BibitemOpen
  \bibfield  {author} {\bibinfo {author} {\bibfnamefont {X.-G.}\ \bibnamefont
  {Wen}},\ }\href {\doibase 10.1093/nsr/nwv077} {\bibfield  {journal} {\bibinfo
   {journal} {Nat. Sci. Rev.}\ }\textbf {\bibinfo {volume} {3}},\ \bibinfo
  {pages} {68} (\bibinfo {year} {2015})},\ \Eprint
  {http://arxiv.org/abs/1506.05768} {arXiv:1506.05768} \BibitemShut {NoStop}%
\bibitem [{\citenamefont {Moore}\ and\ \citenamefont {Seiberg}(1989)}]{MS8916}%
  \BibitemOpen
  \bibfield  {author} {\bibinfo {author} {\bibfnamefont {G.}~\bibnamefont
  {Moore}}\ and\ \bibinfo {author} {\bibfnamefont {N.}~\bibnamefont
  {Seiberg}},\ }\href {\doibase https://doi.org/10.1016/0550-3213(89)90511-7}
  {\bibfield  {journal} {\bibinfo  {journal} {Nuclear Physics B}\ }\textbf
  {\bibinfo {volume} {313}},\ \bibinfo {pages} {16 } (\bibinfo {year}
  {1989})}\BibitemShut {NoStop}%
\bibitem [{\citenamefont {Barkeshli}\ \emph {et~al.}(2019)\citenamefont
  {Barkeshli}, \citenamefont {Bonderson}, \citenamefont {Cheng},\ and\
  \citenamefont {Wang}}]{Barkeshli19}%
  \BibitemOpen
  \bibfield  {author} {\bibinfo {author} {\bibfnamefont {M.}~\bibnamefont
  {Barkeshli}}, \bibinfo {author} {\bibfnamefont {P.}~\bibnamefont
  {Bonderson}}, \bibinfo {author} {\bibfnamefont {M.}~\bibnamefont {Cheng}}, \
  and\ \bibinfo {author} {\bibfnamefont {Z.}~\bibnamefont {Wang}},\ }\href
  {\doibase 10.1103/PhysRevB.100.115147} {\bibfield  {journal} {\bibinfo
  {journal} {Phys. Rev. B}\ }\textbf {\bibinfo {volume} {100}},\ \bibinfo
  {pages} {115147} (\bibinfo {year} {2019})}\BibitemShut {NoStop}%
\bibitem [{\citenamefont {Heinrich}\ \emph {et~al.}(2016)\citenamefont
  {Heinrich}, \citenamefont {Burnell}, \citenamefont {Fidkowski},\ and\
  \citenamefont {Levin}}]{HL160607816}%
  \BibitemOpen
  \bibfield  {author} {\bibinfo {author} {\bibfnamefont {C.}~\bibnamefont
  {Heinrich}}, \bibinfo {author} {\bibfnamefont {F.}~\bibnamefont {Burnell}},
  \bibinfo {author} {\bibfnamefont {L.}~\bibnamefont {Fidkowski}}, \ and\
  \bibinfo {author} {\bibfnamefont {M.}~\bibnamefont {Levin}},\ }\href
  {\doibase 10.1103/physrevb.94.235136} {\bibfield  {journal} {\bibinfo
  {journal} {Phys. Rev. B}\ }\textbf {\bibinfo {volume} {94}},\ \bibinfo
  {pages} {235136} (\bibinfo {year} {2016})},\ \Eprint
  {http://arxiv.org/abs/1606.07816} {arXiv:1606.07816} \BibitemShut {NoStop}%
\bibitem [{\citenamefont {Fuchs}\ \emph {et~al.}(2002)\citenamefont {Fuchs},
  \citenamefont {Runkel},\ and\ \citenamefont {Schweigert}}]{FSh0110133}%
  \BibitemOpen
  \bibfield  {author} {\bibinfo {author} {\bibfnamefont {J.}~\bibnamefont
  {Fuchs}}, \bibinfo {author} {\bibfnamefont {I.}~\bibnamefont {Runkel}}, \
  and\ \bibinfo {author} {\bibfnamefont {C.}~\bibnamefont {Schweigert}},\
  }\href {\doibase 10.1016/S0550-3213(01)00638-1} {\bibfield  {journal}
  {\bibinfo  {journal} {Nucl. Phys. B}\ }\textbf {\bibinfo {volume} {624}},\
  \bibinfo {pages} {452} (\bibinfo {year} {2002})},\ \Eprint
  {http://arxiv.org/abs/hep-th/0110133} {arXiv:hep-th/0110133} \BibitemShut
  {NoStop}%
\bibitem [{\citenamefont {{Fuchs}}\ \emph {et~al.}(2002)\citenamefont
  {{Fuchs}}, \citenamefont {{Runkel}},\ and\ \citenamefont
  {{Schweigert}}}]{FSh0204148}%
  \BibitemOpen
  \bibfield  {author} {\bibinfo {author} {\bibfnamefont {J.}~\bibnamefont
  {{Fuchs}}}, \bibinfo {author} {\bibfnamefont {I.}~\bibnamefont {{Runkel}}}, \
  and\ \bibinfo {author} {\bibfnamefont {C.}~\bibnamefont {{Schweigert}}},\
  }\href {\doibase 10.1016/S0550-3213(02)00744-7} {\bibfield  {journal}
  {\bibinfo  {journal} {Nuclear Physics B}\ }\textbf {\bibinfo {volume}
  {646}},\ \bibinfo {pages} {353} (\bibinfo {year} {2002})},\ \Eprint
  {http://arxiv.org/abs/hep-th/0204148} {arXiv:hep-th/0204148} \BibitemShut
  {NoStop}%
\bibitem [{\citenamefont {{Fuchs}}\ \emph
  {et~al.}(2004{\natexlab{a}})\citenamefont {{Fuchs}}, \citenamefont
  {{Runkel}},\ and\ \citenamefont {{Schweigert}}}]{FSh0306164}%
  \BibitemOpen
  \bibfield  {author} {\bibinfo {author} {\bibfnamefont {J.}~\bibnamefont
  {{Fuchs}}}, \bibinfo {author} {\bibfnamefont {I.}~\bibnamefont {{Runkel}}}, \
  and\ \bibinfo {author} {\bibfnamefont {C.}~\bibnamefont {{Schweigert}}},\
  }\href {\doibase 10.1016/j.nuclphysb.2003.11.026} {\bibfield  {journal}
  {\bibinfo  {journal} {Nuclear Physics B}\ }\textbf {\bibinfo {volume}
  {678}},\ \bibinfo {pages} {511} (\bibinfo {year} {2004}{\natexlab{a}})},\
  \Eprint {http://arxiv.org/abs/hep-th/0306164} {arXiv:hep-th/0306164}
  \BibitemShut {NoStop}%
\bibitem [{\citenamefont {{Fuchs}}\ \emph
  {et~al.}(2004{\natexlab{b}})\citenamefont {{Fuchs}}, \citenamefont
  {{Runkel}},\ and\ \citenamefont {{Schweigert}}}]{FSh0403157}%
  \BibitemOpen
  \bibfield  {author} {\bibinfo {author} {\bibfnamefont {J.}~\bibnamefont
  {{Fuchs}}}, \bibinfo {author} {\bibfnamefont {I.}~\bibnamefont {{Runkel}}}, \
  and\ \bibinfo {author} {\bibfnamefont {C.}~\bibnamefont {{Schweigert}}},\
  }\href {\doibase 10.1016/j.nuclphysb.2004.05.014} {\bibfield  {journal}
  {\bibinfo  {journal} {Nuclear Physics B}\ }\textbf {\bibinfo {volume}
  {694}},\ \bibinfo {pages} {277} (\bibinfo {year} {2004}{\natexlab{b}})},\
  \Eprint {http://arxiv.org/abs/hep-th/0403157} {arXiv:hep-th/0403157}
  \BibitemShut {NoStop}%
\bibitem [{\citenamefont {{Fuchs}}\ \emph {et~al.}(2005)\citenamefont
  {{Fuchs}}, \citenamefont {{Runkel}},\ and\ \citenamefont
  {{Schweigert}}}]{FSh0412290}%
  \BibitemOpen
  \bibfield  {author} {\bibinfo {author} {\bibfnamefont {J.}~\bibnamefont
  {{Fuchs}}}, \bibinfo {author} {\bibfnamefont {I.}~\bibnamefont {{Runkel}}}, \
  and\ \bibinfo {author} {\bibfnamefont {C.}~\bibnamefont {{Schweigert}}},\
  }\href {\doibase 10.1016/j.nuclphysb.2005.03.018} {\bibfield  {journal}
  {\bibinfo  {journal} {Nuclear Physics B}\ }\textbf {\bibinfo {volume}
  {715}},\ \bibinfo {pages} {539} (\bibinfo {year} {2005})},\ \Eprint
  {http://arxiv.org/abs/hep-th/0412290} {arXiv:hep-th/0412290} \BibitemShut
  {NoStop}%
\bibitem [{\citenamefont {{Fjelstad}}\ \emph {et~al.}(2006)\citenamefont
  {{Fjelstad}}, \citenamefont {{Fuchs}}, \citenamefont {{Runkel}},\ and\
  \citenamefont {{Schweigert}}}]{FSh0503194}%
  \BibitemOpen
  \bibfield  {author} {\bibinfo {author} {\bibfnamefont {J.}~\bibnamefont
  {{Fjelstad}}}, \bibinfo {author} {\bibfnamefont {J.}~\bibnamefont {{Fuchs}}},
  \bibinfo {author} {\bibfnamefont {I.}~\bibnamefont {{Runkel}}}, \ and\
  \bibinfo {author} {\bibfnamefont {C.}~\bibnamefont {{Schweigert}}},\
  }\href@noop {} {\bibfield  {journal} {\bibinfo  {journal} {Theor. Appl.
  Categor.}\ }\textbf {\bibinfo {volume} {16}},\ \bibinfo {pages} {342}
  (\bibinfo {year} {2006})},\ \Eprint {http://arxiv.org/abs/hep-th/0503194}
  {arXiv:hep-th/0503194} \BibitemShut {NoStop}%
\bibitem [{\citenamefont {{Runkel}}\ \emph {et~al.}(2007)\citenamefont
  {{Runkel}}, \citenamefont {{Fjelstad}}, \citenamefont {{Fuchs}},\ and\
  \citenamefont {{Schweigert}}}]{RSm0512076}%
  \BibitemOpen
  \bibfield  {author} {\bibinfo {author} {\bibfnamefont {I.}~\bibnamefont
  {{Runkel}}}, \bibinfo {author} {\bibfnamefont {J.}~\bibnamefont
  {{Fjelstad}}}, \bibinfo {author} {\bibfnamefont {J.}~\bibnamefont {{Fuchs}}},
  \ and\ \bibinfo {author} {\bibfnamefont {C.}~\bibnamefont {{Schweigert}}},\
  }\href@noop {} {\bibfield  {journal} {\bibinfo  {journal} {Contemp. Math.}\
  }\textbf {\bibinfo {volume} {431}},\ \bibinfo {pages} {225} (\bibinfo {year}
  {2007})},\ \Eprint {http://arxiv.org/abs/math/0512076} {arXiv:math/0512076}
  \BibitemShut {NoStop}%
\bibitem [{\citenamefont {Kong}\ and\ \citenamefont
  {Runkel}(2009)}]{KR08073356}%
  \BibitemOpen
  \bibfield  {author} {\bibinfo {author} {\bibfnamefont {L.}~\bibnamefont
  {Kong}}\ and\ \bibinfo {author} {\bibfnamefont {I.}~\bibnamefont {Runkel}},\
  }\href {\doibase 10.1007/s00220-009-0901-6} {\bibfield  {journal} {\bibinfo
  {journal} {Commun. Math. Phys.}\ }\textbf {\bibinfo {volume} {292}},\
  \bibinfo {pages} {871} (\bibinfo {year} {2009})},\ \Eprint
  {http://arxiv.org/abs/0807.3356} {arXiv:0807.3356} \BibitemShut {NoStop}%
\bibitem [{\citenamefont {{Kong}}(2011)}]{K11073649}%
  \BibitemOpen
  \bibfield  {author} {\bibinfo {author} {\bibfnamefont {L.}~\bibnamefont
  {{Kong}}},\ }\href@noop {} {\  (\bibinfo {year} {2011})},\ \Eprint
  {http://arxiv.org/abs/1107.3649} {arXiv:1107.3649} \BibitemShut {NoStop}%
\bibitem [{\citenamefont {{Kong}}\ \emph {et~al.}(2013)\citenamefont {{Kong}},
  \citenamefont {{Li}},\ and\ \citenamefont {{Runkel}}}]{KR13101875}%
  \BibitemOpen
  \bibfield  {author} {\bibinfo {author} {\bibfnamefont {L.}~\bibnamefont
  {{Kong}}}, \bibinfo {author} {\bibfnamefont {Q.}~\bibnamefont {{Li}}}, \ and\
  \bibinfo {author} {\bibfnamefont {I.}~\bibnamefont {{Runkel}}},\ }\href@noop
  {} {\  (\bibinfo {year} {2013})},\ \Eprint {http://arxiv.org/abs/1310.1875}
  {arXiv:1310.1875} \BibitemShut {NoStop}%
\bibitem [{\citenamefont {{Kapustin}}\ and\ \citenamefont
  {{Saulina}}(2010)}]{KS10120911}%
  \BibitemOpen
  \bibfield  {author} {\bibinfo {author} {\bibfnamefont {A.}~\bibnamefont
  {{Kapustin}}}\ and\ \bibinfo {author} {\bibfnamefont {N.}~\bibnamefont
  {{Saulina}}},\ }\href@noop {} {\  (\bibinfo {year} {2010})},\ \Eprint
  {http://arxiv.org/abs/1012.0911} {arXiv:1012.0911} \BibitemShut {NoStop}%
\end{thebibliography}%

\end{document}